\documentclass[aps,prd,superscriptaddress,twocolumn,floatfix,showpacs]{revtex4}
\newcommand{\beq}{\begin{equation}}
\newcommand{\eeq}{\end{equation}}
\newcommand{\beqn}{\begin{eqnarray}}
\newcommand{\eeqn}{\end{eqnarray}}
\newcommand{\cB}{{\cal{B}}}
\usepackage{graphicx}
\usepackage{epsf}
\begin{document}
\title{General relativistic simulations of magnetized binary neutron 
star mergers}

\author{Yuk Tung Liu}
\affiliation{Department of Physics, University of Illinois at
  Urbana-Champaign, Urbana, IL 61801}

\author{Stuart~L.~Shapiro}
\altaffiliation{Also at Department of Astronomy and NCSA, University of
  Illinois at Urbana-Champaign, Urbana, IL 61801}
\affiliation{Department of Physics, University of Illinois at
  Urbana-Champaign, Urbana, IL 61801}

\author{Zachariah B. Etienne}
\affiliation{Department of Physics, University of Illinois at
  Urbana-Champaign, Urbana, IL 61801}

\author{Keisuke Taniguchi}
\affiliation{Department of Physics, University of Illinois at
  Urbana-Champaign, Urbana, IL 61801}

\begin{abstract}
Binary neutron stars (NSNS) are expected to be among the
leading sources of gravitational waves observable by ground-based
laser interferometers and may be the progenitors of short-hard 
gamma ray bursts. We present a series of general relativistic
NSNS coalescence simulations both for unmagnetized and magnetized stars.
We adopt quasiequilibrium initial data for circular, irrotational  
binaries constructed in the conformal thin-sandwich (CTS) 
framework. We adopt the BSSN 
formulation for evolving the metric and a high-resolution shock-capturing 
scheme to handle the magnetohydrodynamics. Our simulations of 
unmagnetized binaries agree with the results of 
Shibata, Taniguchi and 
Ury\=u~\cite{stu03}. 
In cases in which the mergers result in a prompt collapse to a 
black hole, we are able to use puncture gauge conditions to extend the 
evolution and determine the mass of the material that forms a disk. We find 
that the disk mass is less than 2\% of the total mass in all cases studied. 
We then add a small poloidal magnetic field to the initial configurations 
and study the subsequent evolution. 
For cases in which the remnant is a hypermassive neutron star, we see 
measurable differences in both the amplitude and phase of the gravitational  
waveforms following the merger. For cases in which the remnant is a black hole 
surrounded by a disk, the disk mass and the 
gravitational waveforms are about the same as the unmagnetized cases. 
Magnetic fields substantially affect the long-term, secular 
evolution of a hypermassive neutron star (driving `delayed collapse') 
and an accretion disk around a nascent black hole.
\end{abstract}


\maketitle

\section{Introduction}

There is great interest in studying the effects of magnetic fields 
in relativistic astrophysics. Magnetic fields are always present in 
astrophysical plasmas, which are usually highly conducting. 
Even if the initial magnetic field is small, 
it can be amplified via magnetic winding and other magnetic 
instabilities (see, e.g.,~\cite{spruit99,bh98} for reviews). 
Neutron stars (NSs) have the strongest observed magnetic fields 
(up to $\sim 10^{15}$G) among astrophysical objects~\cite{kou98}. 
The strong magnetic fields result from 
gravitational collapse, which amplifies the magnetic fields in the 
core of the progenitor, and from various dynamo processes after 
the collapse (see, e.g., \cite{spruit08} for a review). Strong 
magnetic fields in a NS may trigger observable events such as 
pulsar glitches and the emission of large bursts of gamma rays and X-rays 
as in a soft gamma-ray repeater. Magnetic fields in a binary neutron star 
(NSNS) system may also influence the dynamics of the remnant after 
the NSs merge.

Mergers of binary neutron stars (NSNSs) are expected to be among the
leading sources of gravitational waves observable by ground-based
laser interferometers. Observations of short-hard gamma-ray bursts 
(GRBs) suggest that a substantial fraction of them may be formed from 
mergers of NSNSs or mergers of neutron star-black hole binaries (BHNSs). 
Many theoretical models of GRB engines consist of magnetized accretion 
disks around a spinning black hole~\cite{macfadyen99,king05,piran05}. 
General relativistic magnetohydrodynamics (GRMHD) 
is necessary to model such systems. 

The first two GRMHD codes capable of evolving the 
GRMHD equations in {\it dynamical} spacetimes were 
developed by Duez et al.~\cite{dlss05} (hereafter DLSS) and 
Shibata \& Sekiguchi~\cite{ss05} (hereafter SS). These codes 
are based on the BSSN (Baumgarte-Shapiro-Shibata-Nakamura) 
scheme to integrate the Einstein field equations, a high-resolution 
shock-capturing (HRSC) scheme to integrate the MHD and induction 
equations, and a constrained transport scheme to enforce the 
``no-monopole'' magnetic constraint.
Subsequently, Giacomazzo \& Rezzolla~\cite{whiskyMHD} and Anderson 
et al.~\cite{ahln06} developed similar codes. Our code (DLSS)
and the code of SS have been used to study 
magnetic fields in 
hypermassive neutron stars~\cite{dlsss06a,sdlss06,dlsss06b}, 
magnetorotational collapse of massive stellar cores to 
neutron stars~\cite{slss06}, and (unmagnetized) coalescing 
BHNSs~\cite{SU1,SU2,ST07,eflstb07}. We have also used our code 
to study the magnetorotational collapse of massive stellar cores to 
black holes~\cite{lss07} as well as coalescing BHBHs~\cite{efls07}, 
which are pure vacuum simulations. Shibata et al.\ have also 
performed simulations of (unmagnetized) coalescing 
NSNSs~\cite{su00,su02,stu03,stu05,st06}. 

Recently, Anderson et al.\ have used their code to study the coalescence of 
both unmagnetized and magnetized NSNSs~\cite{ahllmnpt08a,ahllmnpt08b}. 
In the unmagnetized cases, they find an initial configuration 
in~\cite{ahllmnpt08a} that leads 
to prompt collapse to a black hole following the merger. 
The total (baryon) rest mass of their initial configuration is 
$M_0 \approx 1.03 M_0^{\rm (TOV)}$~\cite{fn}, where $M_0^{\rm (TOV)}$ is the 
maximum rest mass of a nonrotating neutron star for the $n$=1 
polytropic equation of state (EOS) adopted in their simulation. 
This result seems to contradict the earlier finding of Shibata, 
Taniguchi \& Ury\=u~\cite{stu03} that prompt black hole 
formation is possible for this EOS only if $M_0 \gtrsim 1.7 M_0^{\rm (TOV)}$. 
We note, however, that the initial data used by Shibata et al.\ 
and Anderson et al.\ are different. Shibata et al.\ use 
quasiequilibrium initial data for binaries in nearly circular 
orbits constructed using the conformal 
thin-sandwich (CTS) method. In contrast, Anderson et al.\ set 
up the initial data by superposing boosted metrics of 
two spherical neutron stars. Anderson et al.'s initial data results in 
an orbital eccentricity of about 0.2~\cite{ahllmnpt08a}, 
whereas the eccentricities of the CTS initial data are $\lesssim 
0.015$ according to a post-Newtonian analysis~\cite{BIW}. 
We point out that a quasi-circular orbit is more realistic
because gravitational radiation would have circularized the 
orbit long before the binary separation reaches a few 
NS radii. Also, a NS in a quasiequilibrium circular orbit will be 
distorted by tidal effects.

Anderson et al.\ report in~\cite{ahllmnpt08b} that a NSNS 
with the same EOS and masses as in~\cite{ahllmnpt08a} 
but with different initial separation leads to a hypermassive 
neutron star after the merger. Later, the star undergoes 
`delayed collapse' to 
a black hole due to gravitational radiation. This finding is also 
surprising. The total rest mass $M_0 \approx 1.03 M_0^{\rm (TOV)}$ 
is smaller than the maximum mass of a uniformly rotating star (the 
supramassive limit), $M_0^{\rm (sup)} = 1.15M_0^{\rm (TOV)}$~\cite{cst94}. 
Hence the remnant cannot be a hypermassive NS, i.e.\ a NS whose mass 
exceeds the supramassive limit. As a result, the 
star will be unstable to gravitational collapse only if a very large 
amount of angular momentum is removed. A priori, the expected outcome 
is that gravitational radiation removes only some of the angular momentum, 
the remnant acquires some differential rotation, 
and the star settles down to a stable, stationary, rotating configuration.

To better understand the coalescence of NSNSs and the role of 
magnetic fields, we perform a new series of simulations using our 
DLSS code. In this paper, we consider 
three models, using the same CTS initial data as in~\cite{stu03}. 
Specifically, we study the irrotational binary models M1414, M1616 and M1418 
in~\cite{stu03}. In models M1414 and M1616, the NSs 
are of equal mass. In model M1418, the ratio of the rest masses 
of the two NSs are $q$=0.855.

We first repeat the calculations of Shibata, Taniguchui \& 
Ury\=u~\cite{stu03} for unmagnetized 
NSNS mergers. Our results agree with those reported in~\cite{stu03}. 
Model M1414 results in a dynamically stable, differentially rotating 
hypermassive NS. 
For models M1616 and M1418, the mergers lead to prompt collapse 
to a black hole. The simulations in~\cite{stu03} are terminated 
soon after black hole formation because of grid stretching. 
We are able to follow the evolution using 
puncture gauge conditions (see, e.g.~\cite{God1,RIT1}) until the system settles 
down to a quasi-equilibrium state. This allows us to 
estimate the disk mass around the black hole more accurately, and 
our results are again consistent with those estimated in~\cite{stu03}. 
We next consider the magnetized cases. We add a 
poloidal field with strength $B\sim 10^{16}$G inside each NS of the 
three NSNS models and follow the evolution. While such interior field 
strength may be larger than the value expected for a typical NS, it is 
comparable to the strength inferred for magnetars~\cite{dt92} 
and is high enough to demonstrate the dynamical effects of a magnetic field, 
if any. For model M1414, the merger again
results in a differentially rotating, hypermassive 
neutron star. We see observable difference in 
the magnetized case after the merger as magnetic fields are 
amplified. For model M1616, the system collapses to a black hole 
after the merger as before, leaving negligible 
amount of material to form a disk. This result is unaffected by the presence 
of the magnetic field. For model M1418, about 1.5\% of rest mass 
is left to form a disk for both magnetized and unmagnetized cases. 
Gravitational waveforms for models M1616 and M1418 show only a slight 
difference in amplitude during the entire simulations. This is because 
the remnants quickly collapse to a black hole after the merger and 
magnetic fields do not have enough time to amplify and alter the dynamics 
of the fluid substantially. 
This result is not surprising since the ratio of magnetic to gravitational 
potential energy is $E_M/|W|\sim 10^{-4}$ initially, 
and hence the magnetic fields 
are not expected to have an impact on the dynamics before they are 
amplified.

For models M1414 and M1418, magnetic fields are expected to affect 
the {\it long-term} secular evolution of the remnants. For the cases where the 
remnant is a hypermassive neutron star (M1414), magnetic fields 
are crucial for driving `delayed collapse' of the 
star~\cite{dlsss06a,dlsss06b}, and the resulting remnant 
could be a central engine for a short-hard GRB~\cite{sdlss06}. 
The effect of a magnetic field may be diminished whenever the 
merged hypermassive remnants develop a bar~\cite{stu05,st06}.
The bar leads to dissipation of angular 
momentum by gravitational radiation and may result in delayed 
collapse on a timescale faster than that of magnetic field amplification. 
However, a bar does not develop for model M1414. 
In general, the development of a bar depends on the NS EOS. 
For cases in which the remnant consists of a black hole surrounded 
by a disk (M1418), magnetic fields may produce turbulence in the disk 
via MHD instabilities and may generate ultra-relativistic 
jets~\cite{mg04,dvhk03,dvhkh05,dvso05,mk06,mnkhf06,kskm02}. 
In this paper, we are primarily interested in studying the 
effect of the magnetic field 
during the late inspiral, merging and the early post-merger phases, so 
we do not follow the long-term evolution of the remnants. 
We have previously studied the long-term secular evolution 
of magnetized hypermassive NSs in~\cite{dlsss06a,dlsss06b} 
and the evolution of magnetized disks around rotating black holes 
in~\cite{ssl08}.

This paper is organized as follows.  In Sec~\ref{sec:basic_eqns}, 
we briefly summarize the basic equations and their
specific implementation in our GRMHD scheme. 
In Sec.~\ref{sec:results}, we present the results of 
our simulations and compare them with those in~\cite{stu03}. 
We summarize our results in Sec.~\ref{sec:discussion} 
and comment on future directions.

\section{Formulation}
\label{sec:basic_eqns}

\subsection{Basic equations and numerical methods}

The formulation and numerical scheme for our GRMHD simulations are
the same as those reported in~\cite{dlss05,eflstb07}, to which the reader
may refer for details. Here we briefly summarize the
method and introduce our notation. We adopt geometrized units 
($G = c = 1$) except where stated explicitly.

We use the 3+1 formulation of general relativity and decompose
the metric into the following form:
\beq
  ds^2 = -\alpha^2 dt^2
+ \gamma_{ij} (dx^i + \beta^i dt) (dx^j + \beta^j dt) \ .
\eeq
The fundamental variables for the metric evolution are the spatial
three-metric $\gamma_{ij}$ and extrinsic curvature $K_{ij}$. We adopt
the BSSN
formalism~\cite{BS,SN} to evolve $\gamma_{ij}$ and $K_{ij}$. In this
formalism, the evolution variables are the conformal exponent $\phi
\equiv \ln \gamma/12$, the conformal 3-metric $\tilde
\gamma_{ij}=e^{-4\phi}\gamma_{ij}$, three auxiliary functions
$\tilde{\Gamma}^i \equiv -\tilde \gamma^{ij}{}_{,j}$, the trace of
the extrinsic curvature $K$, and the tracefree part of the conformal extrinsic
curvature $\tilde A_{ij} \equiv e^{-4\phi}(K_{ij}-\gamma_{ij} K/3)$.
Here, $\gamma={\rm det}(\gamma_{ij})$. The full spacetime metric $g_{\mu \nu}$
is related to the three-metric $\gamma_{\mu \nu}$ by $\gamma_{\mu \nu}
= g_{\mu \nu} + n_{\mu} n_{\nu}$, where the future-directed, timelike
unit vector $n^{\mu}$ normal to the time slice can be written in terms
of the lapse $\alpha$ and shift $\beta^i$ as $n^{\mu} = \alpha^{-1}
(1,-\beta^i)$. As for the gauge conditions, we adopt an advective 
``1+log'' slicing condition for the lapse and a second-order 
``non-shifting-shift''~\cite{RIT1,GodGauge} as in our BHNS 
simulations~\cite{eflstb07}. 

The BSSN equations
are evolved with fourth-order accurate spatial differencing and
upwinding on the shift advection terms. We apply Sommerfeld outgoing 
wave boundary conditions to all BSSN fields, as in~\cite{eflstb07}. 
Our code is embedded in the
Cactus parallelization framework~\cite{Cactus}, whereby our
second-order iterated Crank-Nicholson time-stepping is managed by the
{\tt MoL}, or method of lines, thorn. We use the moving puncture 
technique to handle any black hole that may form after the merger 
of the NSNS. The apparent horizon of the black hole is computed using the {\tt
ahfinderdirect} Cactus thorn~\cite{ahfinderdirect}.
Before an apparent horizon appears, we find that adding a 
Hamiltonian constraint term to the evolution equation of $\phi$ 
as in~\cite{dmsb03} leads to smaller constraint violation 
during the evolution. However, when a black hole appears, 
we remove this term as it sometimes leads to unstable evolution.

The fundamental variables in ideal MHD are the rest-mass density
$\rho_0$, specific internal energy $\epsilon$, pressure $P$, four-velocity
$u^{\mu}$, and magnetic field $B^{\mu}$ measured by a normal
observer moving with a 4-velocity $n^{\mu}$ (note that $B^{\mu} n_{\mu}=0$).
The ideal MHD condition is written as $u_{\mu} F^{\mu\nu}=0$,
where $F^{\mu\nu}$ is the electromagnetic (Faraday) tensor. The tensor
$F^{\mu\nu}$ and its dual in the ideal MHD approximation are
given by
\beqn
&&F^{\mu\nu}=\epsilon^{\mu\nu\alpha\beta}u_{\alpha}b_{\beta}, \label{eqFF}\\
&&F^*_{\mu\nu} \equiv {1 \over 2}\epsilon_{\mu\nu\alpha\beta} F^{\alpha\beta}
=b_{\mu} u_{\nu}- b_{\nu} u_{\mu},
\eeqn
where $\epsilon_{\mu\nu\alpha\beta}$ is the Levi-Civita tensor.
Here we have introduced an auxiliary magnetic 4-vector
$b^{\mu}=B^{\mu}_{(u)}/\sqrt{4\pi}$, where $B^{\mu}_{(u)}$ is the
magnetic field measured by an observer comoving with the fluid and
is related to $B^{\mu}$ by
\beq
  B^{\mu}_{(u)} = -\frac{(\delta^{\mu}{}_{\nu} + u^{\mu} u_{\nu}) B^{\nu}}
  {n_{\lambda}u^{\lambda}} \ .
\eeq

The energy-momentum tensor is written as
\beqn
T_{\mu\nu}=T_{\mu\nu}^{\rm Fluid} + T_{\mu\nu}^{\rm EM},
\eeqn
where $T_{\mu\nu}^{\rm Fluid}$ and $T_{\mu\nu}^{\rm EM}$ denote the
fluid and electromagnetic contributions to the stress-energy
tensor. They are given by
\beq
T_{\mu\nu}^{\rm Fluid}= \rho_0 h u_{\mu} u_{\nu} + P g_{\mu\nu}, 
\eeq
and 
\beqn
&&T_{\mu\nu}^{\rm EM}= \frac{1}{4\pi} \left(
F_{\mu\sigma} F^{~\sigma}_{\nu}-{1 \over 4}g_{\mu\nu}
F_{\alpha\beta} F^{\alpha\beta} \right) \nonumber \\
&&~~~~~~=\biggl({1 \over 2}g_{\mu\nu}+u_{\mu}u_{\nu}\biggr)b^2
-b_{\mu}b_{\nu},
\eeqn
where $h\equiv 1+\epsilon+P/\rho_0$ is the specific enthalpy, and
$b^2\equiv b^{\mu}b_{\mu}$. Hence, the total stress-energy tensor becomes
\beq
  T_{\mu\nu}= (\rho_0 h + b^2) u_{\mu} u_{\nu} + \left( P + \frac{b^2}{2}
\right) g_{\mu\nu} - b_{\mu} b_{\nu} \ .
\label{eq:mhdTab}
\eeq

In our numerical implementation of the GRMHD and magnetic
induction equations,
we evolve the densitized density $\rho_*$,
densitized momentum density $\tilde{S}_i$,
densitized energy density $\tilde{\tau}$,
and densitized magnetic field $\cB^i$. They are defined as
\beqn
&&\rho_* \equiv - \sqrt{\gamma}\, \rho_0 n_{\mu} u^{\mu},
\label{eq:rhos} \\
&& \tilde{S}_i \equiv -  \sqrt{\gamma}\, T_{\mu \nu}n^{\mu} \gamma^{\nu}_{~i}, \\
&& \tilde{\tau} \equiv  \sqrt{\gamma}\, T_{\mu \nu}n^{\mu} n^{\nu} - \rho_*,
\label{eq:S0} \\
&& \cB^i \equiv  \sqrt{\gamma}\, B^i.
\eeqn
During the evolution, we also need the three-velocity $v^i = u^i/u^t$.

The MHD and induction equations are written in conservative form for
variables $\rho_*$, $\tilde{S}_i$, $\tilde{\tau}$, and $\cB^i$ and
evolved using an HRSC scheme. Specifically, we
use the monotonized central (MC) scheme~\cite{vL77} for data reconstruction
and the HLL (Harten, Lax and van-Leer) scheme~\cite{HLL} to compute
the flux. The
magnetic field $\cB^i$ has to satisfy the ``no monopole'' constraint
$\partial_i \cB^i=0$. We adopt the flux-interpolated constrained
transport (flux-CT) scheme~\cite{t00,gmt03} to impose this constraint. 
In this scheme, the induction
equation is differenced in such a way that a second order, corner-centered
representation of the divergence is preserved as a numerical
identity. 

At each timestep, the hydrodynamic 
``primitive'' variables $(\rho_0,P,v^i)$ must be computed from the 
``conservative'' 
variables $(\rho_*,\tilde{\tau},\tilde{S}_i)$. This is done by
numerically solving the
algebraic equations~(\ref{eq:rhos})--(\ref{eq:S0}) together with an
EOS $P=P(\rho_0,\epsilon)$. In this paper, we adopt a $\Gamma$-law 
EOS, $P=(\Gamma-1)\rho_0 \epsilon$, with $\Gamma=2$.

As in many hydrodynamic simulations in
astrophysics, we add a tenuous ``atmosphere'' that covers the
computational grid outside the star. The atmospheric rest-mass density
is set to $10^{-10} \rho_{\rm max}(0)$, 
where $\rho_{\rm max}(0)$ is the initial maximum rest-mass 
density of the stars. As in~\cite{dlsss06b},
we apply outer boundary conditions on the primitive variables $\rho,P,v^i,$
and $B^i$.  Outflow boundary conditions are imposed on the hydrodynamic
variables, and 
the magnetic field is linearly extrapolated onto the boundaries. Finally,
the evolution variables $\rho_*$, $\tilde{S}_i$, and $\tilde{\tau}$ \
are recomputed on the boundary. 

Our GRMHD code (DLSS) has been thoroughly tested by passing a robust 
suite of tests. These tests include
maintaining stable rotating stars in stationary equilibrium, reproducing
the exact Oppenheimer-Snyder solution for collapse to a black hole, 
and reproducing analytic solutions for MHD shocks, nonlinear
MHD wave propagation, magnetized Bondi accretion, and MHD waves induced
by linear gravitational waves~\cite{dlss05}. Our DLSS code has also been
compared with SS's GRMHD code~\cite{ss05} by
performing identical simulations of the evolution of magnetized hypermassive
NSs~\cite{dlsss06a,dlsss06b} and of magnetorotational 
collapse of stellar
cores~\cite{slss06}. We obtain good agreement between these two independent
codes. Our code has also been used to study the collapse of very 
massive, magnetized, rotating stars to black hole~\cite{lss07}, 
evolution of merging BHBH~\cite{junk-ID} and
BHNS binaries~\cite{eflstb07}, and the
evolution of relativistic hydrodynamic matter in the presence of
puncture black holes~\cite{fbest07}. Recently, our code has been 
generalized to incorporate (optically thick) radiation transport 
and its feedback on hydrodynamic matter~\cite{flls08}.

\subsection{Initial data}

We adopt the same irrotational, quasi-circular NSNS initial data 
as in~\cite{stu03}. These initial data set were generated by 
Taniguchi \& Gourgoulhon~\cite{tg02,tg03} 
by numerically solving the constraint equations of general relativity 
in the CTS framework. We consider three models studied in~\cite{stu03}:  
M1414, M1616 and M1418. 

\begin{table*}
\caption{Irrotational, quasiequilibrium NSNS models in circular orbit. 
Here $(M_*/R)_\infty$ 
is the neutron star compaction, $\bar{\rho}_0^{\max}$ is the maximum 
nondimensional rest-mass density of a neutron star, 
$\bar{M}_0$ is the nondimensional 
total rest mass of the binary, $\bar{M}$ is the nondimensional 
ADM mass of the system, $J$ is the ADM angular momentum, 
$q=M_0^{(1)}/M_0^{(2)}$ is the ratio of the rest masses of the stars, 
$\Omega_0$ is the quasi-circular orbital angular velocity, and 
$P_0=2\pi/\Omega_0$ is the orbital period.}
\label{tab:models}
\begin{tabular}{ccccccccc}
\hline
 Model & $(M_*/R)_\infty$ & $\bar{\rho}_0^{\max}$ & $\bar{M}_0$ & 
 $\bar{M}$ & $J/M^2$ & $q$ & $M\Omega_0$ & $P_0/M$ \\ 
\hline
M1414 & 0.14, 0.14 & 0.118, 0.118 & 0.292 & 0.269 & 0.951 & 1.00  & 0.0326 & 193 \\
M1616 & 0.16, 0.16 & 0.151, 0.151 & 0.320 & 0.292 & 0.914 & 1.00  & 0.0395 & 158 \\
M1418 & 0.14, 0.18 & 0.118, 0.195 & 0.317 & 0.290 & 0.933 & 0.855 & 0.0345 & 182 \\
\hline
\end{tabular}
\end{table*}

All models assume an $n$=1 polytropic EOS for the neutron stars: 
$P$=$\kappa \rho_0^2$. The compaction, $(M_*/R)_\infty$,
is defined as the ratio of the ADM (Arnowitt-Deser-Misner) mass $M_*$ to
the areal radius $R$ of a spherical neutron star in isolation. For an 
$n$=1 polytropic EOS, the compaction uniquely specifies the neutron star. 
We thus label the NSNS models by the compaction of each neutron star.  
Model M1418 means the compactions of the two neutron stars 
are 0.14 and 0.18. Hence the two neutron stars do not have the same rest 
masses.  For models M1414 and M1616, the two neutron stars are of equal 
rest mass and their compactions are 0.14 (for model M1414) 
and 0.16 (for 
model M1616). It is convenient to rescale all quantities
with respect to $\kappa$. Since $\kappa^{1/2}$ has dimensions of length,
we can define the nondimensional variables~\cite{cst92} 
$\bar{M}$=$\kappa^{-1/2} M$, $\bar{R}$=$\kappa^{-1/2} R$, 
and $\bar{\rho}_0$=$\kappa \rho_0$. Here $M$ is the ADM mass of the 
binary. The relationship between these  
nondimensional variables and quantities in cgs units are 
\beqn
  M &=& 10 M_\odot \left( \frac{\kappa}{1.455\times 10^5 {\rm cgs}}
\right)^{1/2} \bar{M} \\
  R &=& 14.8{\rm km} \left( \frac{\kappa}{1.455\times 10^5 {\rm cgs}}
\right)^{1/2} \bar{R} \\
 \rho_0 &=& 6.18\times 10^{15} {\rm g}{\rm cm}^{-3} 
\left( \frac{\kappa}{1.455\times 10^5 {\rm cgs}}\right) \bar{\rho}_0 \ ,
\eeqn
where the value $\kappa$=$1.455\times 10^5$cgs is used by~\cite{font02}.
The maximum rest mass for a spherical neutron star 
for this EOS is $\bar{M}_0^{\rm (TOV)}$=0.180, and the maximum ADM mass 
is $\bar{M}_*^{\max}$=0.164. Table~\ref{tab:models} 
summarizes the characteristics of our models.

\begin{table*}
\caption{Parameters and results of various runs. The label HNS stands 
for ``hypermassive neutron star,'' BH stands for ``black hole,'' 
$M_0^{\rm disk}$ is the rest mass of the disk around the black hole, 
and $J_H/M_H^2$ is the spin parameter of the black hole.}
\label{tab:runs}
\begin{tabular}{cccccccc}
\hline
\hline
 Run & Model & B-field & $n_p$ & $P_{\rm cut}/P_{\rm max}$ & 
Result & $M_0^{\rm disk}/M_0$ & $J_H/M_H^2$ \\
\hline
 M1414B0 & M1414 & no & -- & -- & HNS & -- & -- \\
 M1414B1 &       & yes & 0 & 0.04 & HNS & -- & -- \\
 M1616B0 & M1616 & no & -- & -- & BH & $<10^{-6}$ & $\approx$0.85 \\
 M1616B1 &       & yes & 0 & 0.04 & BH & $< 10^{-4}$ & $\approx$0.85 \\
 M1616B2 &       & yes & 3 & 0.001 & BH & $< 2\times 10^{-4}$ & $\approx$0.85 \\ 
 M1418B0 & M1418 & no & -- & -- & BH & $\approx$ 0.013 & $\approx$0.8 \\
 M1418B1 &       & yes & 0 & 0.04 & BH & $\lesssim 0.018$ & $\approx$0.8 \\
\hline
\end{tabular}
\end{table*}

\begin{figure}
\vspace{-4mm}
\begin{center}
\epsfxsize=6cm
\leavevmode
\epsffile{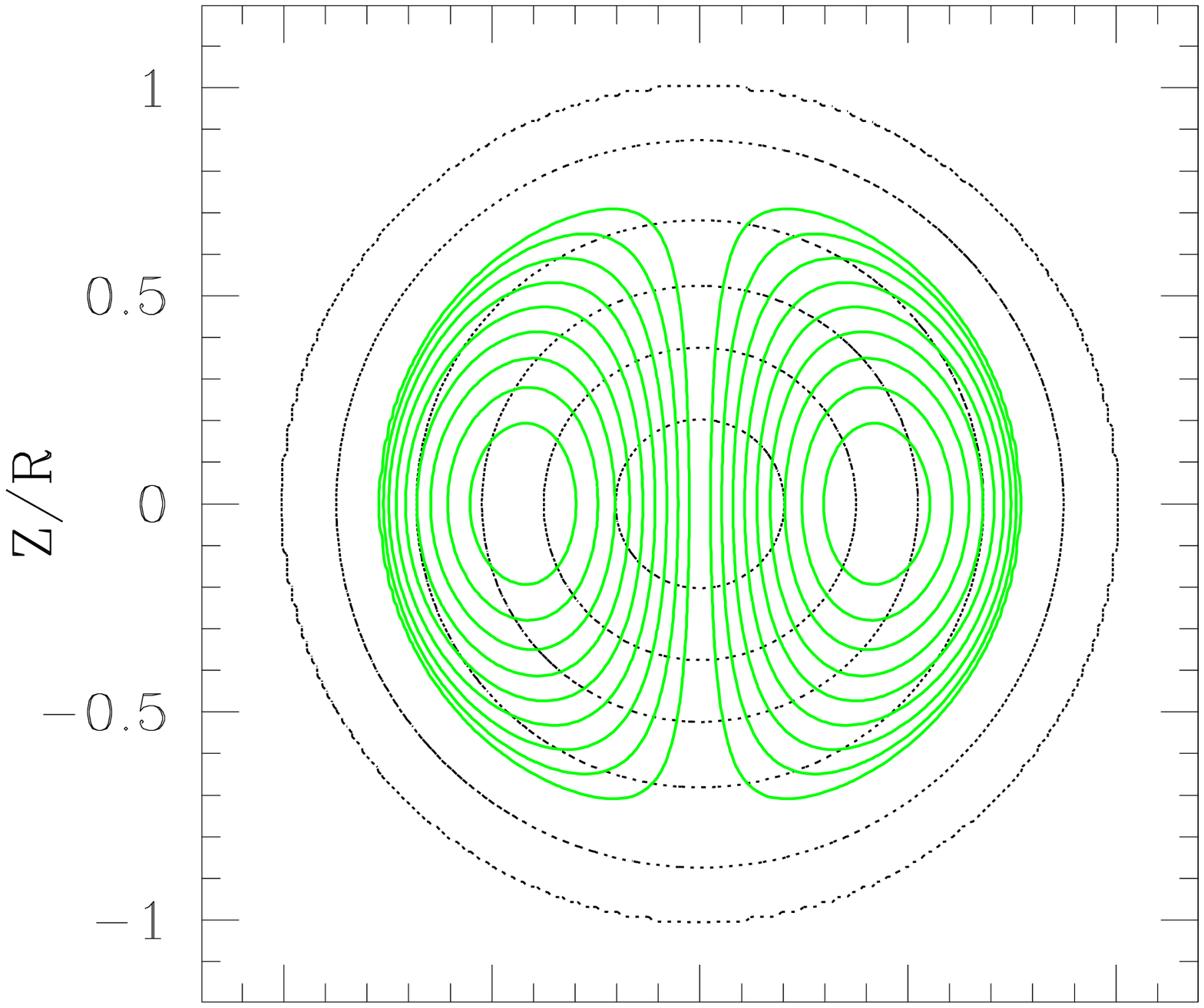}\\
\vspace{-3mm}
\epsfxsize=6cm
\leavevmode
\epsffile{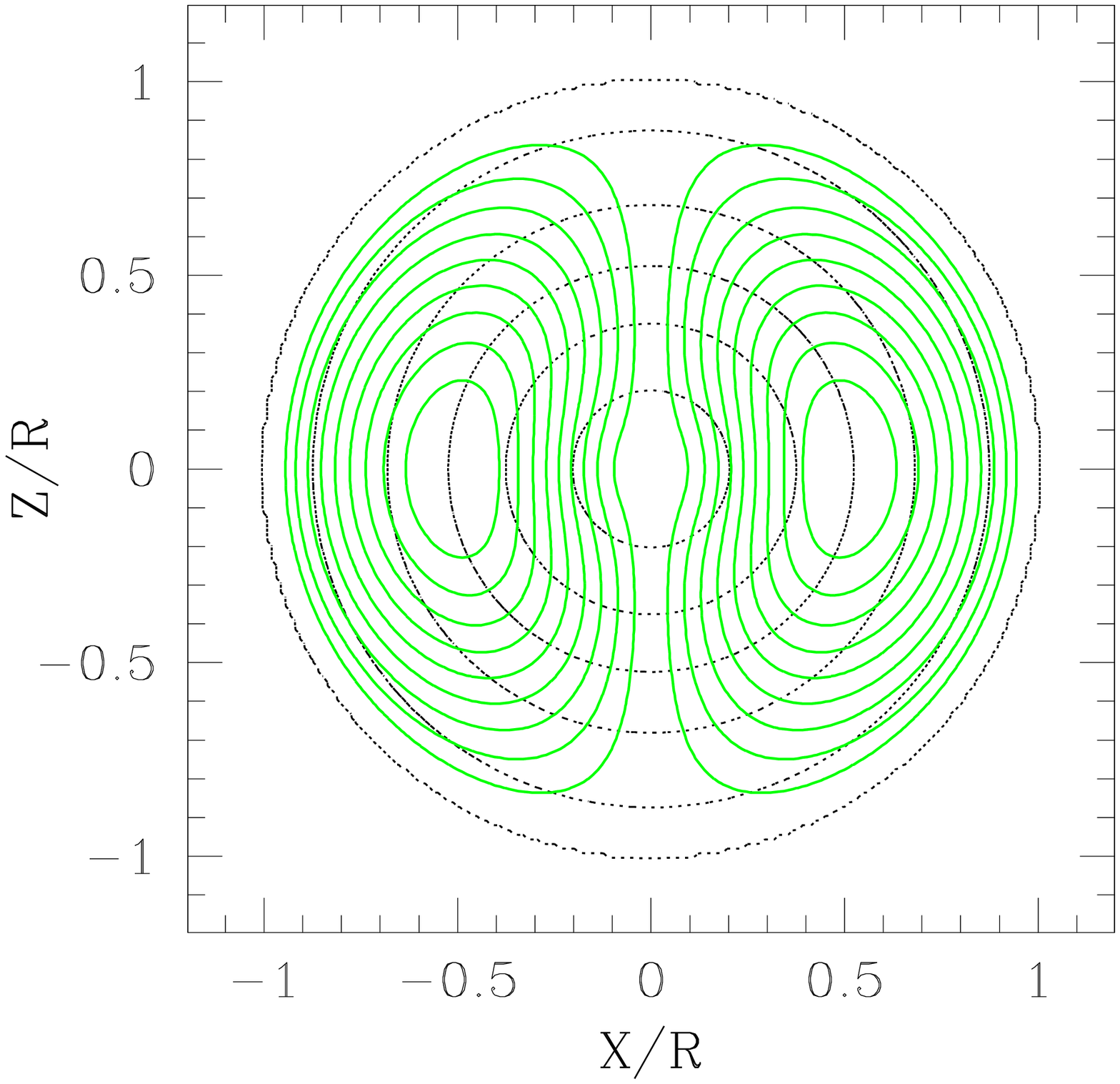}
\caption{Magnetic field configurations in a widely separated, 
spherical NS companion constructed from 
the vector potential in Eq.~(\ref{eq:Aphi}) for $n_p$=0 and 
$P_{\rm cut}$=0.04$P_{\rm max}$ (upper panel), and for $n_p$=3 and 
$P_{\rm cut}$=0.001$P_{\rm max}$ (lower panel). Here $P_{\rm max}$ 
is the maximum pressure. The compaction of this NS is $M_*/R$=0.14. 
Dotted (black) concentric circles are rest-mass density contours 
drawn for $\rho_0/\rho_0^{\rm max}$= 0.9, 0.7, 0.5, 0.3, 0.1 and 0.001. 
Solid (green) lines are contours of $A_\varphi$, which coincide 
with the magnetic field lines in axisymmetry.
Contour levels of $A_{\rm \varphi}$ are drawn for 
$A_\varphi=(A_\varphi^{\rm max} - A_\varphi^{\rm min}) (i/10)^2
+A_\varphi^{\rm min}$ with $i$=1, 2, ... 9, 
where $A_\varphi^{\rm max}$ and $A_\varphi^{\rm min}$ 
are the maximum and minimum value of $A_\varphi$, respectively.}
\label{fig:B-lines}
\end{center}
\end{figure}

To study the effects of magnetic fields, we add a small poloidal magnetic 
field to the quasi-equilibrium model. We orient our coordinates so that 
the initial maximum densities of the two neutron stars are located 
at $(x_1,0,0)$ and 
$(x_2,0,0)$ with $x_1<0$ and $x_2>0$. For each neutron star $l$, we specify 
the magnetic vector potential 
\beqn
  A_x^{(l)} &=& -(y/\varpi_l^2) A_{\varphi}^{(l)} \ , \ \ \ 
  A_y^{(l)} =  (x/\varpi_l^2) A_{\varphi}^{(l)} \ , \\
  A_z^{(l)} &=& 0 \ , \ \ \ (l=1,2) 
\eeqn
where $\varpi_l \equiv \sqrt{(x-x_l)^2+y^2}$ and we set 
\beq
  A_{\varphi}^{(l)} = A_b \varpi_l^2 \left( 1-\frac{\rho_0}{\rho_0^{\max}} 
\right)^{n_p} \max(P-P_{\rm cut},0) \ .
\label{eq:Aphi}
\eeq
Here $A_b$, $n_p$ and $P_{\rm cut}$ are free parameters. The magnetic 
field in the $l$-th neutron star is then computed by 
$B^i_{(l)} = n_\mu \epsilon^{\mu ijk} \partial_j A_k^{(l)}$. This guarantees 
that the magnetic constraint $\partial_j \cB^j=0$ is automatically 
satisfied. The parameter $A_b$ determines the strength of the B-field. 
The cutoff pressure parameter $P_{\rm cut}$ confines the B-field 
inside the neutron star to reside 
within $P>P_{\rm cut}$. The parameter $n_p$ shifts the location of the maximum 
B-field in the star. A larger value of $n_p$ results in the maximum B-field 
located in the lower density region of the star. In this paper, we set 
$P_{\rm cut}$ 
to be 4\% or 0.1\% of the maximum pressure, $n_p$=0 or 3, and set $A_b$ so that 
the volume-averaged magnetic field is $10^{16}{\rm G} (M_0/2.8M_\odot)$, 
where $M_0$ is the total rest mass of the stars. There is no initial 
exterior magnetic field. Figure~\ref{fig:B-lines} shows the initial magnetic 
field configurations for a widely separated NS companion with 
compaction $M_*/R$=0.14. 
The magnetic field lines, as well as density distribution, for each NS in 
our binary at $t=0$ are slightly distorted from those shown in 
Fig.~\ref{fig:B-lines} due to tidal effects.
Table~\ref{tab:runs} lists the simulations performed in this paper with 
a brief summary of the outcomes. We note that Anderson et 
al.~\cite{ahllmnpt08b} also set up the magnetic fields in a very similar 
way. Their setup corresponds to setting the parameters 
$n_p$=0, $P_{\rm cut}$ to the atmosphere value, and 
$A_b$ such that the maximum B-field strength in each NS is 
9.6$\times 10^{15}$G$(1.78M_\odot/M)$.

We note that the interior magnetic field strength of $10^{16}$G 
is large compared to values inferred for the surface of a typical 
pulsar ($B\sim 10^{12}$G), but it is comparable to 
the strength in a magnetar~\cite{dt92}. We find that for this B-field strength, 
the magnetic pressure, $P_{\rm mag}=b^2/2$, is about 0.1\% of the 
gas pressure, and the 
total magnetic energy is $\sim 10^{-5}$ of the ADM mass. We note that 
the accuracy of the ADM mass of our CTS initial data is also of order 
$10^{-5}$~\cite{tg02}. Hence, adding a B-field of this strength 
induces negligible constraint violation and causes only a small perturbation 
to the equilibrium of the stars and their orbit. 
In addition, the magnetic field profile inside a neutron star is not 
known. Our profile is to make the ratio $P_{\rm mag}/P$ small
initially in most regions. As a result, the magnetic field introduces 
only a small perturbation to the fluid and no ``magnetic wind'' is generated 
in the low-density regions of the star. 

In nature, magnetic fields are not confined inside the NS, but extend 
out to the NS exterior. The exterior fields of the two NSs in the binary will 
interact and modify the dynamics during the inspiral phase. This effect 
has been studied analytically in Newtonian and post-Newtonian 
calculations in~\cite{it00,hun03,hun05}. As a rough estimate, 
we approximate the NS by a sphere of radius $R$. Consider a  
pure dipole exterior field and a nearly uniform interior field 
alligned with the orbital angular momentum. The magnetic dipole moment 
$\mu$ is related to the interior field strength $B$ by $\mu\approx BR^3/2$. 
The accumulated gravitational wave cycles during the inspiral phase 
from gravitational wave frquencies $f_{\rm min}$ to $f_{\rm max}$
due to magnetic dipole interaction is estimated to be~\cite{it00} 
\beq
  \delta N_{\rm mag} \sim \left. -\frac{25}{64\pi} \frac{B_1 B_2 R_1^3 R_2^3}
{\eta^2 M^4}(\pi Mf)^{-1/3} \right|_{f_{\rm min}}^{f_{\rm max}} \ ,
\eeq
where $\eta=M_1 M_2/(M_1+M_2)^2$ is the symmetric 
mass ratio of the two NSs. To estimate the upper bound starting from 
our initial data,  
we set $f_{\rm max}=\infty$ and $f_{\rm min}$ to twice of 
the initial orbital frequency of the binaries in our models.
We find $\delta N_{\rm mag} < 0.02$ for all three models. 
Hence the effect of magnetic dipole interaction during the 
inspiral phase is negligible, even assuming our large adopted 
field strength. For this reason, any appreciable dynamical effects of 
the magnetic fields will occur only during and after the merger phase, 
for which our confined field model will be adequate.

\subsection{Grid setup}

Even though the dynamics of the system is mainly concentrated in the 
central region with radius $r \lesssim 10M$, we set our computational 
grid to $r \approx 50M $ in order to extract gravitational radiation. 
To reduce computational resources, we employ fisheye 
coordinates~\cite{fisheye,RIT2} 
to allocate the grid more effectively. Fisheye coordinates 
$\bar{x}^i$ are related to the
original coordinates $x^i$ through the following transformation:
\beqn
  x^i &=& \frac{\bar{x}^i}{\bar{r}} r(\bar{r}) ,
\label{eq:fisheyet1}  \\
  r(\bar{r}) &=& a \bar{r} + \frac{(1-a)s}{2\tanh(\bar{r}_{0}/s)} \ln
\frac{\cosh [(\bar{r}+\bar{r}_{0})/s]}{\cosh [(\bar{r}-\bar{r}_{0})/s]},
\label{eq:fisheyet2}  
\eeqn
where $r=\sqrt{x^2+y^2+z^2}$, $\bar{r} = \sqrt{\bar{x}^2 + \bar{y}^2
+ \bar{z}^2}$. The quantities $a$, $\bar{r}_{0}$ and $s$ are constant 
parameters, which are set to $a=3$, $\kappa^{-1/2} \bar{r}_{0}$=2.4, and 
$\kappa^{-1/2} s$=0.6 in all of our simulations. With this choice, the 
grid spacing in the region $\kappa^{-1/2} r >2.4$ is increased 
by a factor of 3.

We use a cell-centered grid with size $2N\times 2N\times N$ in 
$\bar{x}$-$\bar{y}$-$\bar{z}$ (assuming equatorial symmetry), 
covering a computational domain $\bar{x}\in (-N\Delta,N\Delta)$, 
$\bar{y}\in (-N\Delta,N\Delta)$, and $\bar{z}\in (0,N\Delta)$. 
Here $N$ is an integer, $\Delta$ is the grid spacing and 
$\bar{z}$ is the rotation axis. 
We set $N=150$, $\kappa^{-1/2} \Delta=0.04$ for models M1414 and M1616, and 
$N=200$, $\kappa^{-1/2} \Delta=0.03$ for models M1418. These values of $\Delta$ 
are chosen so that the diameter of each neutron star in the equatorial plane 
is covered by $\gtrsim 40$ grid points. This is the same resolution used 
in~\cite{stu03}. We have performed a simulation with 75\% of the grid spacing 
(but with closer outer boundary) for the unmagnetized M1414 case and find that 
the result is very close to our standard resolution.

\section{Results}
\label{sec:results}

\subsection{Model M1414}

\begin{figure}
\includegraphics[width=8cm]{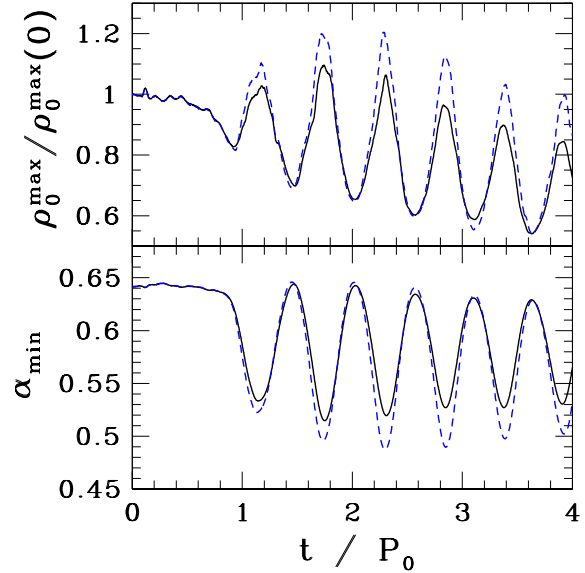}
\caption{Evolution of maximum density $\rho_0^{\max}$ and minimum 
lapse $\alpha_{\rm min}$ for unmagnetized (black solid line) and magnetized 
(blue dash line) runs of model M1414. The time $t$ is normalized by 
the initial orbital period $P_0$=193$M$=$1.3\times 10^{-5}s\, 
(M_0/2.8M_\odot)$. The initial maximum rest-mass density is 
$\rho_0^{\max}(0)$=$0.118/\kappa$=$7.9\times 10^{14}{\rm g}\, {\rm cm}^{-3}
(2.8M_\odot /M_0)^2$. The merger occurs at $t\approx 1P_0$.}
\label{fig:M1414_rho_alp}
\end{figure}

\begin{figure*}
\begin{center}
\epsfxsize=8cm
\leavevmode
\epsffile{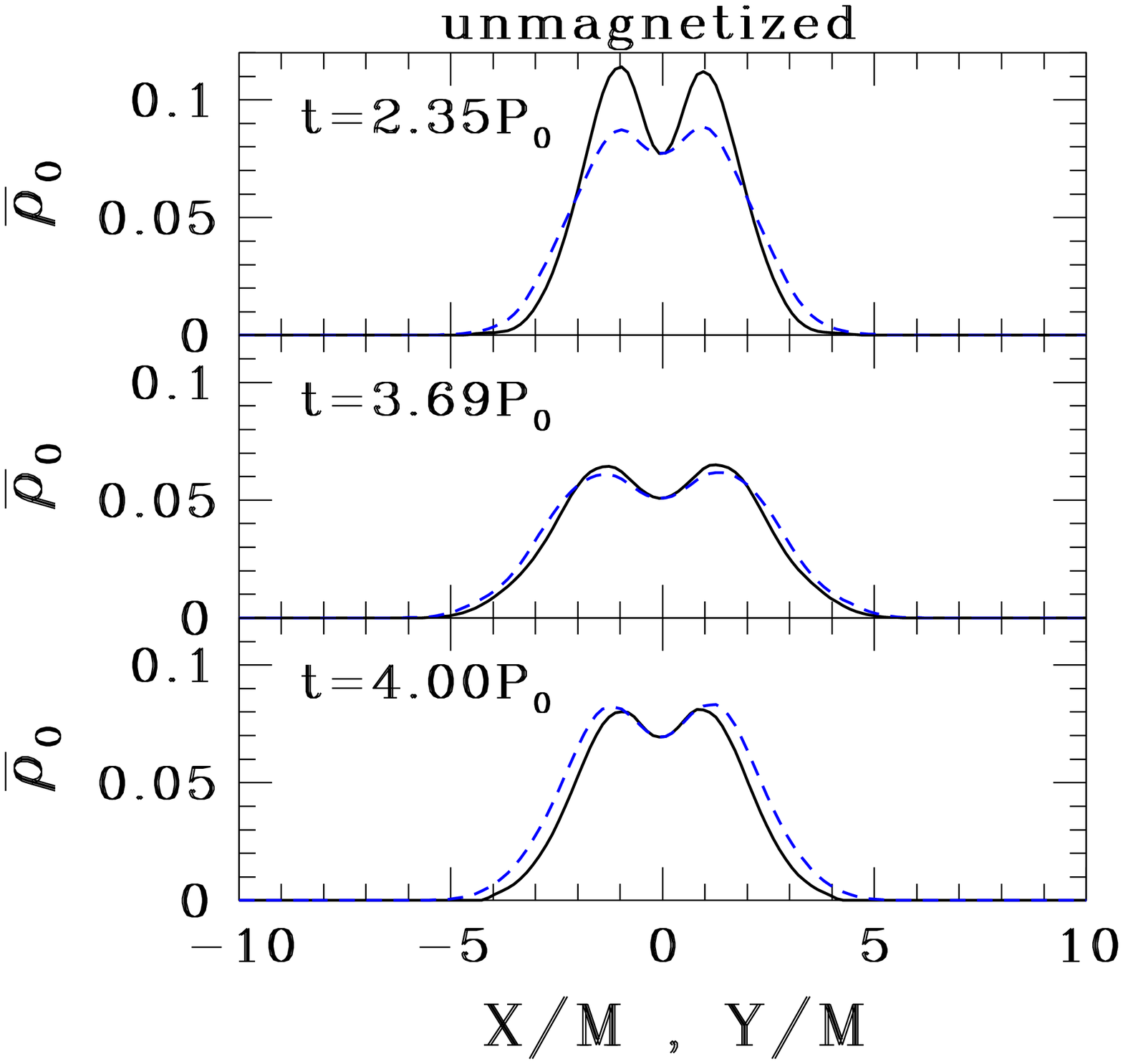}
\epsfxsize=8cm
\leavevmode
\epsffile{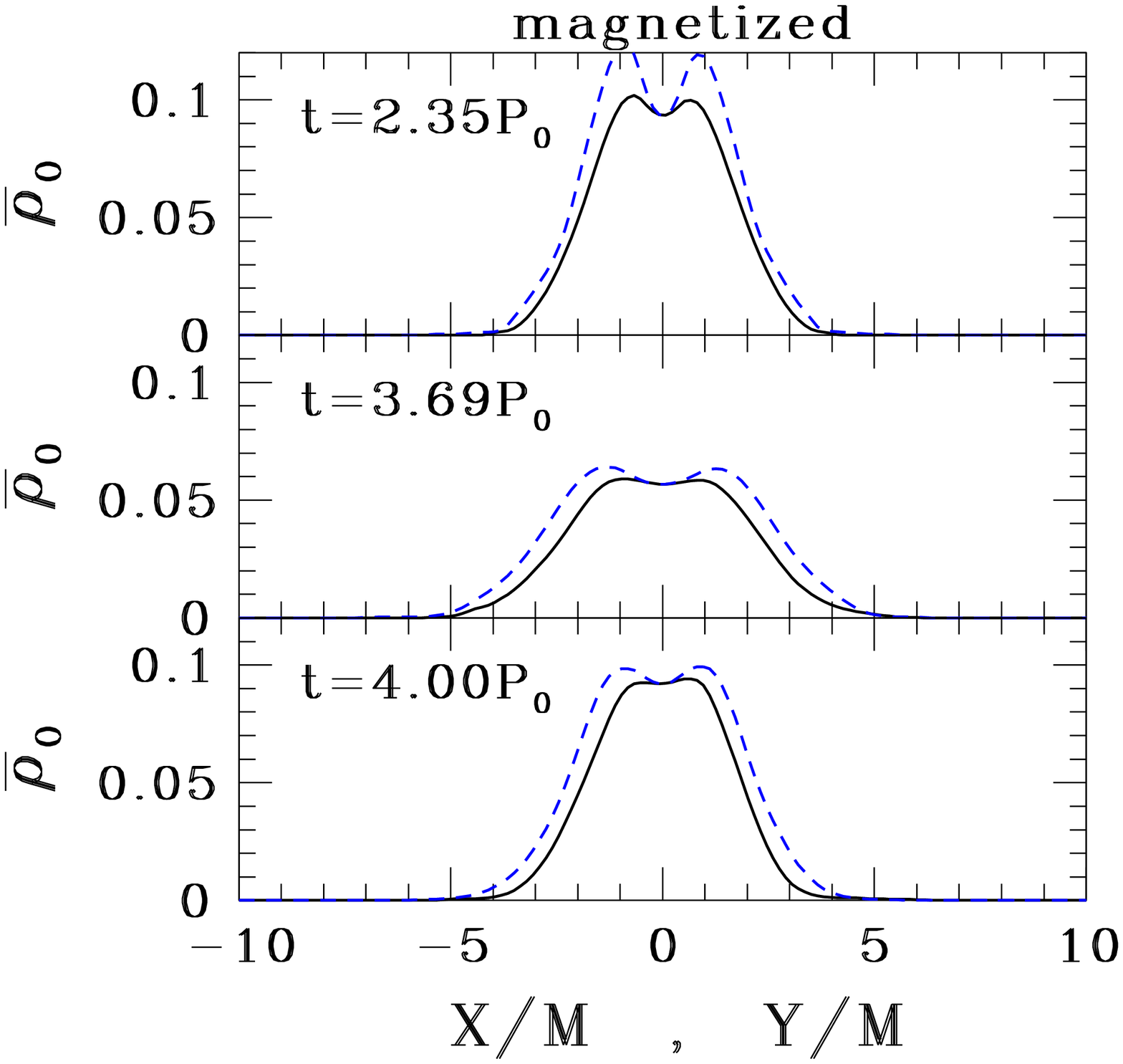}
\caption{Nondimensional rest-mass density $\bar{\rho}_0=\kappa \rho_0$
along the $x$-axis (black solid lines) and $y$-axis
(blue dash lines) in the equatorial plane at three different times for
the unmagnetized (left) and magnetized (right) cases.}
\label{fig:M1414_rho_xy}
\end{center}
\end{figure*}

\begin{figure*}
\vspace{-4mm}
\begin{center}
\epsfxsize=2.15in
\leavevmode
\epsffile{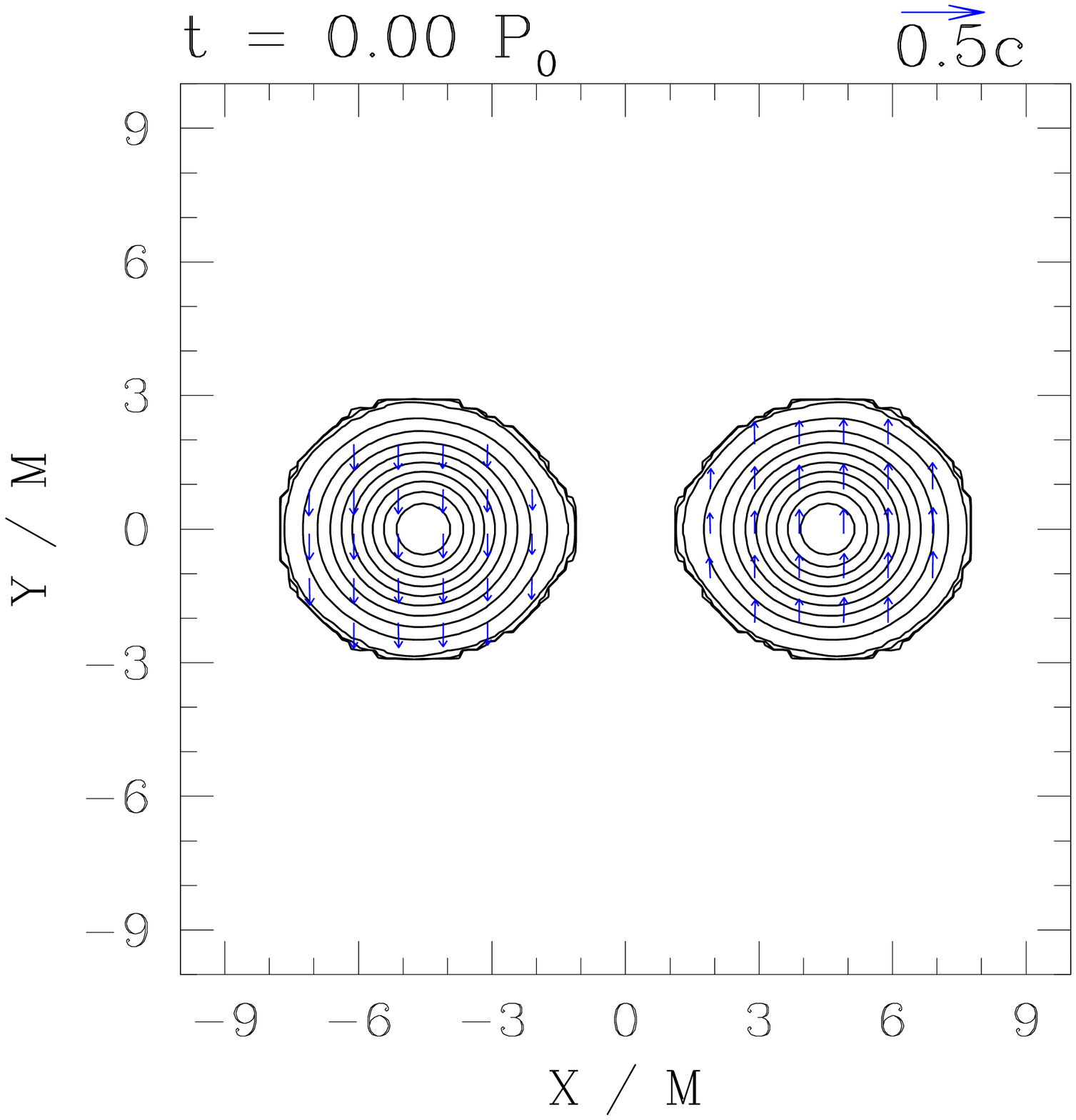}
\epsfxsize=2.15in
\leavevmode
\epsffile{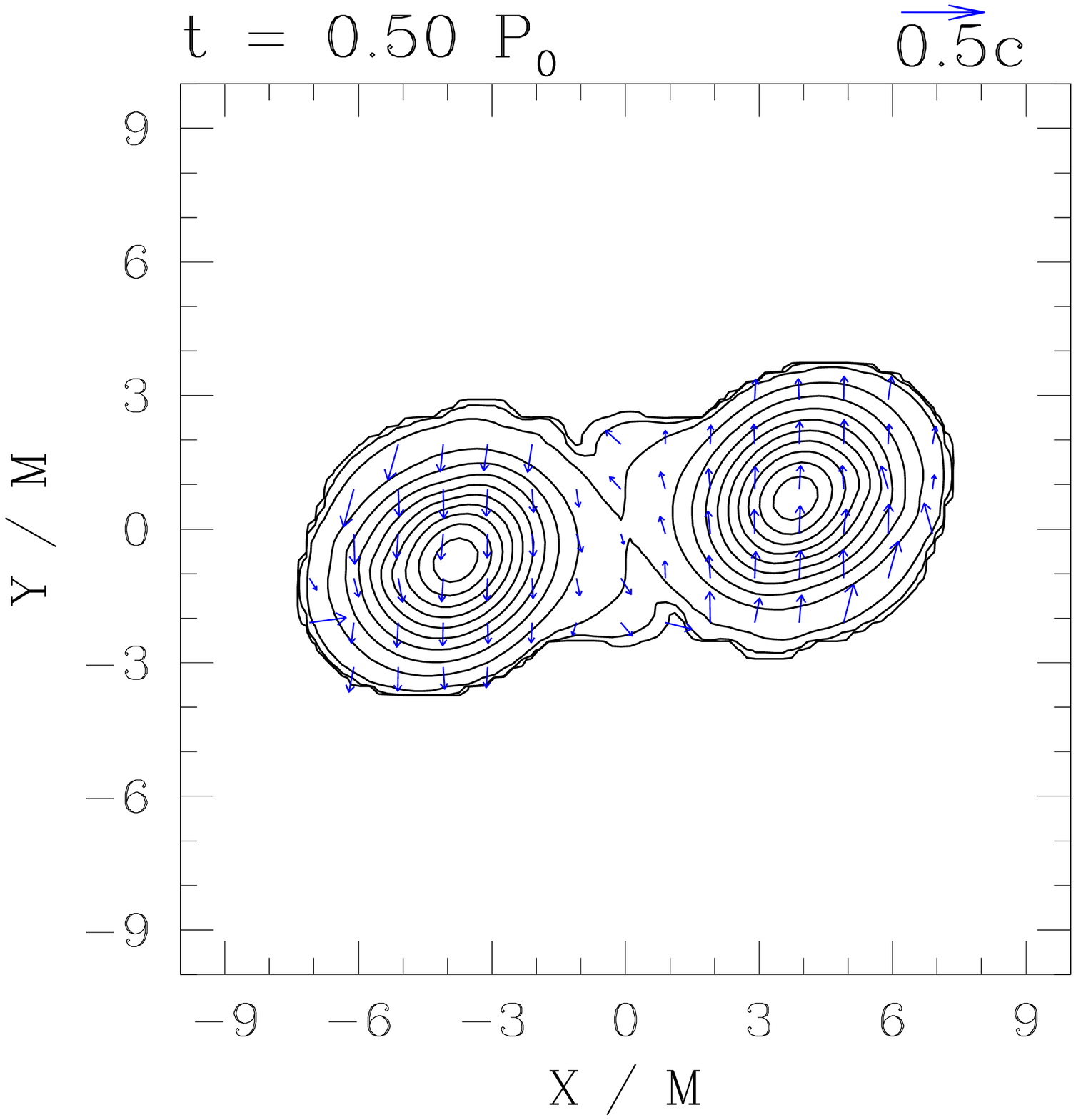} 
\epsfxsize=2.15in
\leavevmode
\epsffile{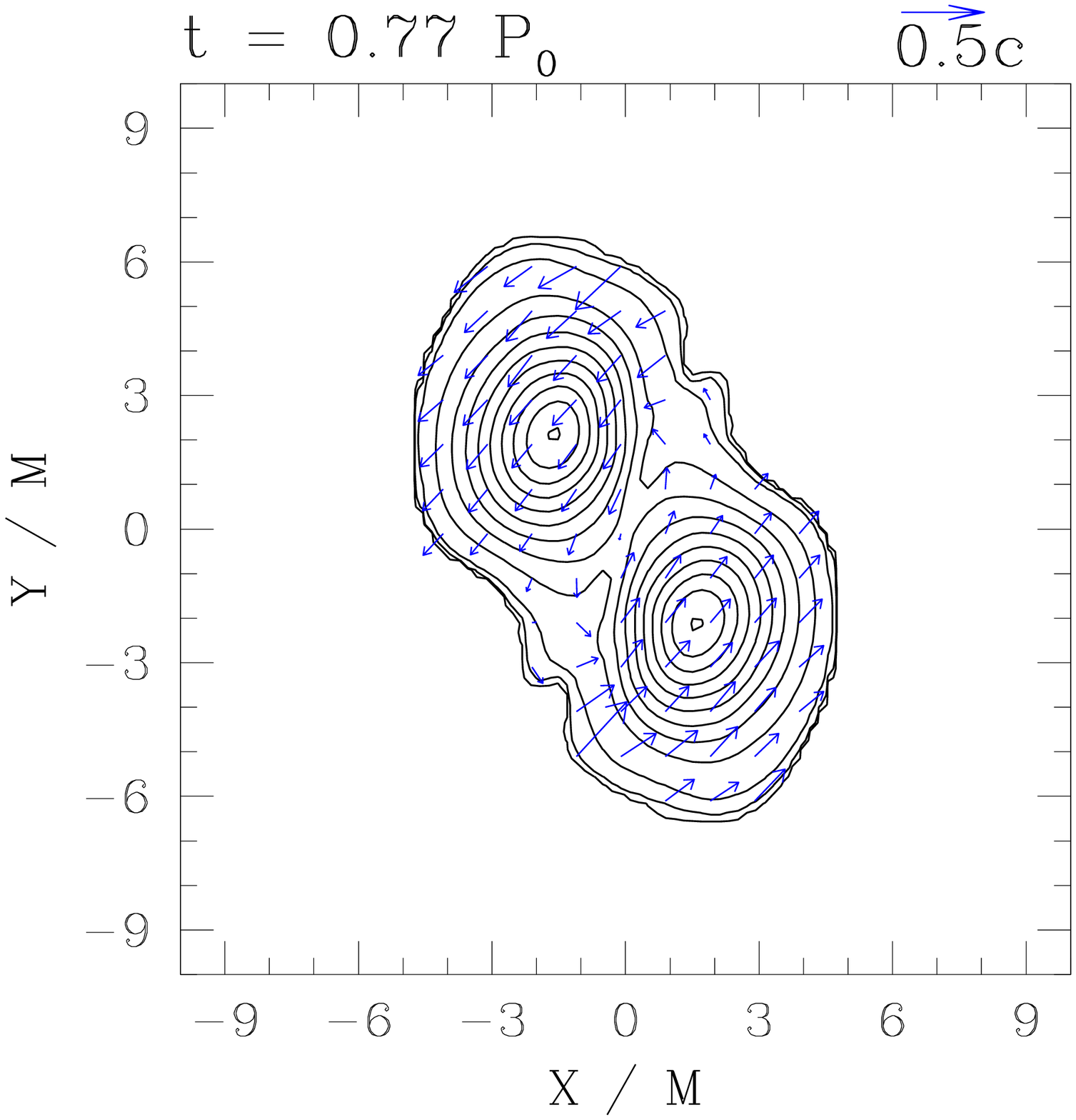}\\
\epsfxsize=2.15in
\leavevmode
\epsffile{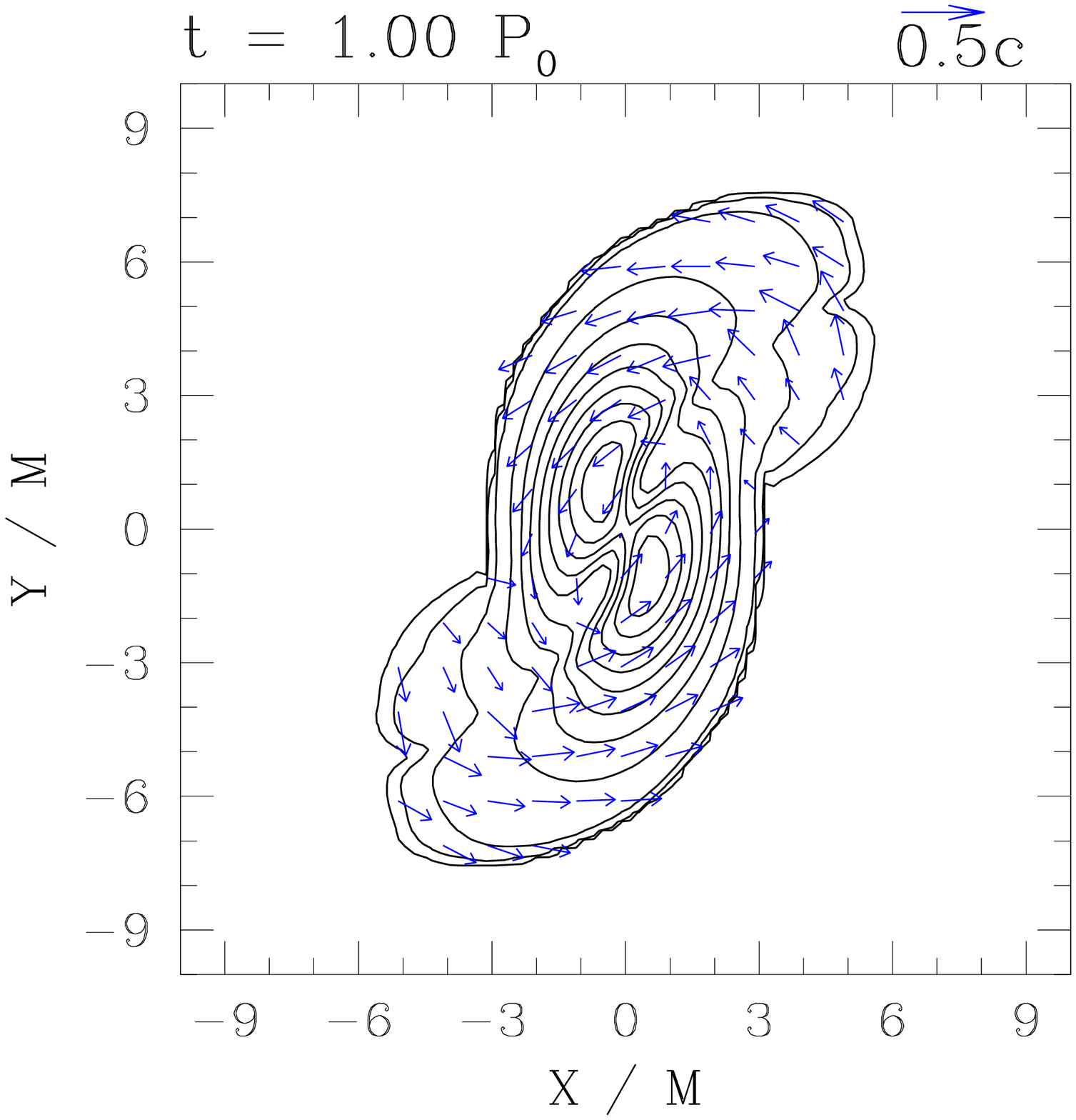}
\epsfxsize=2.15in
\leavevmode
\epsffile{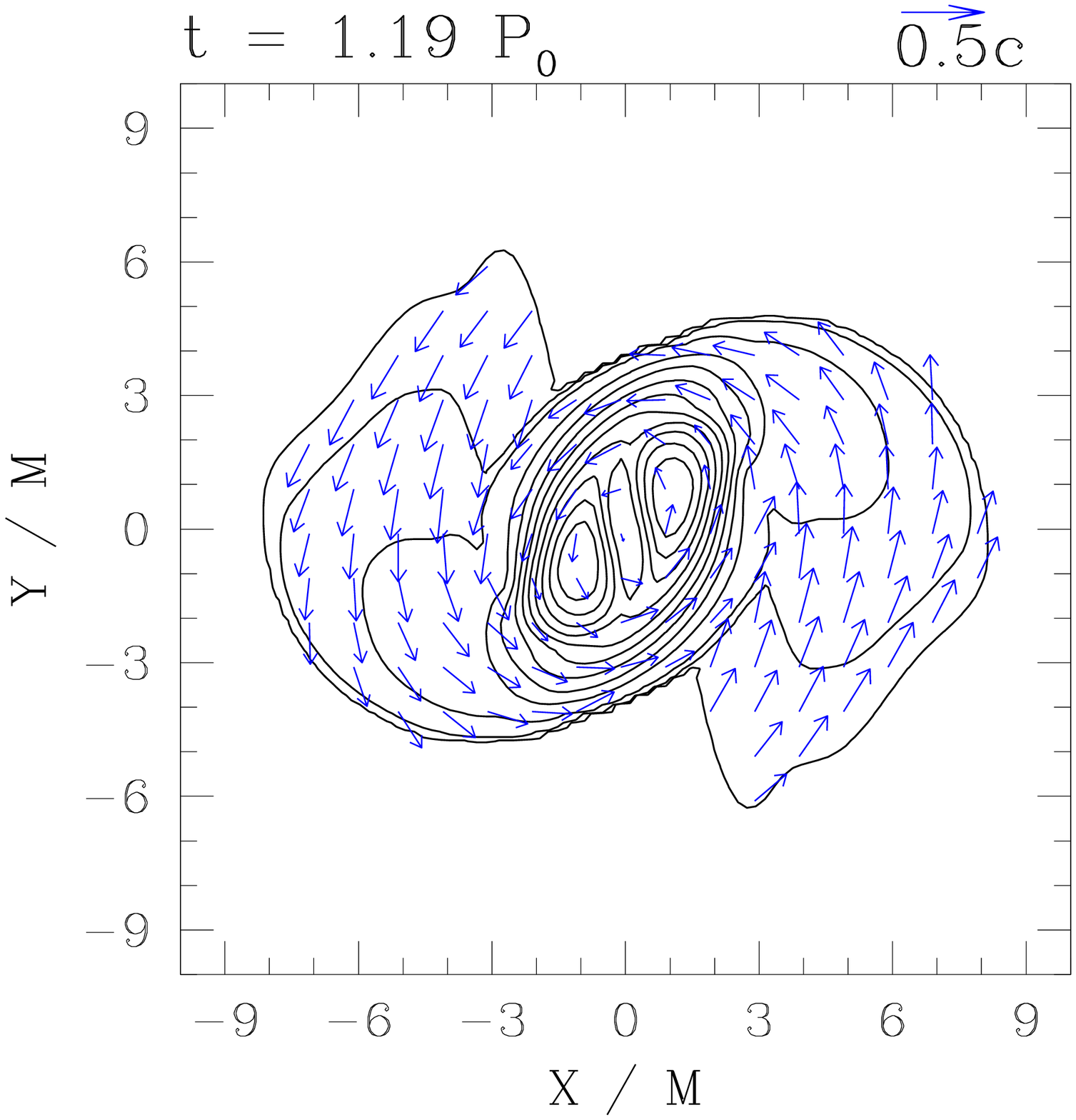}
\epsfxsize=2.15in
\leavevmode
\epsffile{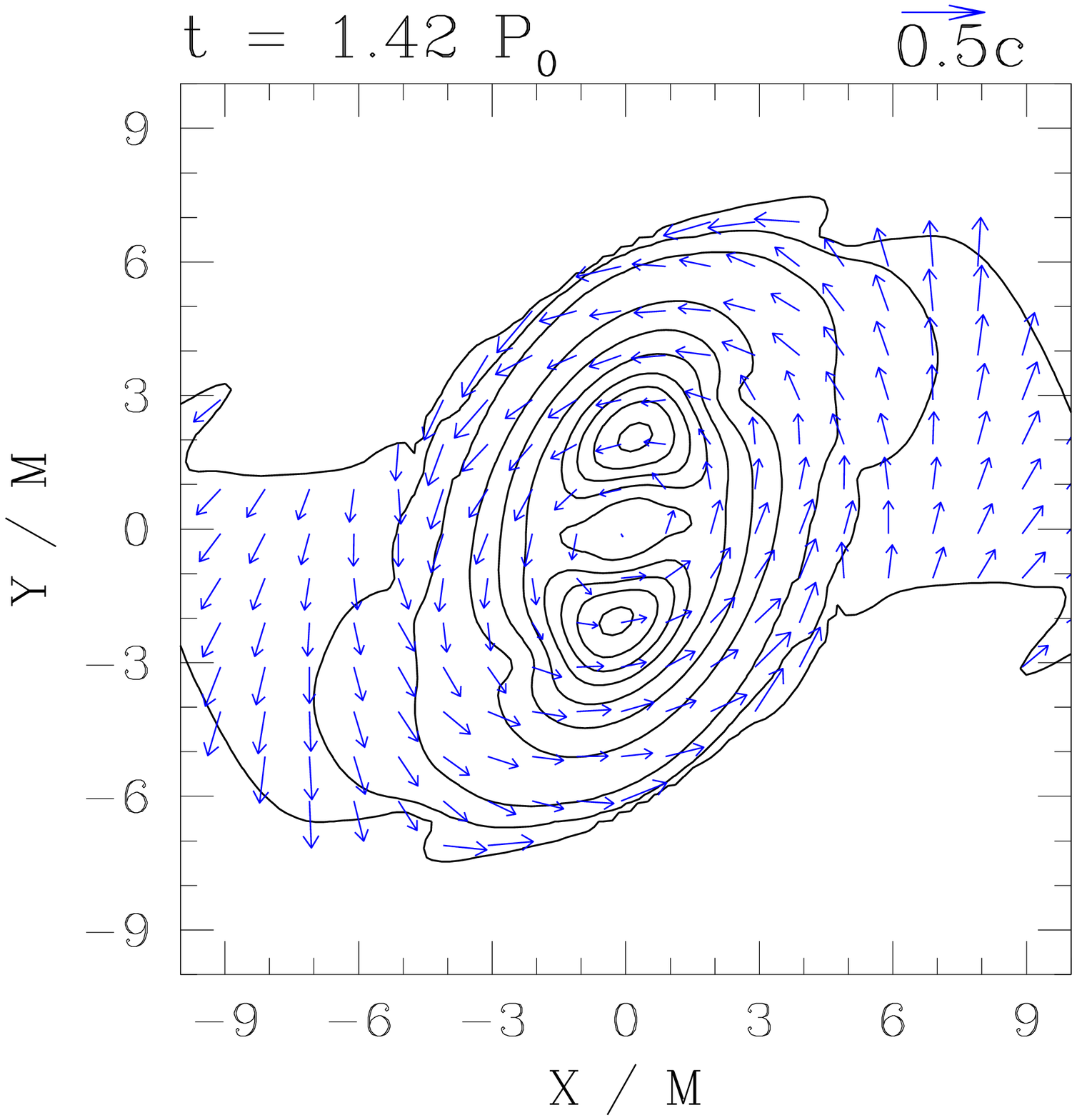}\\
\epsfxsize=2.15in
\leavevmode
\epsffile{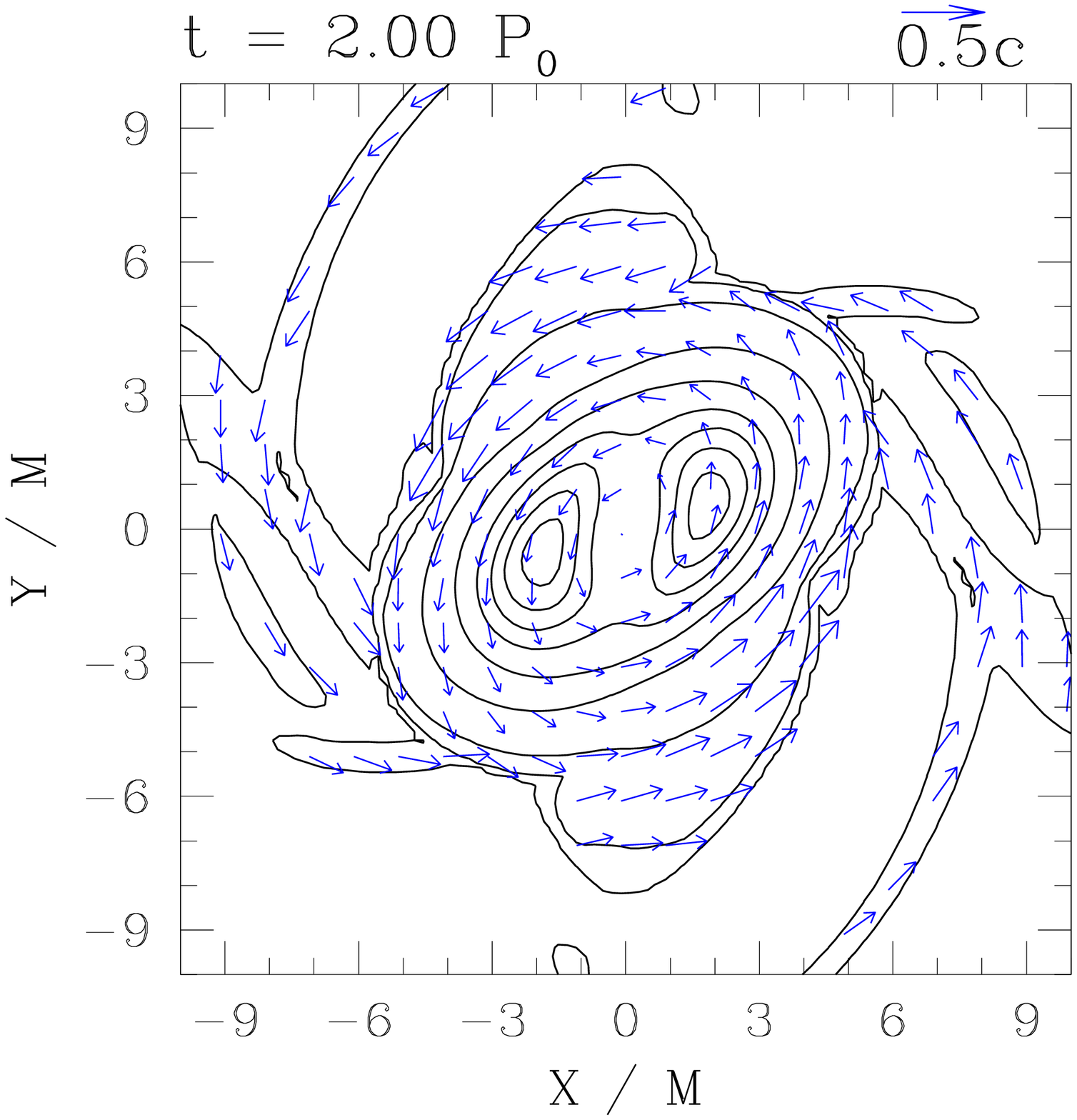}
\epsfxsize=2.15in
\leavevmode
\epsffile{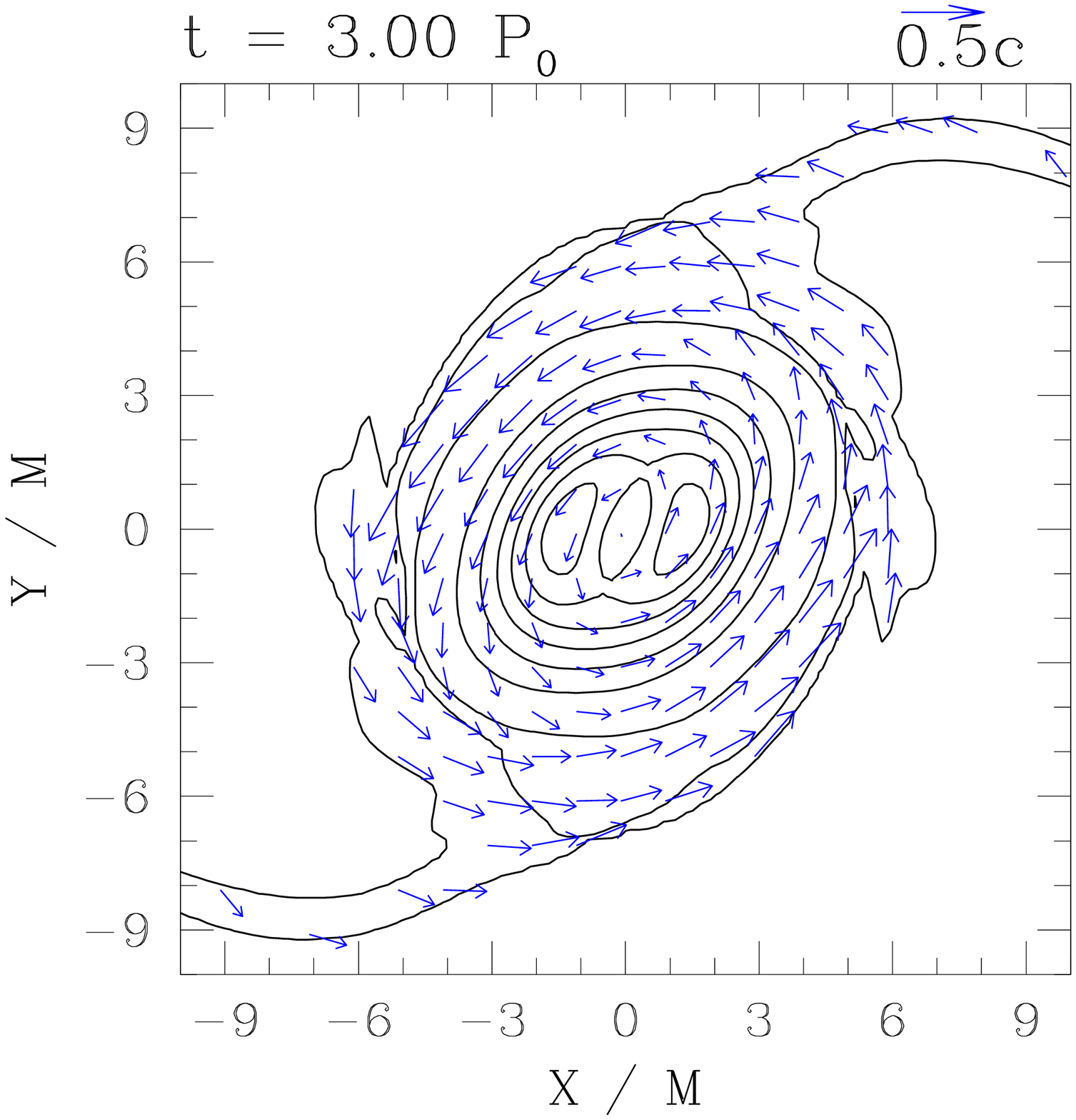}
\epsfxsize=2.15in
\leavevmode
\epsffile{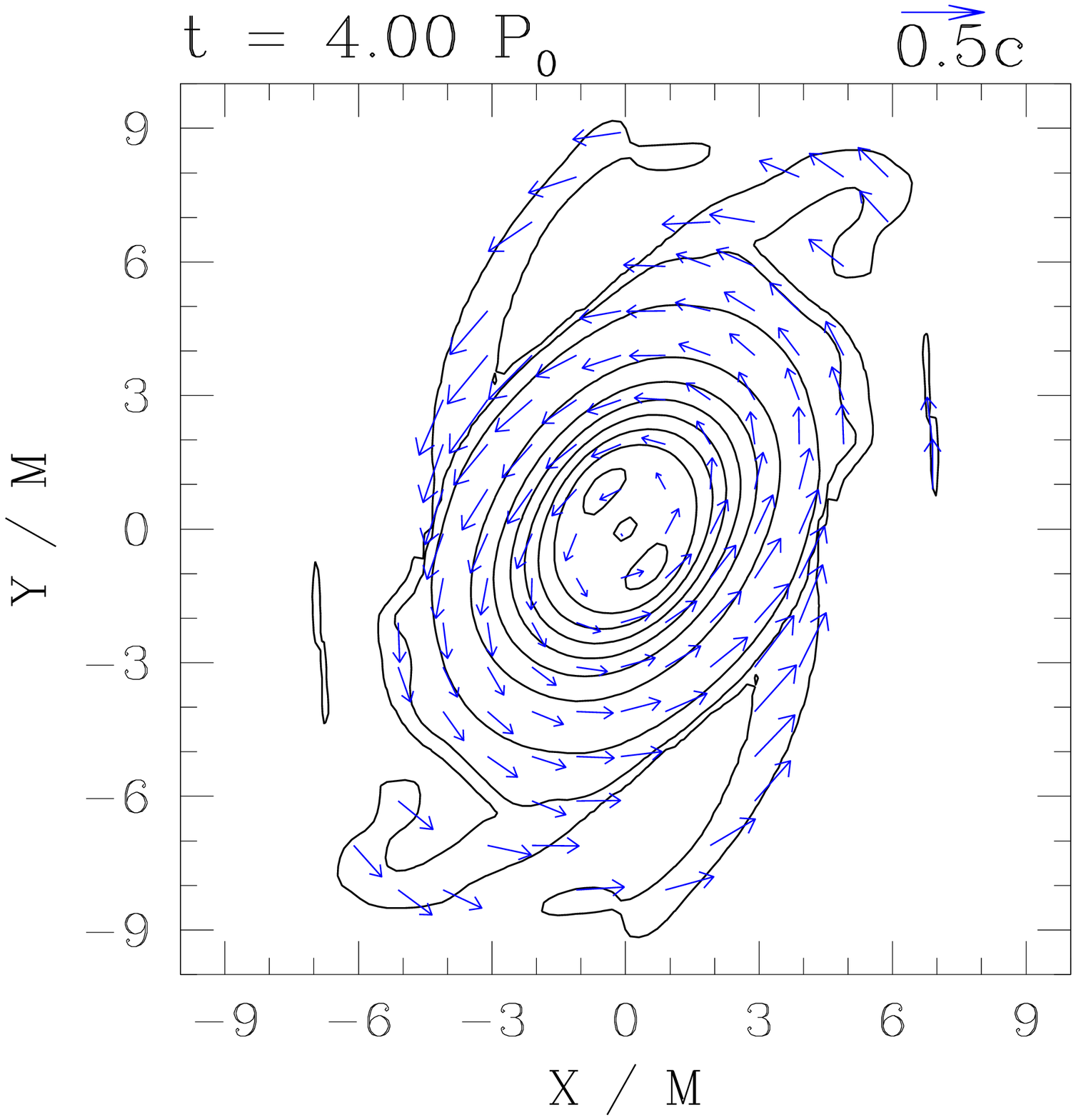}
\caption{Snapshots of density and velocity field in the equatorial plane 
from run M1414B0 (unmagnetized run). Density contours are drawn for 
$\rho_0/\rho_0(0)$=0.9, 0.8, 0.7, 0.6, 0.5, 0.4, 0.3, 0.2, 
0.1, 0.01, 0.001, and 0.0001.}
\label{fig:M1414_m0_snapshots}
\end{center}
\end{figure*}

\begin{figure*}
\vspace{-4mm}
\begin{center}
\epsfxsize=2.15in
\leavevmode
\epsffile{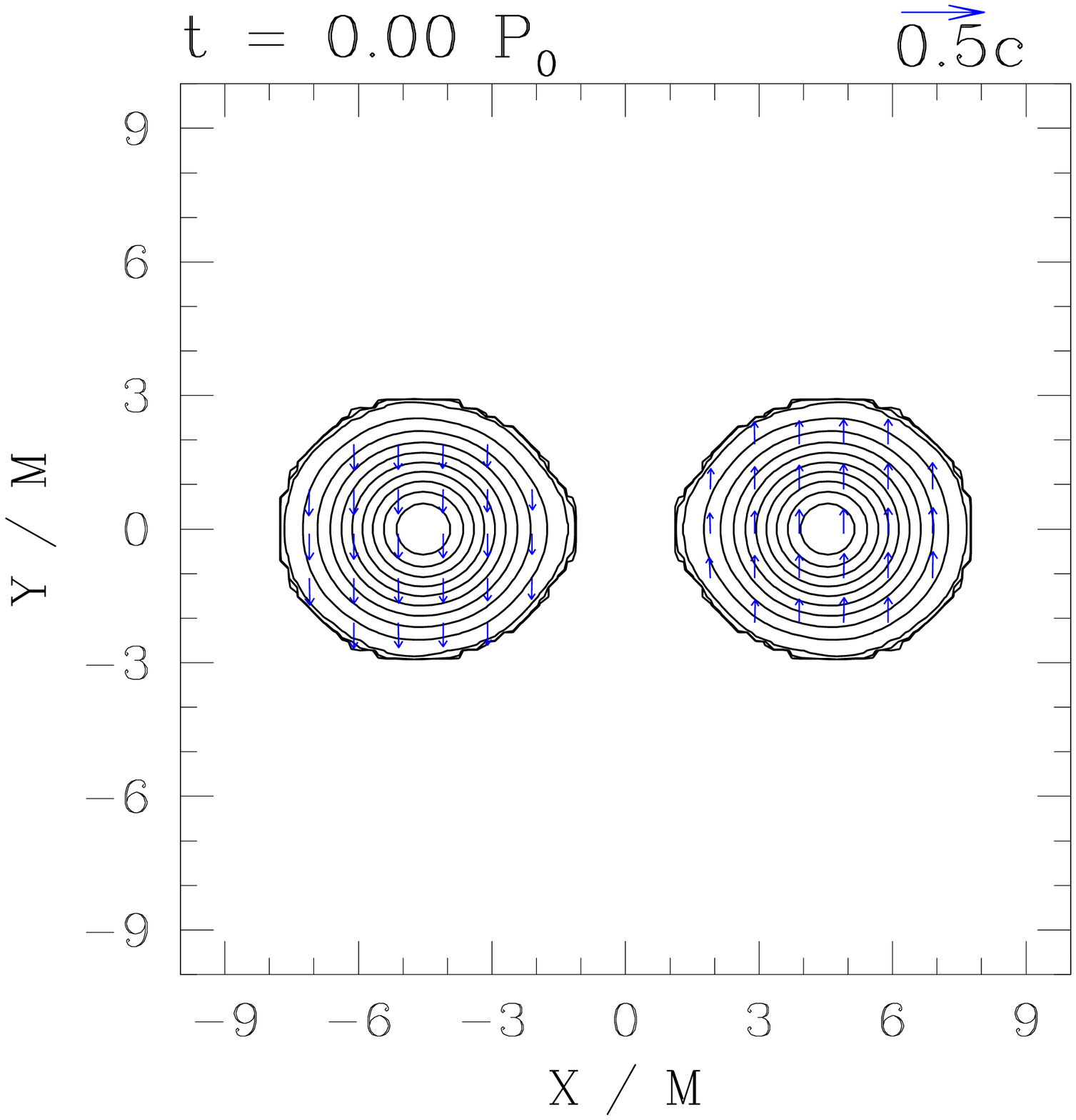}
\epsfxsize=2.15in
\leavevmode
\epsffile{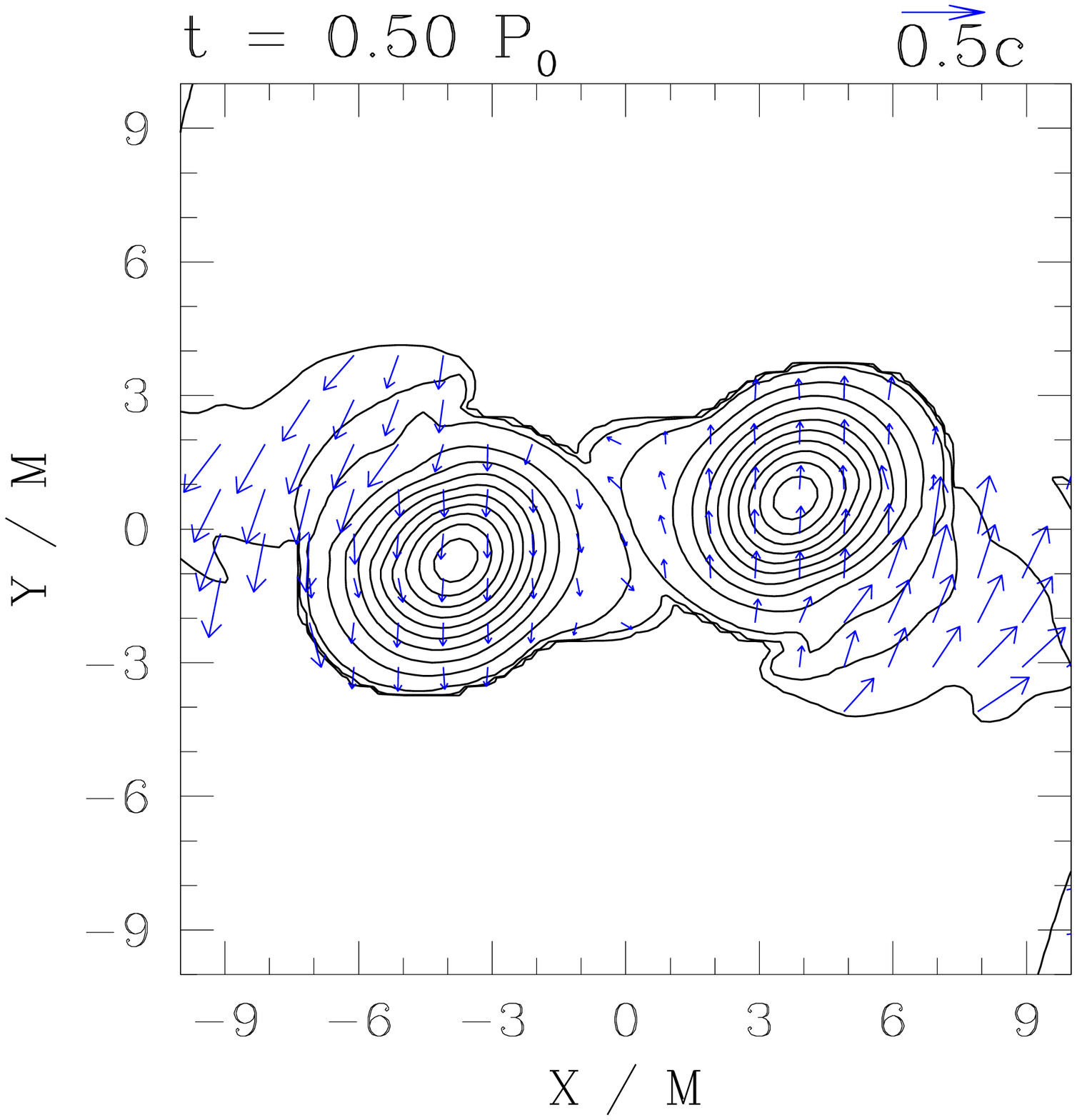}
\epsfxsize=2.15in
\leavevmode
\epsffile{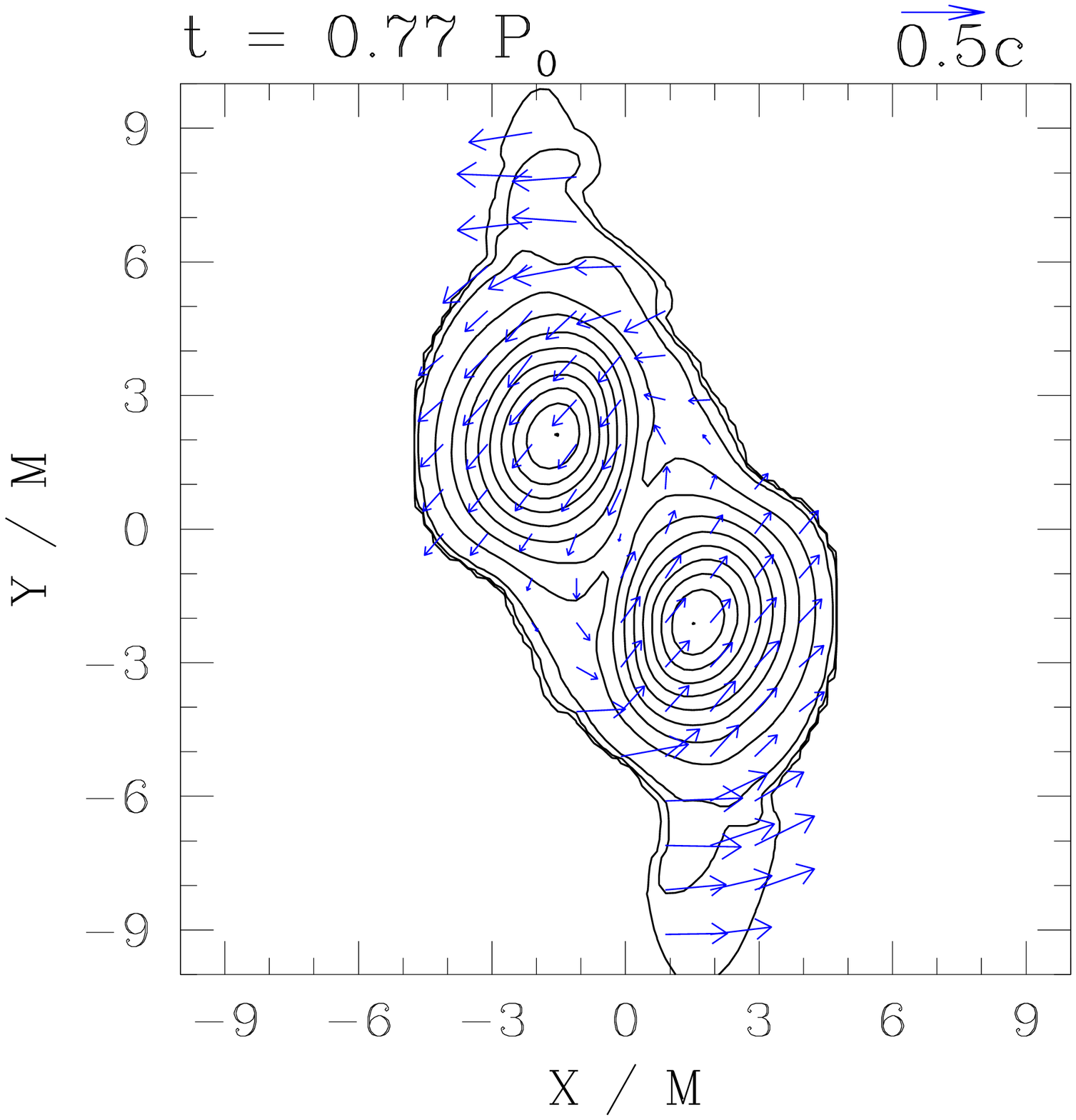}\\
\epsfxsize=2.15in
\leavevmode
\epsffile{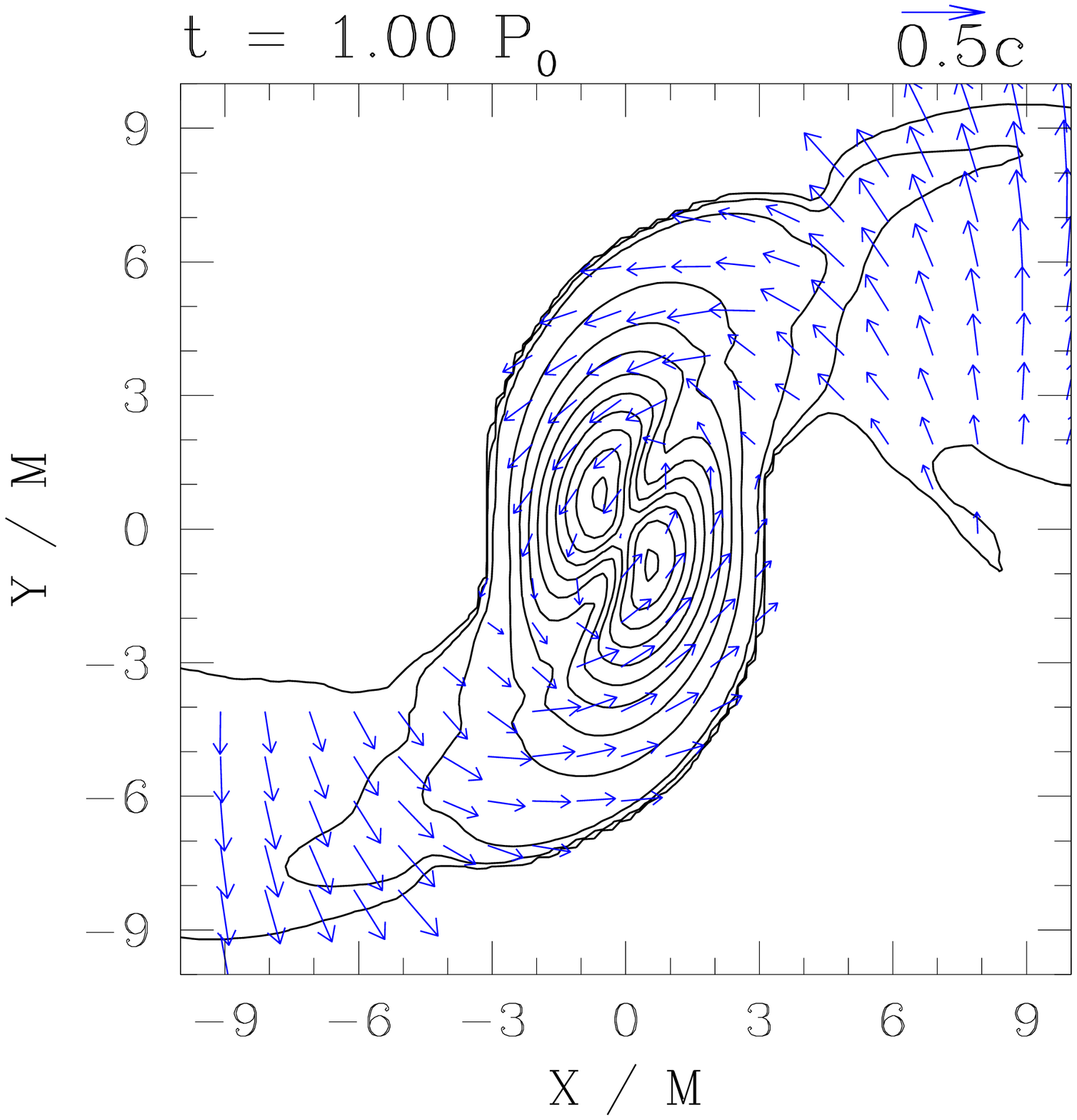}
\epsfxsize=2.15in
\leavevmode
\epsffile{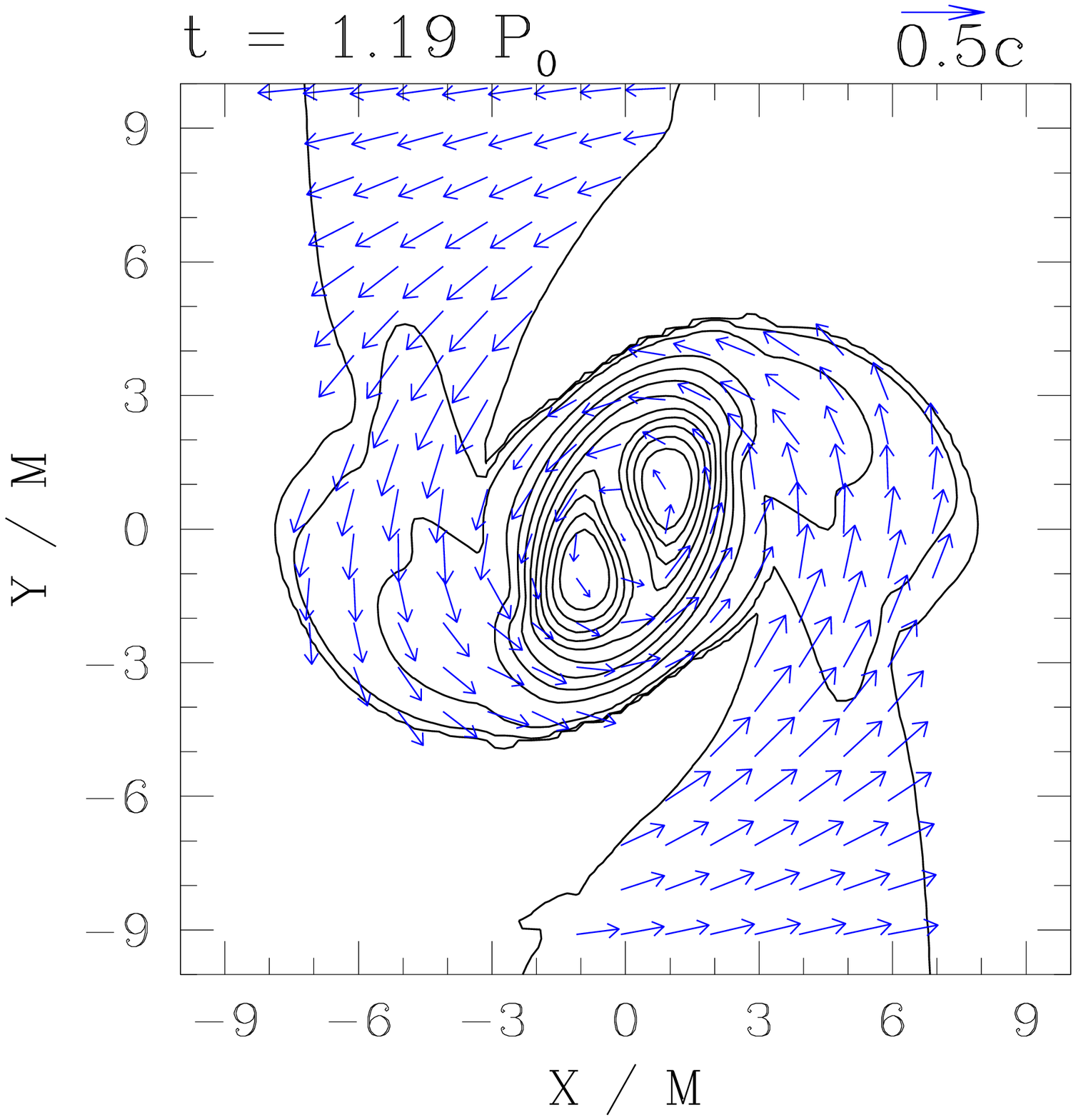}
\epsfxsize=2.15in
\leavevmode
\epsffile{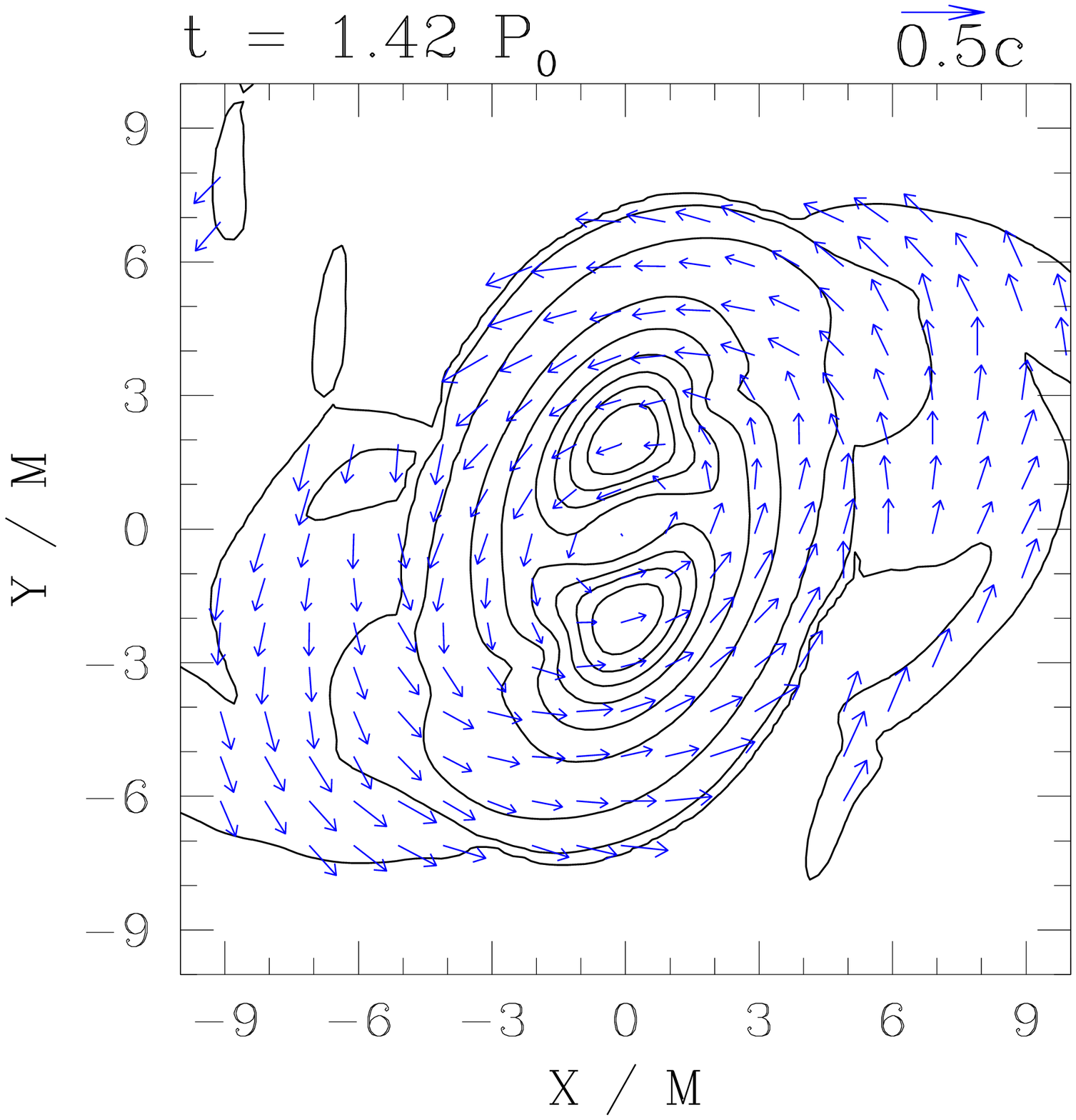}\\
\epsfxsize=2.15in
\leavevmode
\epsffile{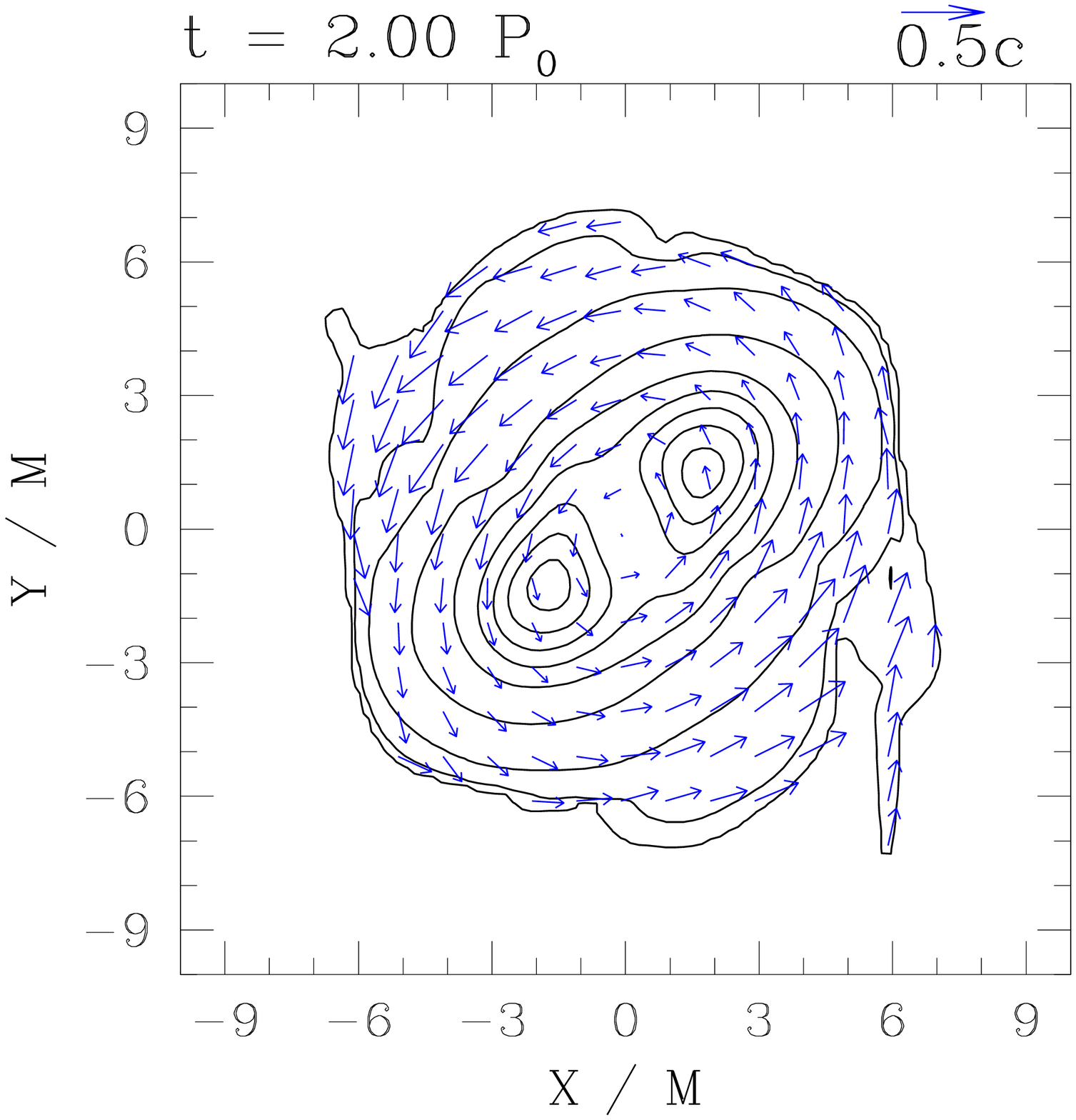}
\epsfxsize=2.15in
\leavevmode
\epsffile{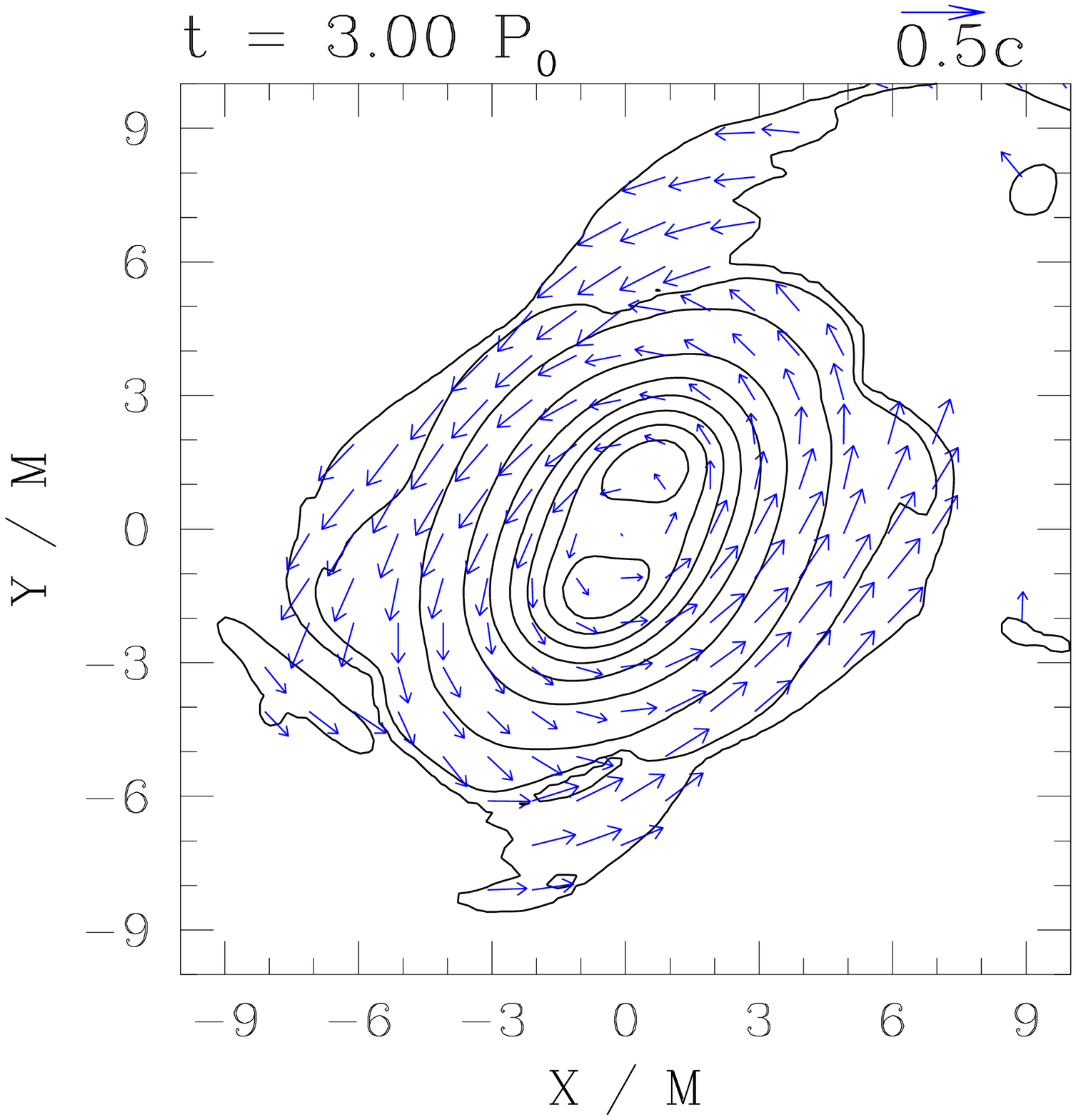}
\epsfxsize=2.15in
\leavevmode
\epsffile{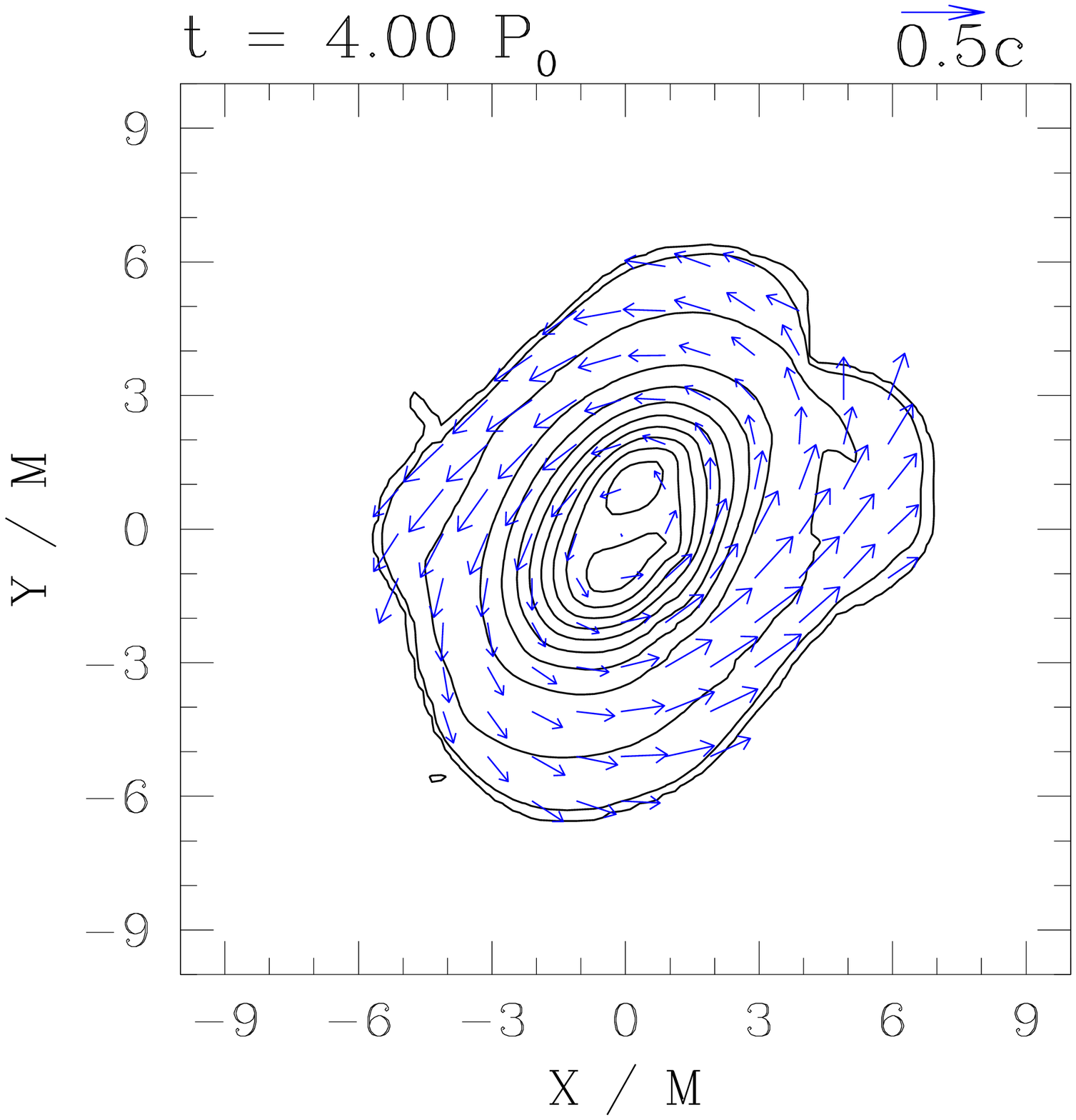}
\caption{Same as Fig.~\ref{fig:M1414_m0_snapshots} but for run 
M1414B1 (magnetized run).}
\label{fig:M1414_m1_snapshots}
\end{center}
\end{figure*}

\begin{figure}
\includegraphics[width=8cm]{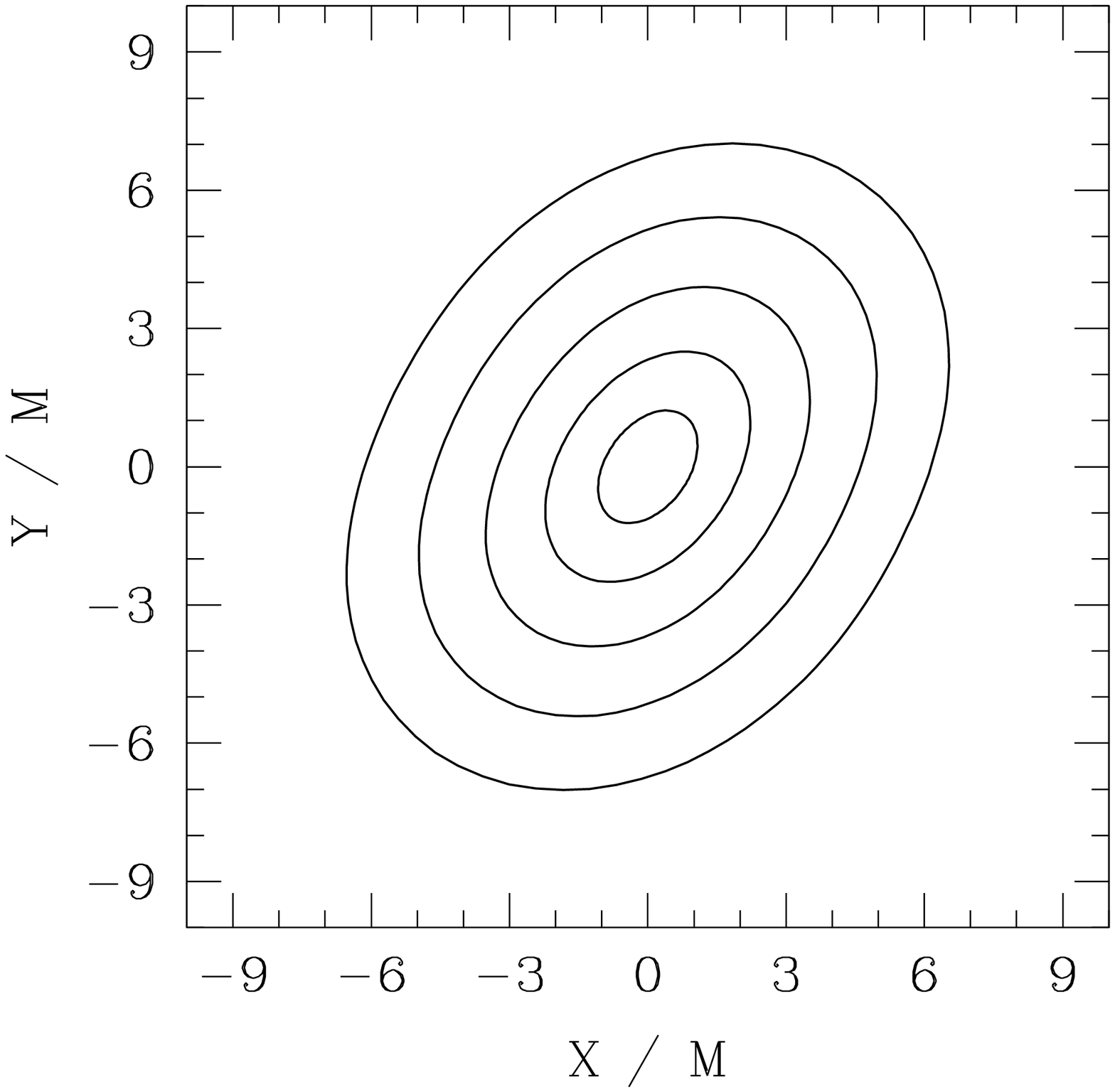}
\caption{Constant geodesic proper distance from the center, $D$, in the 
equatorial plane at $t=4P_0$ for run M1414B0. 
Contours are drawn for $D/M$=1.86$i$ ($i$=1, 2, 3, 4, 5). }
\label{fig:geo}
\end{figure}

Model M1414 is an equal-mass NSNS binary. The total rest mass 
of the system is $M_0=1.6M_0^{\rm (TOV)}$. In the absence of 
magnetic field (run M1414B0), the neutron stars merge after about 
one orbit ($\approx 190M$), consistent with the result in~\cite{stu03}. 
We find that 
magnetic field does not change the result. After the merger, the star 
becomes a hypermassive NS. 
Figure~\ref{fig:M1414_rho_alp}
shows the evolution of maximum density $\rho_0^{\max}$ and minimum
lapse $\alpha_{\rm min}$ for both unmagnetized and magnetized cases. 
Figure~\ref{fig:M1414_rho_xy} shows the density profile along the 
$x$-axis and $y$-axis in the equatorial plane at three different times. 
We see that the unmagnetized case is very close to the simulation 
in~\cite{stu03} (see their Fig.~6a and Fig.~7b). 

Figures~\ref{fig:M1414_m0_snapshots} and \ref{fig:M1414_m1_snapshots}
show the density contours and velocity field in the equatorial
plane. We see that magnetic field causes some mass-shedding in the 
low-density region. After the merger, we see double cores rotating 
around the center, as in~\cite{stu03}. The star is also pulsating with 
a large amplitude (see Figs.~\ref{fig:M1414_rho_alp} and \ref{fig:M1414_rho_xy}). These motions 
give rise to gravitational wave signals after the merger (see below). 
We note that the apparent bar-like structure seen in 
Figs.~\ref{fig:M1414_m0_snapshots} and \ref{fig:M1414_m1_snapshots} 
in some of the density contours at late times is a coordinate (gauge) 
artifact. In our coordinates, the bar is nonrotating and stationary 
for several periods.
Figure~\ref{fig:geo} shows the contours of constant geodesic proper distance 
from the center of the grid. (The geodesic proper distance between $P$ and 
$Q$ in a spatial slice is defined as the proper length of the (spatial) 
geodesic joining $P$ and $Q$). We have verified that the bar-like 
density contours roughly coincide with the constant geodesic proper distance 
contours, indicating that the density distribution in the bar-like 
region is actually close to axisymmetric.

Magnetic fields do not affect the dynamics of the system prior to 
merger, as expected. After the merger, we see a larger amplitude 
of pulsation in the magnetized case. However, we expect that the 
main effects of the magnetic field occur on a longer (secular) 
timescale when the field is amplified by differential rotation. 
Magnetic winding will occur on an Alfv\'en timescale $t_A$$\sim$$R/v_A$
$\sim$10 rotation periods, where $v_A = \sqrt{b^2/(\rho_0 h +b^2)}$ is 
the Alfv\'en velocity. Additional amplification will be triggered 
by the magnetorotational instability (MRI)~\cite{dlsss06a,dlsss06b}.
This amplification will lead to transport of angular momentum 
and may trigger a `delayed' collapse~\cite{dlsss06a,dlsss06b}.  
We do not follow the evolution for that long in this
paper. We note that the grid resolution in our simulation is 
insufficient to resolve some of the MHD instabilities such as 
the MRI~\cite{bh98}. However, our current resolution 
is adequate to capture the main effects of magnetic fields on 
our simulation timescale (a few oscillation periods after the
merger). We note, however, that we have already performed 
high-resolution simulations in axisymmetry to study the long-term 
secular evolution 
of magnetized hypermassive NS remnants~\cite{dlsss06a,dlsss06b}. 
We confirm that the magnetic fields cause transport of angular momentum, 
resulting in a delayed collapse of the star to a black hole surrounded 
by a hot, massive disk. 

\begin{figure}
\includegraphics[width=8cm]{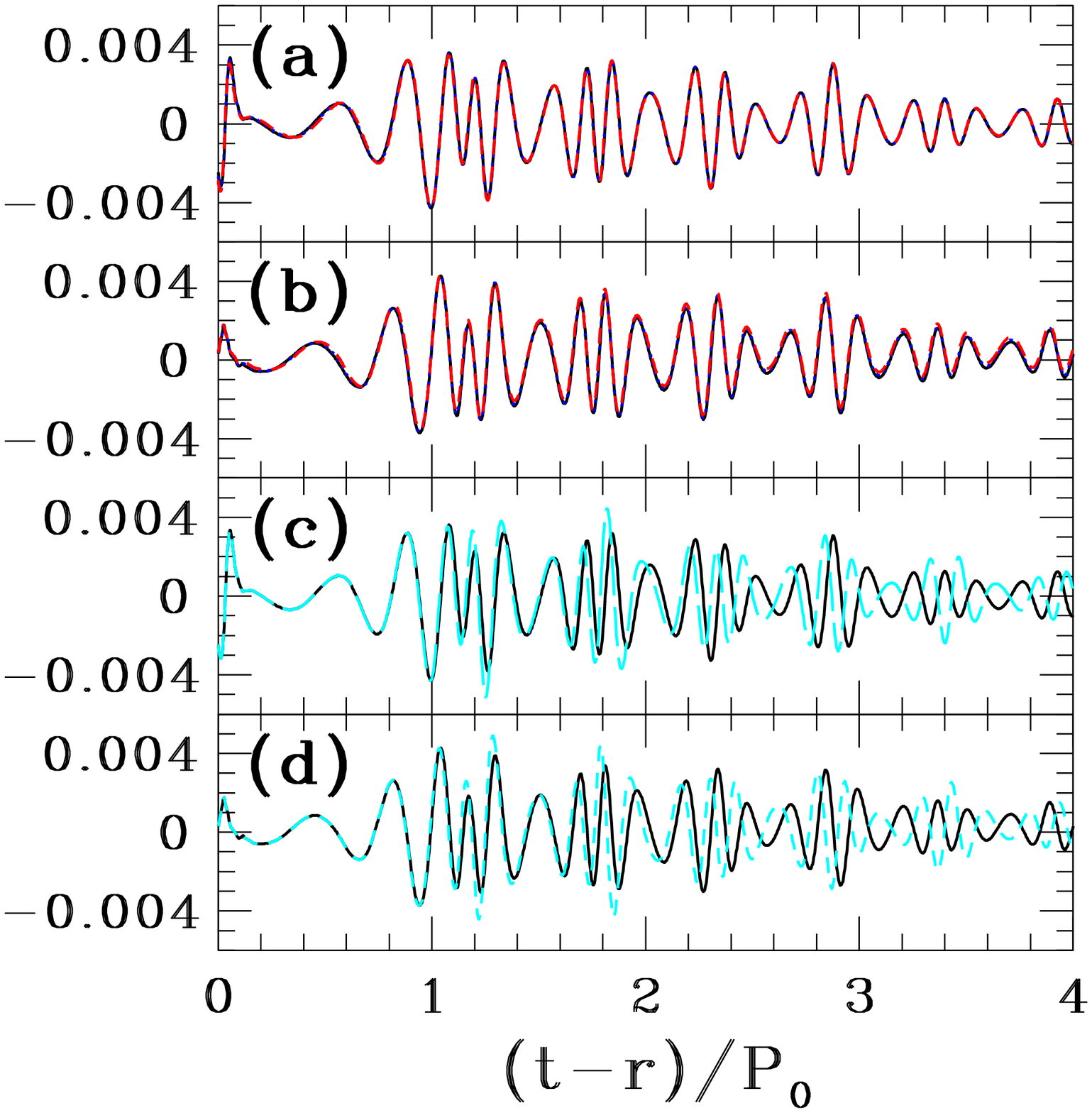}
\caption{Gravitation waveforms for model M1414. (a) Re($rM\psi^{22}_4$)
extracted at $r=43M$ (black solid line), $r=38M$ 
(blue dotted line), and $r=32M$ (red dash line). (b) Same as (a) but for 
Im($rM\psi^{22}_4$). 
(c) Re($rM\psi^{22}_4$) extracted at $r=43M$ for the unmagnetized 
(black solid line) and magnetized (cyan dash line) cases. 
(d) Same as (c) but for Im($rM\psi^{22}_4$). 
Note that in (a) and (b), the lines are hardly distinguishable, 
showing good agreement of waveforms at various extraction radii.}
\label{fig:M1414_psi4}
\end{figure}

Figure~\ref{fig:M1414_psi4} shows the gravitational waveforms for 
both the magnetized and unmagnetized cases. We compute the $l=2$, 
$m=2$, $s=-2$ spin-weighted spherical harmonics of the Weyl tensor 
$\psi_4$ at three radii $32M$, $38M$, and $43M$. As shown in 
Fig.~\ref{fig:M1414_psi4}a and \ref{fig:M1414_psi4}b, the computed 
$\psi_4^{22}$ at these three radii are hardly distinguishable 
when properly scaled, indicating that the extracted waveforms are 
being measured in the wave zone and are reliable. We see again 
that the waveforms of  
the magnetized and unmagnetized cases show negligible differences  
before the merger. After the merger, the magnetic field significantly 
affects the motion of the fluid, and the waveforms exhibit observable  
differences in both the amplitude and phase. Notice that there are still 
gravitational wave signals after the merger. This is mainly caused by the 
rotation of the double cores and the pulsation of the merged remnant.

\begin{figure}
\includegraphics[width=8cm]{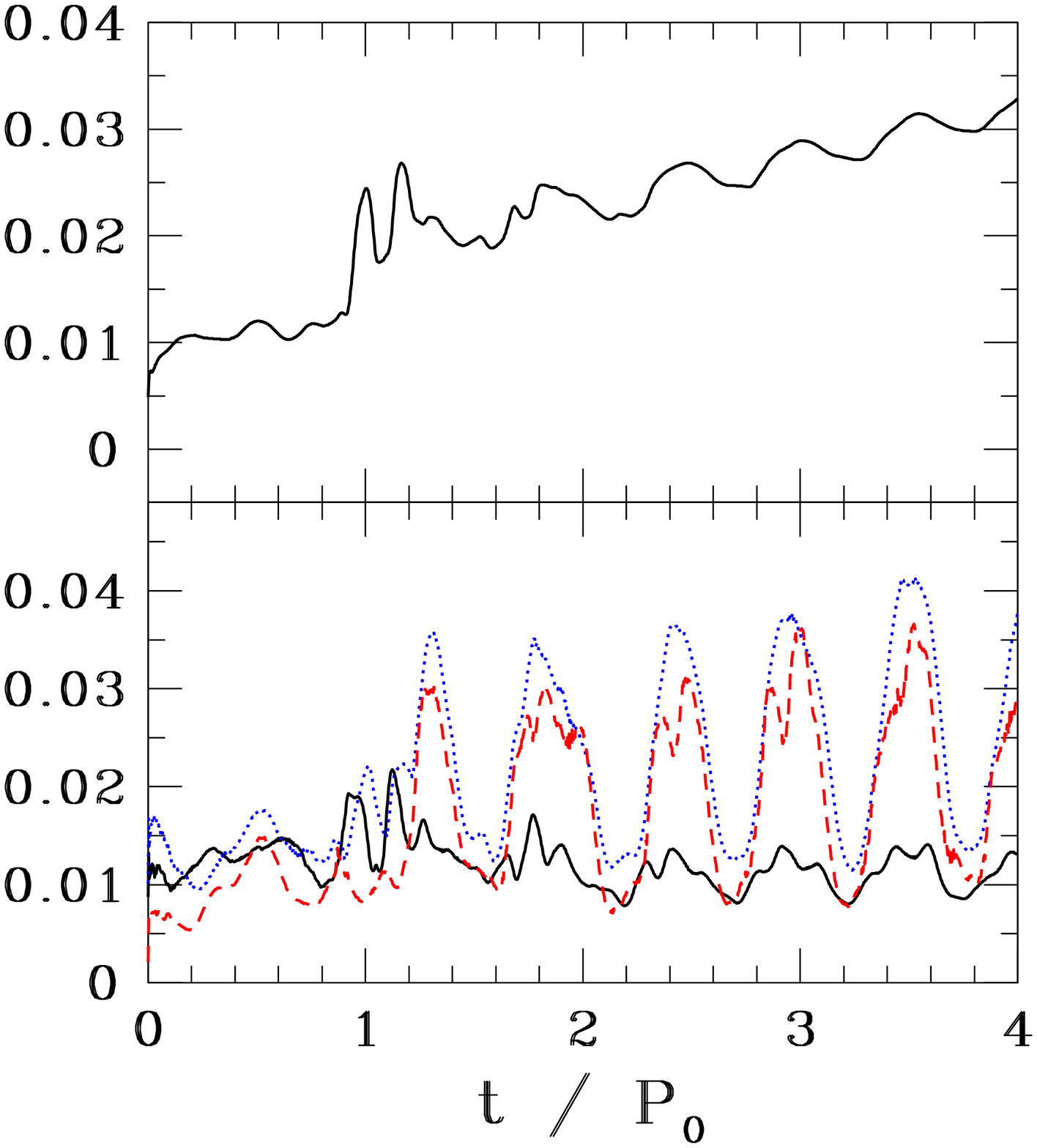}
\caption{Constraint violations for the unmagnetized run of model M1414.
Upper panel: Normalized L2 norm of the Hamiltonian constraint. 
Lower panel: Normalized L2 
norm of the $x$ (black solid line), $y$ (blue dotted line) and $z$ (red 
dash line) components of the momentum constraint.}
\label{fig:M1414_cons_m0}
\end{figure}

\begin{figure}
\includegraphics[width=8cm]{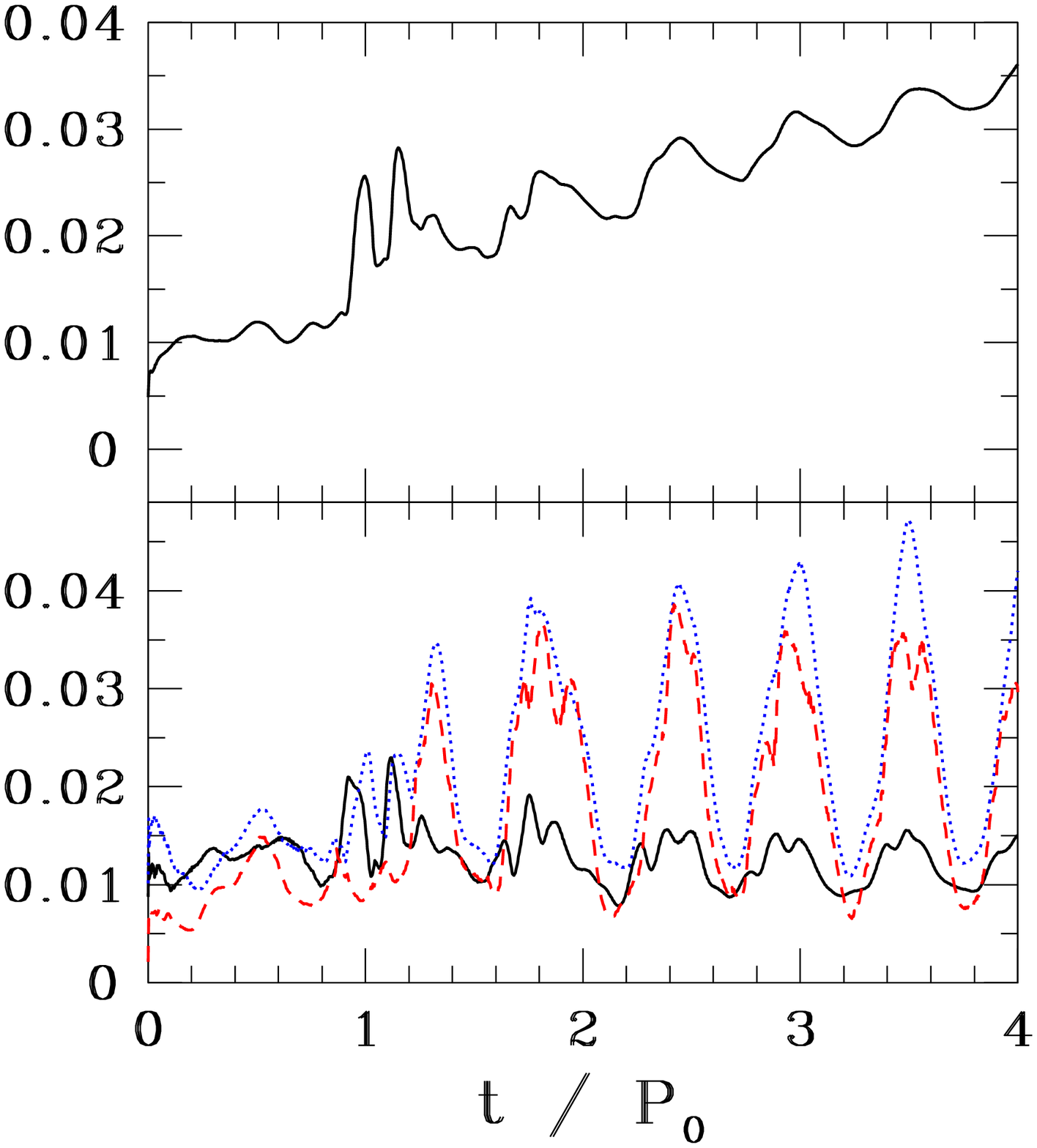}
\caption{Same as Fig.~\ref{fig:M1414_cons_m0} but for the magnetized 
run of model M1414.}
\label{fig:M1414_cons_m1}
\end{figure}

Figures~\ref{fig:M1414_cons_m0} and \ref{fig:M1414_cons_m1} show the 
L2 norms of the Hamiltonian and momentum constraint violations. The 
L2 norms are normalized as in~\cite{eflstb07}. We see that the 
constraint violations are less than 5\% during the entire simulations 
($4P_0=772M$).

\subsection{Model M1616}

\begin{figure}
\includegraphics[width=8cm]{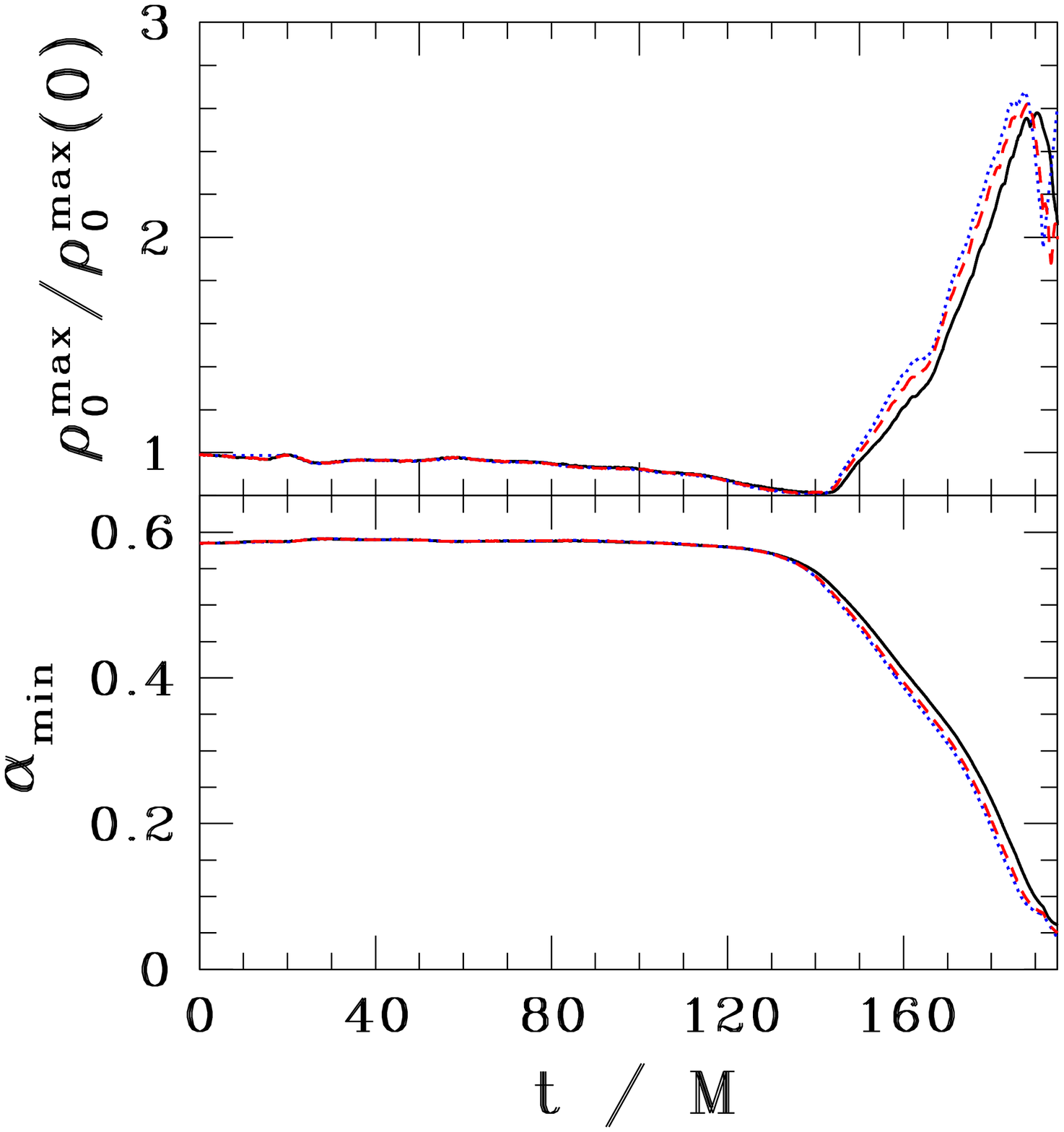}
\caption{Evolution of maximum density $\rho_0^{\max}$ and minimum
lapse $\alpha_{\rm min}$ for runs M1616B0 (black solid line), M1616B1
(blue dotted line) and M1616B2 (red dash line). The merger occurs at 
$t \approx 150M$, and an apparent horizon appears at $t=192M$ 
for both magnetized and unmagnetized cases.}
\label{fig:M1616_rho_alp}
\end{figure}

\begin{figure*}
\vspace{-4mm}
\begin{center}
\epsfxsize=2.15in
\leavevmode
\epsffile{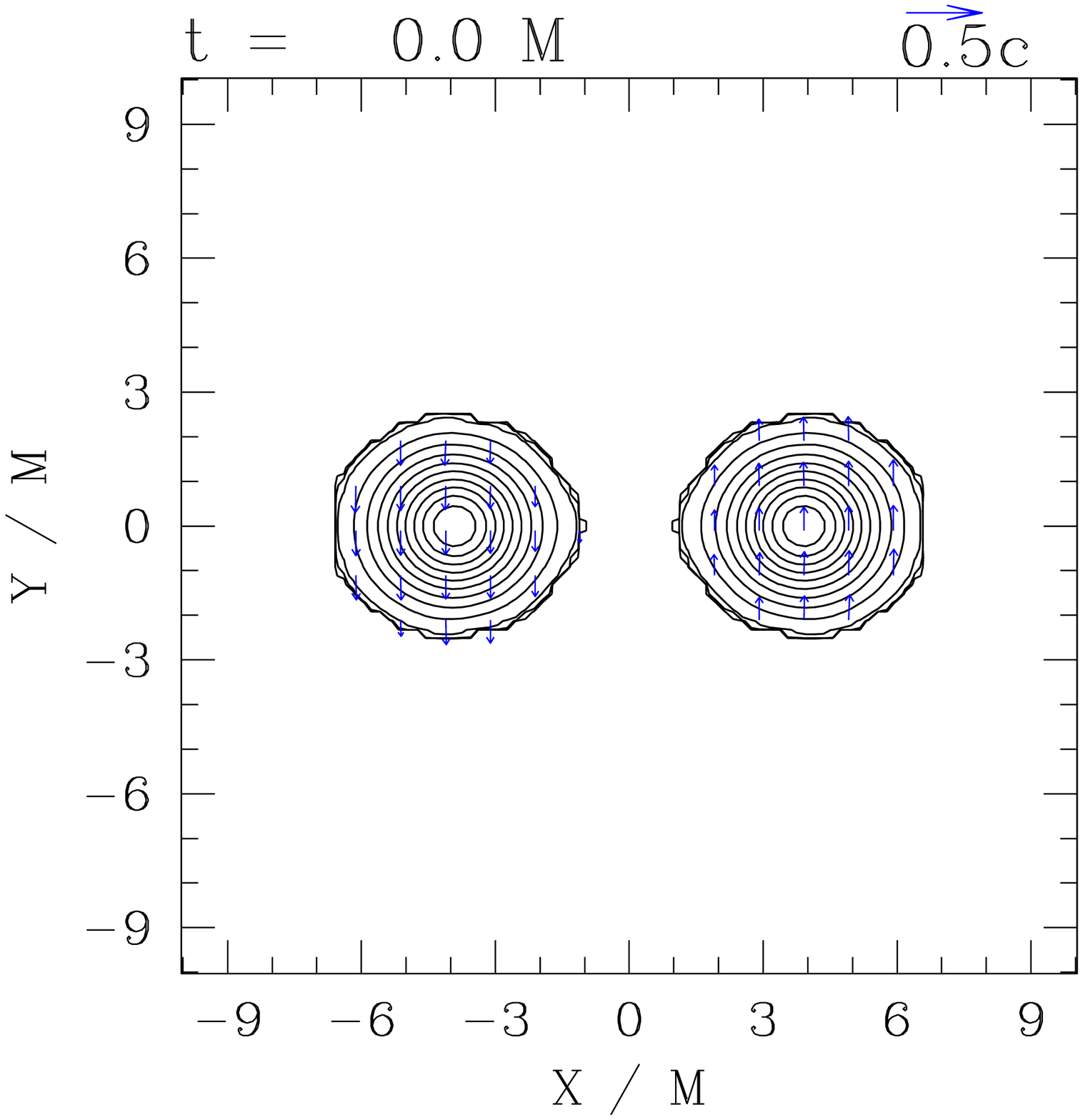}
\epsfxsize=2.15in
\leavevmode
\epsffile{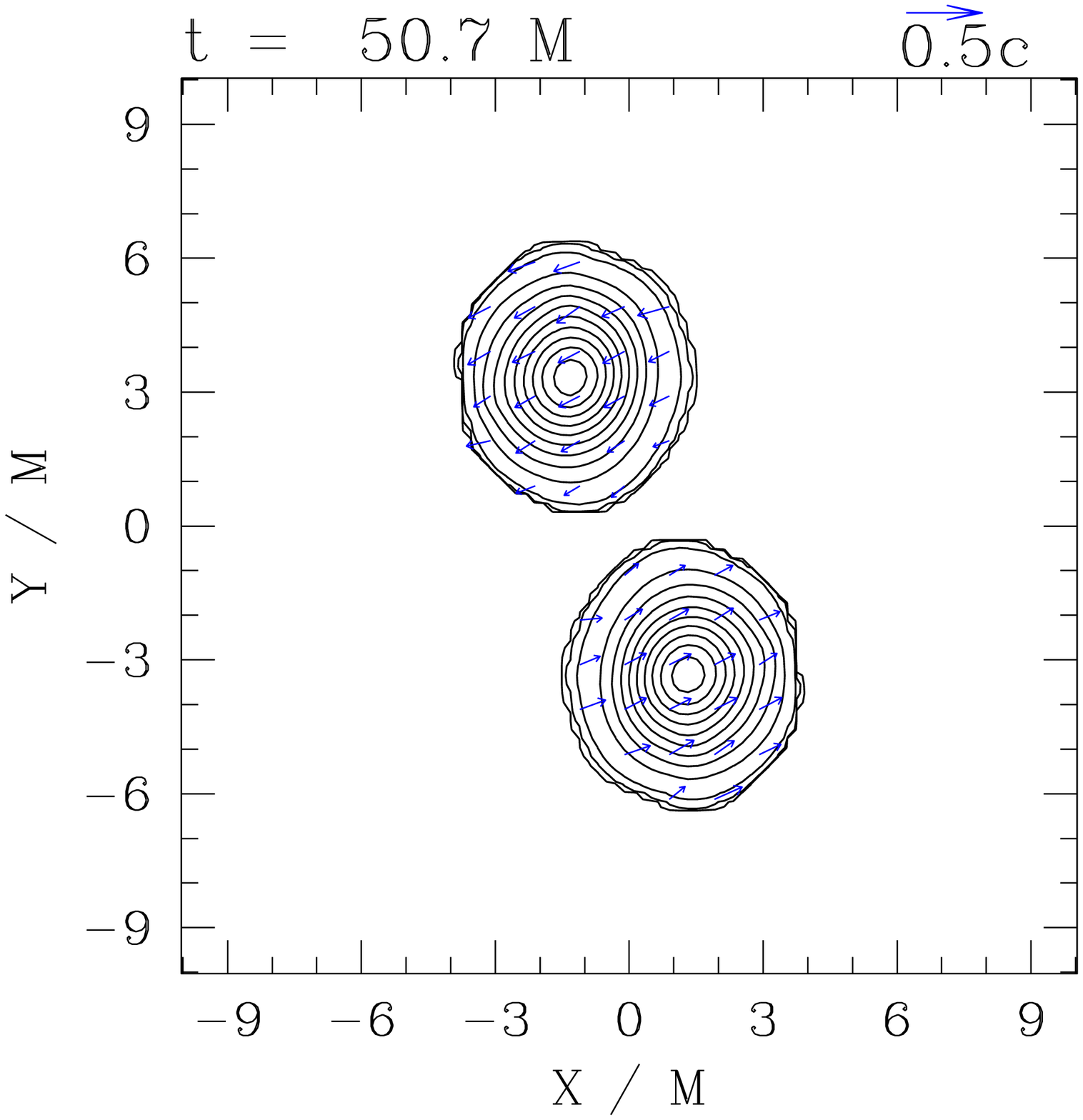}
\epsfxsize=2.15in
\leavevmode
\epsffile{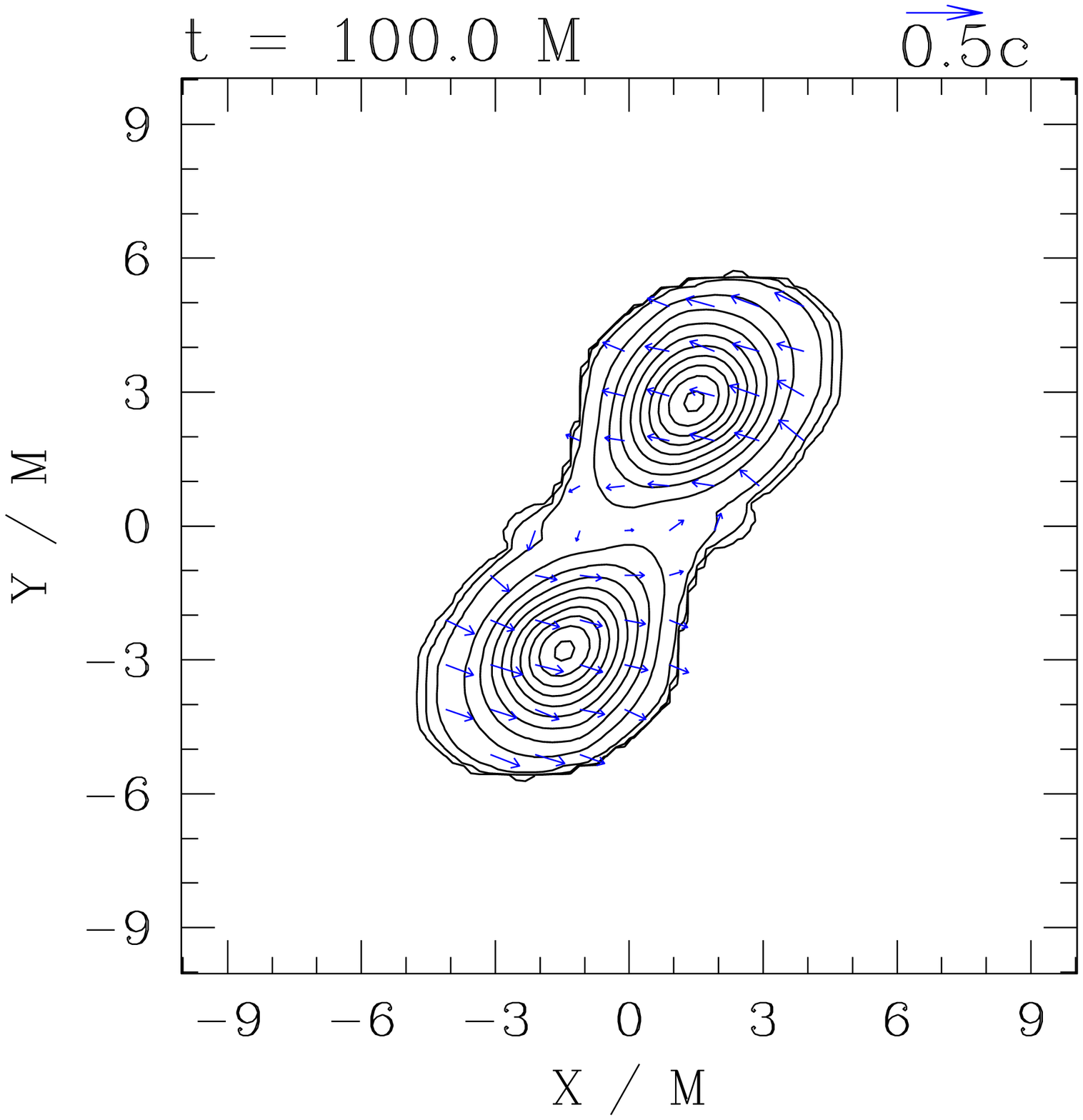}\\
\epsfxsize=2.15in
\leavevmode
\epsffile{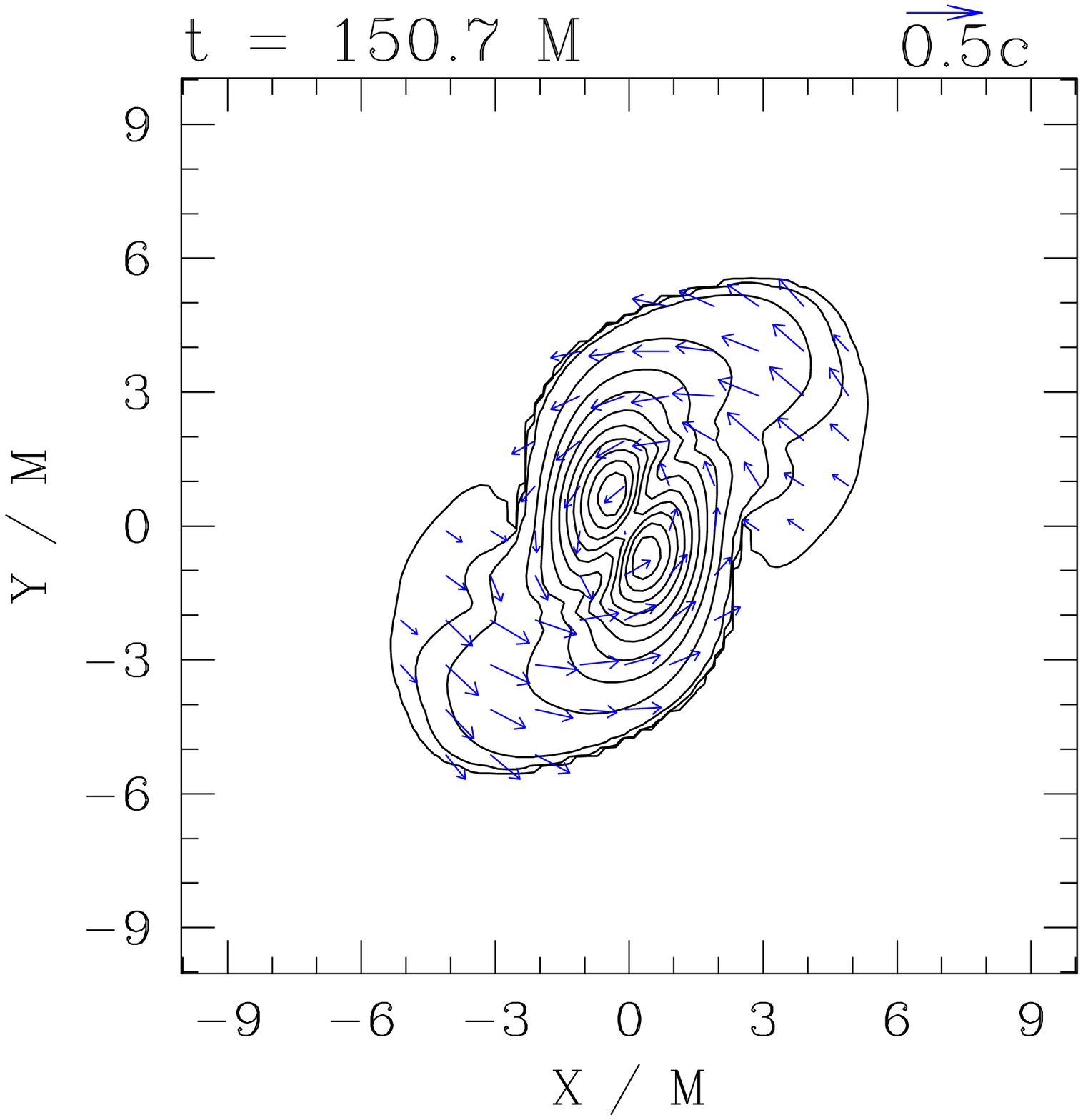}
\epsfxsize=2.15in
\leavevmode
\epsffile{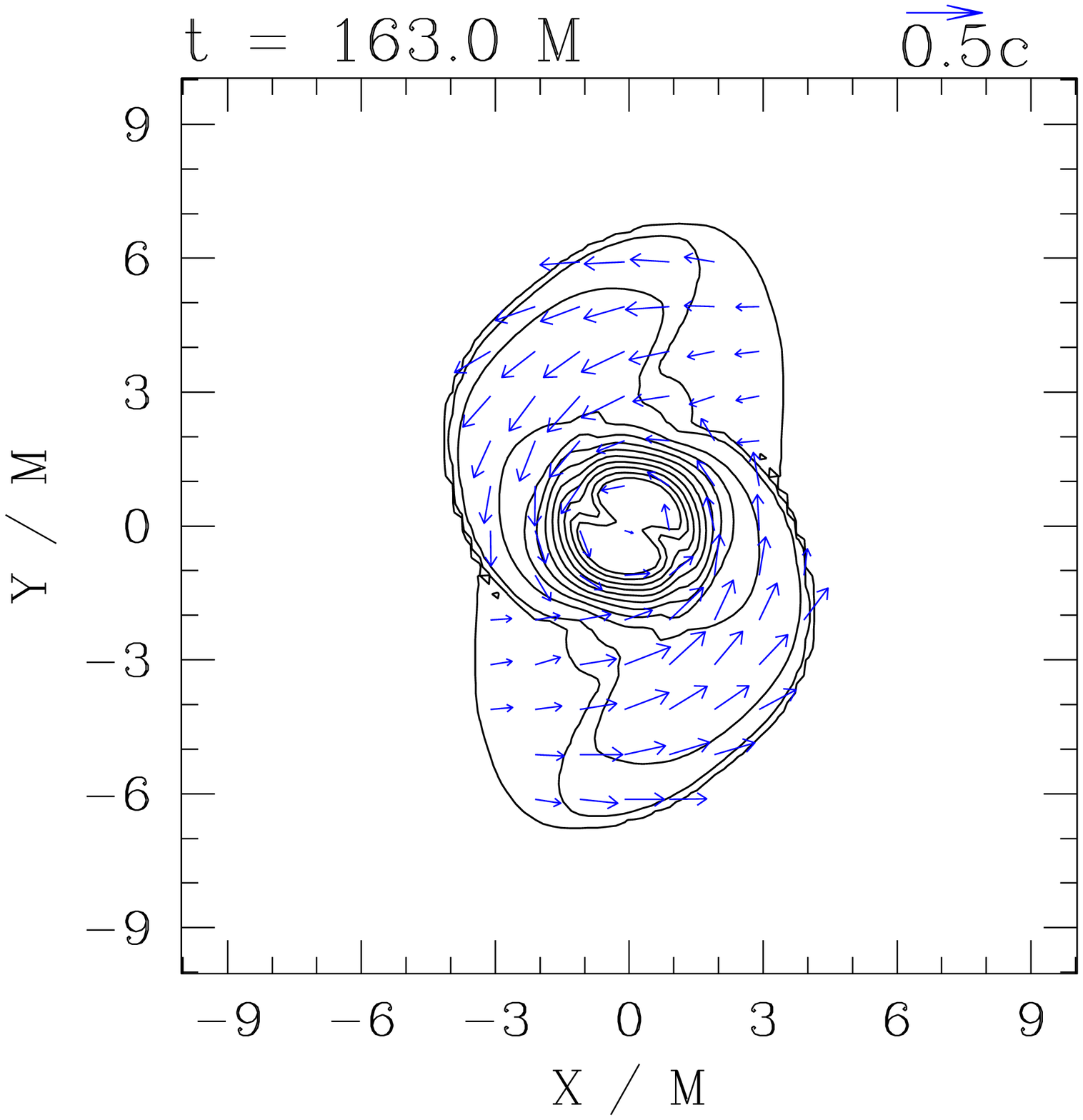}
\epsfxsize=2.15in
\leavevmode
\epsffile{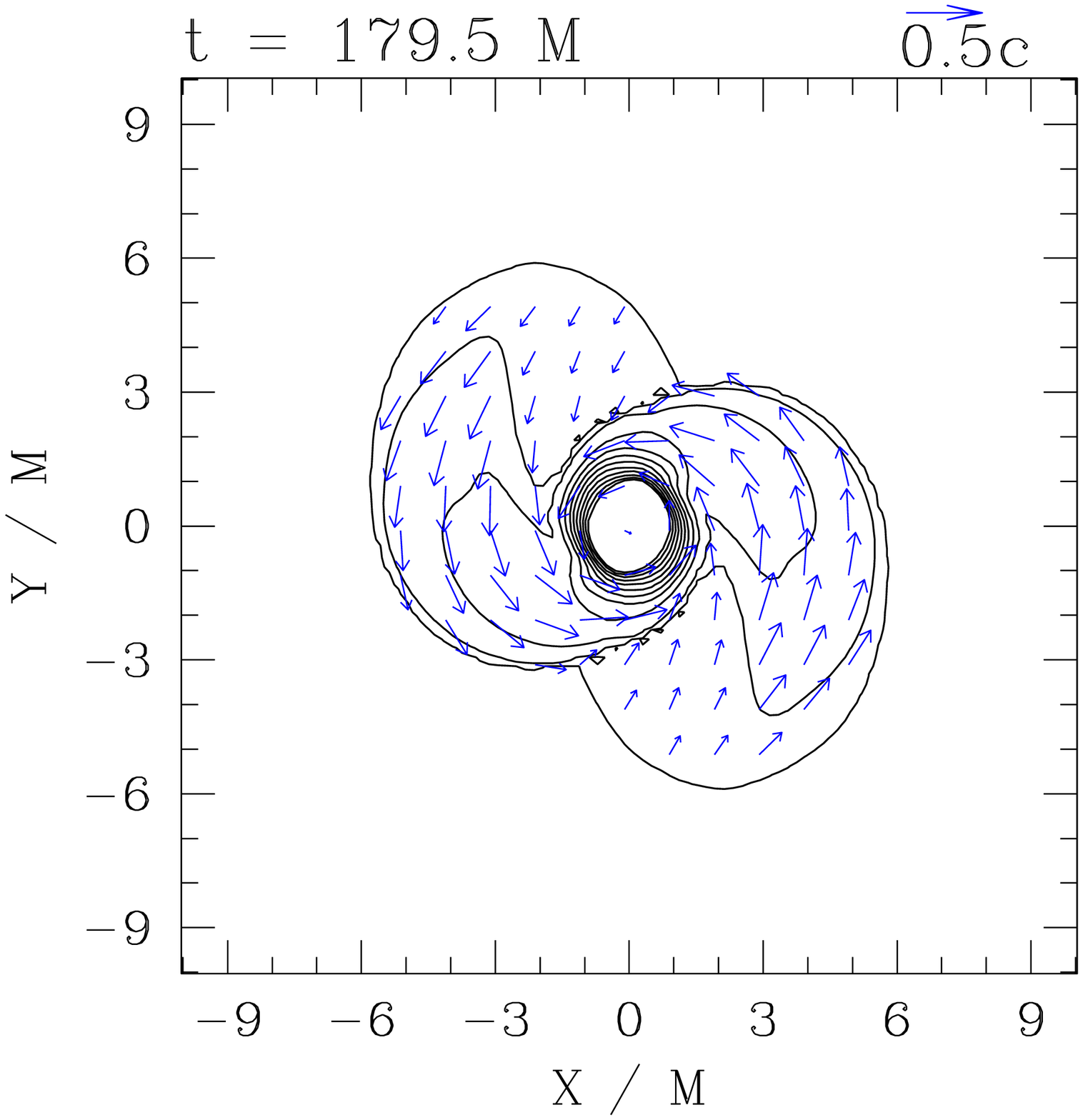}\\
\epsfxsize=2.15in
\leavevmode
\epsffile{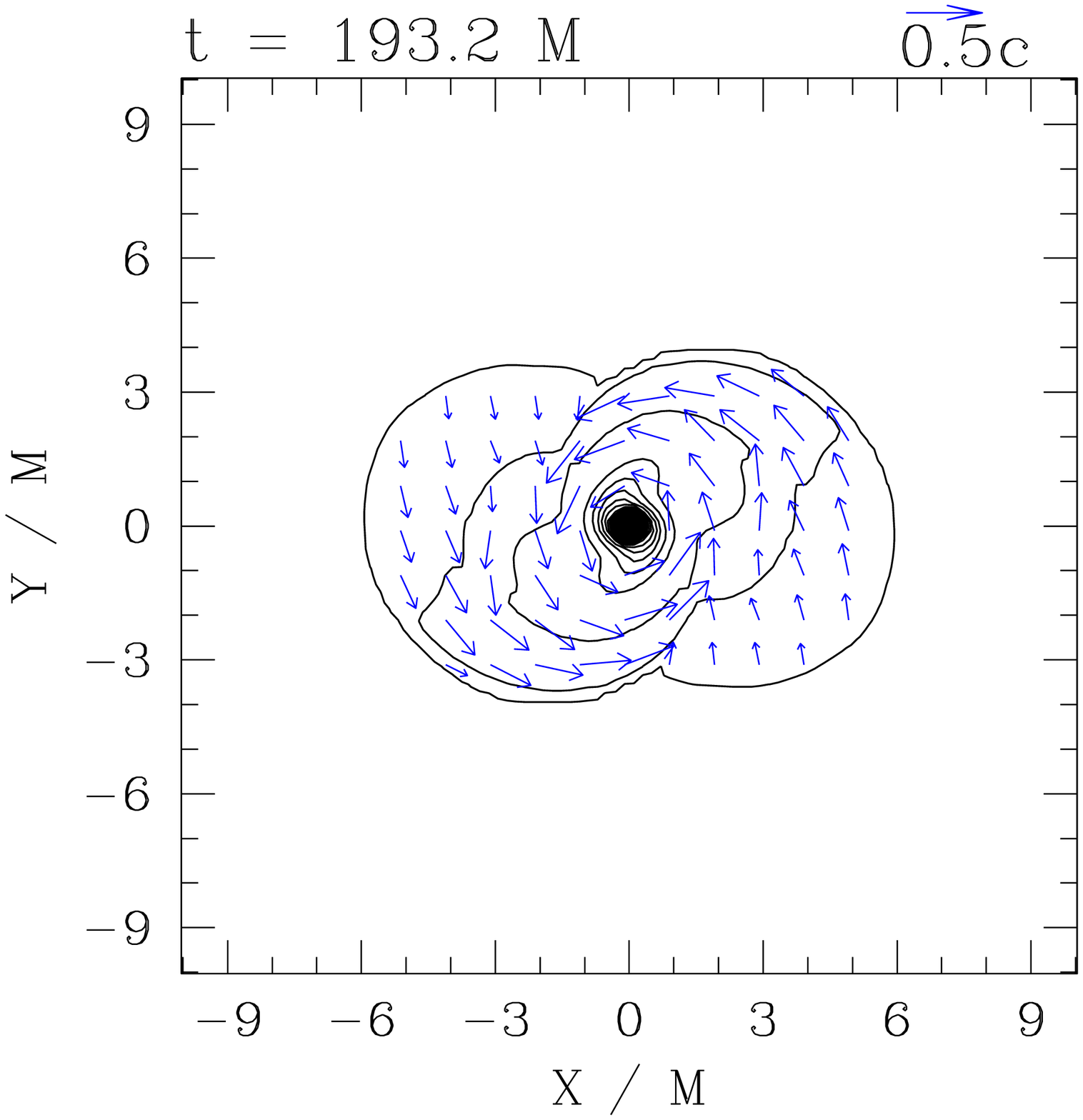}
\epsfxsize=2.15in
\leavevmode
\epsffile{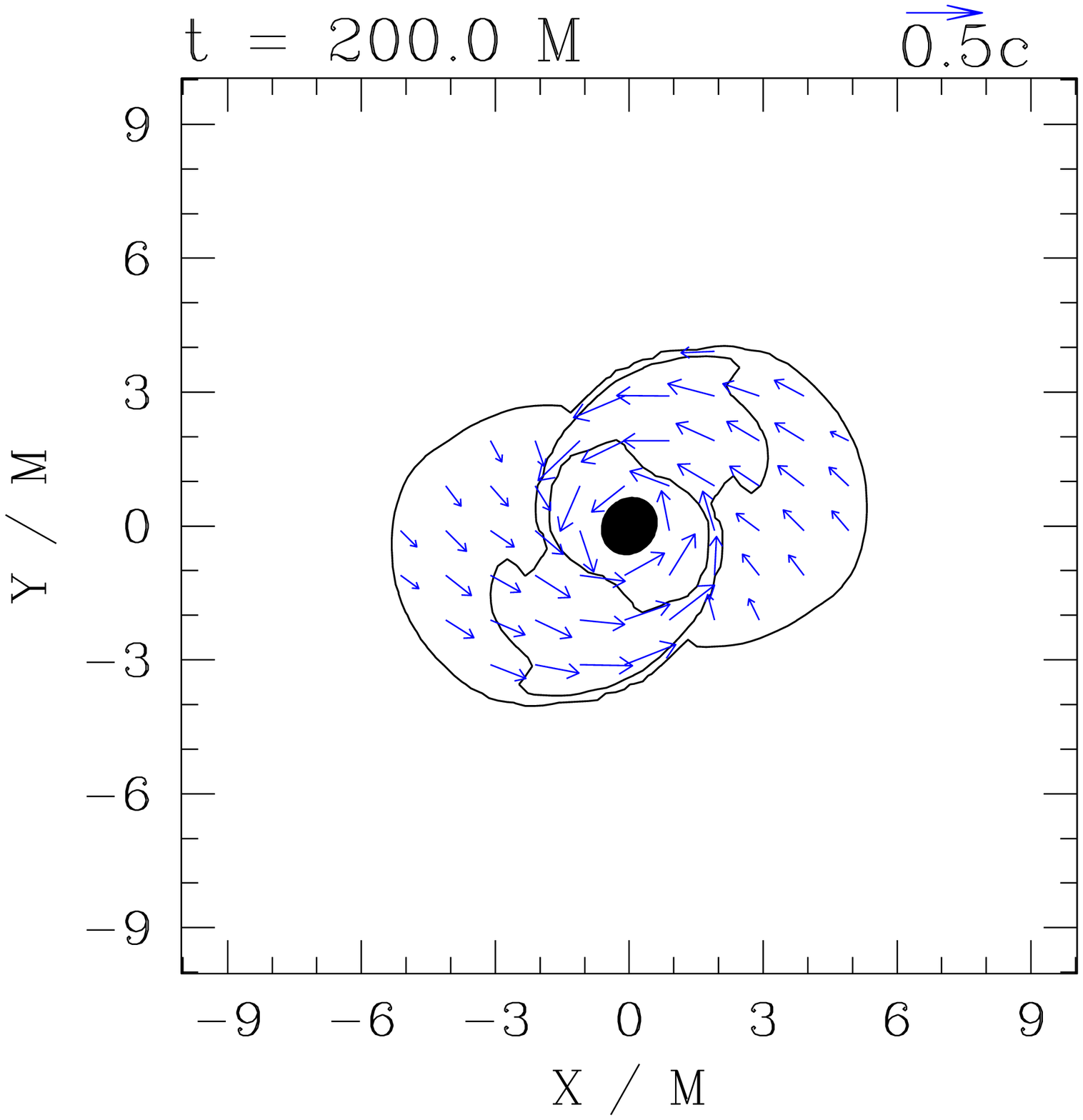}
\epsfxsize=2.15in
\leavevmode
\epsffile{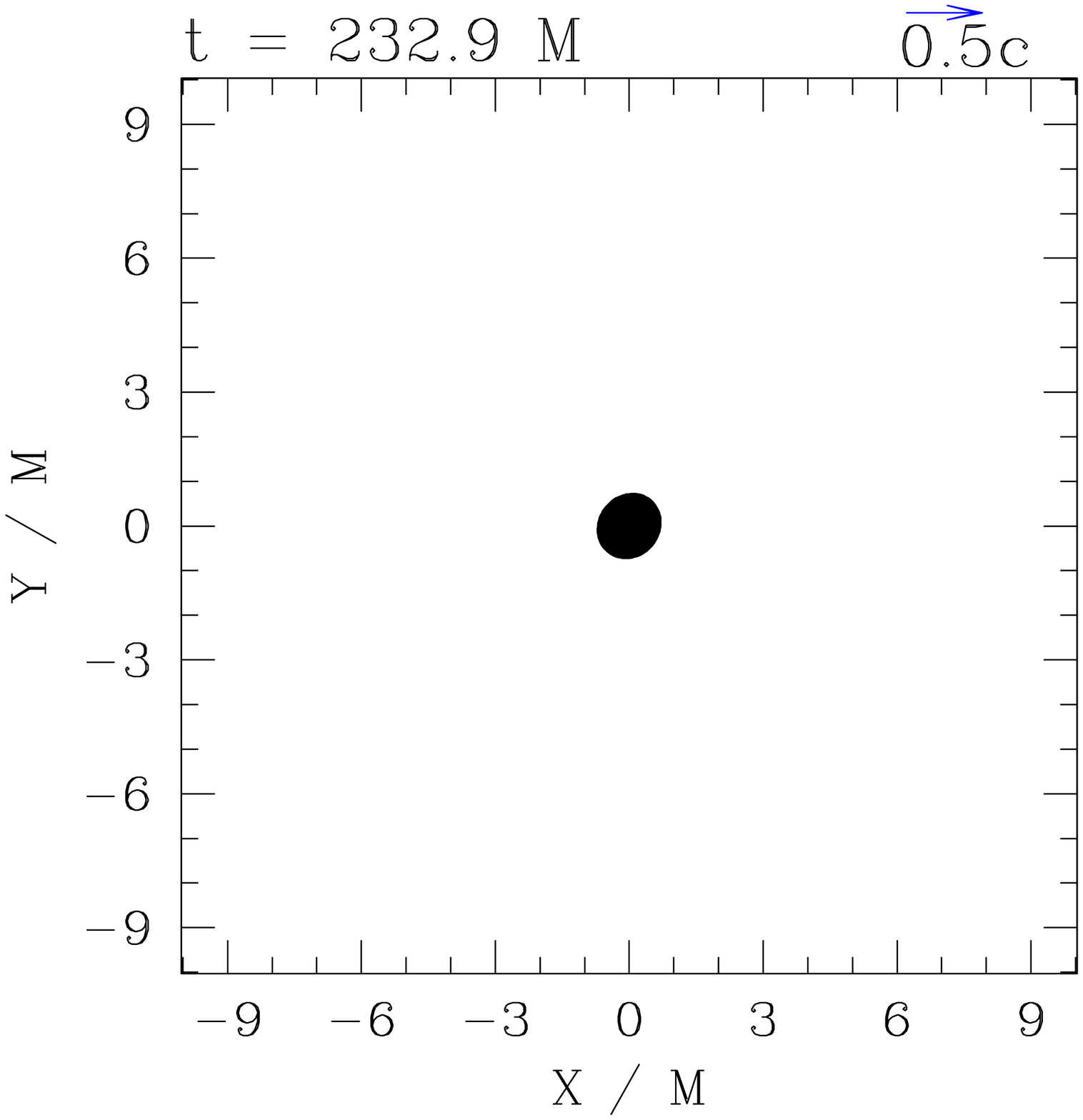}
\caption{Same as Fig.~\ref{fig:M1414_m0_snapshots} but for run
M1616B0 (unmagnetized run). The black region near the center in 
the last three panels denotes the apparent horizon.}
\label{fig:M1616_m0_snapshots}
\end{center}
\end{figure*}

\begin{figure*}
\vspace{-4mm}
\begin{center}
\epsfxsize=2.15in
\leavevmode
\epsffile{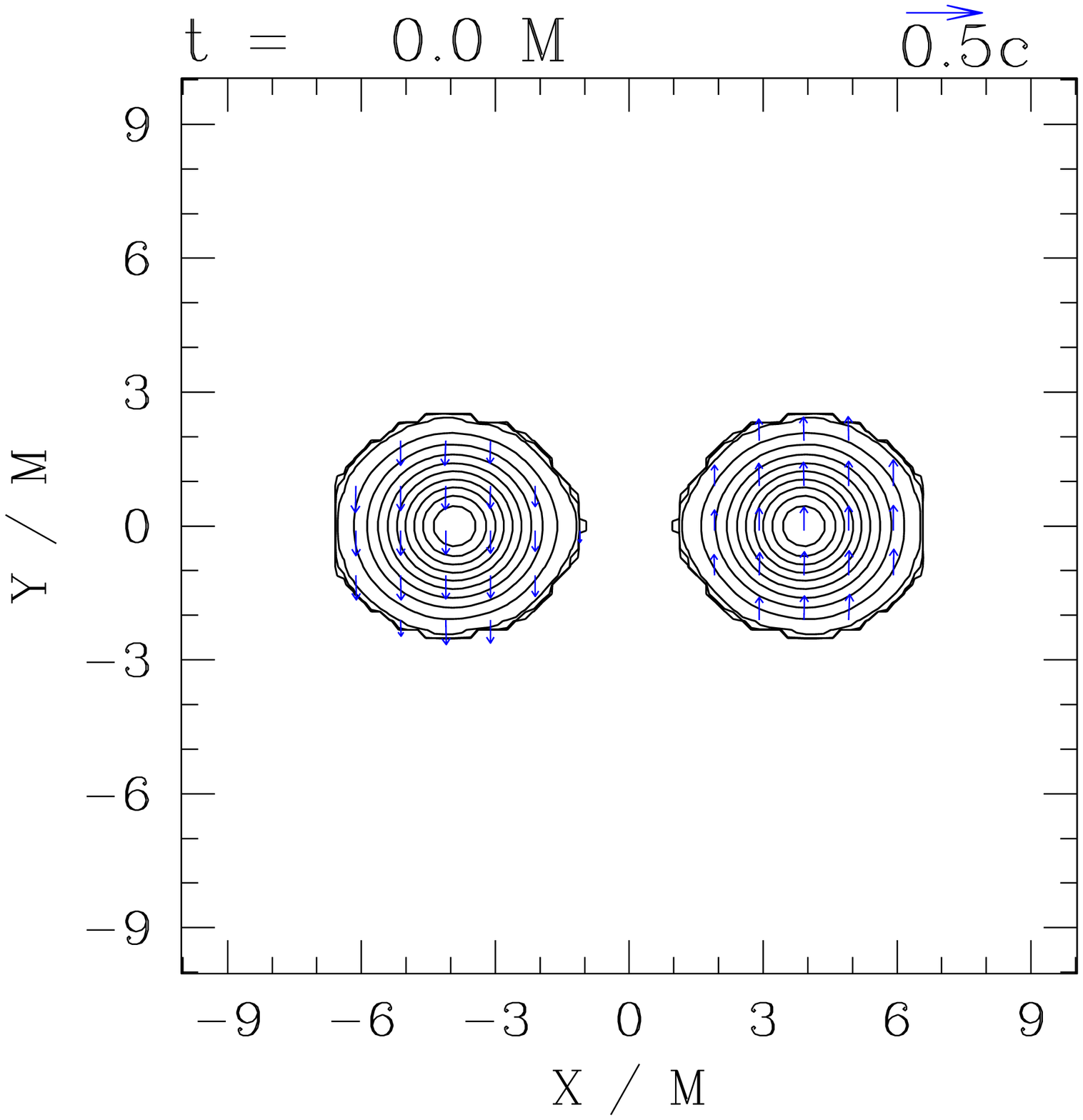}
\epsfxsize=2.15in
\leavevmode
\epsffile{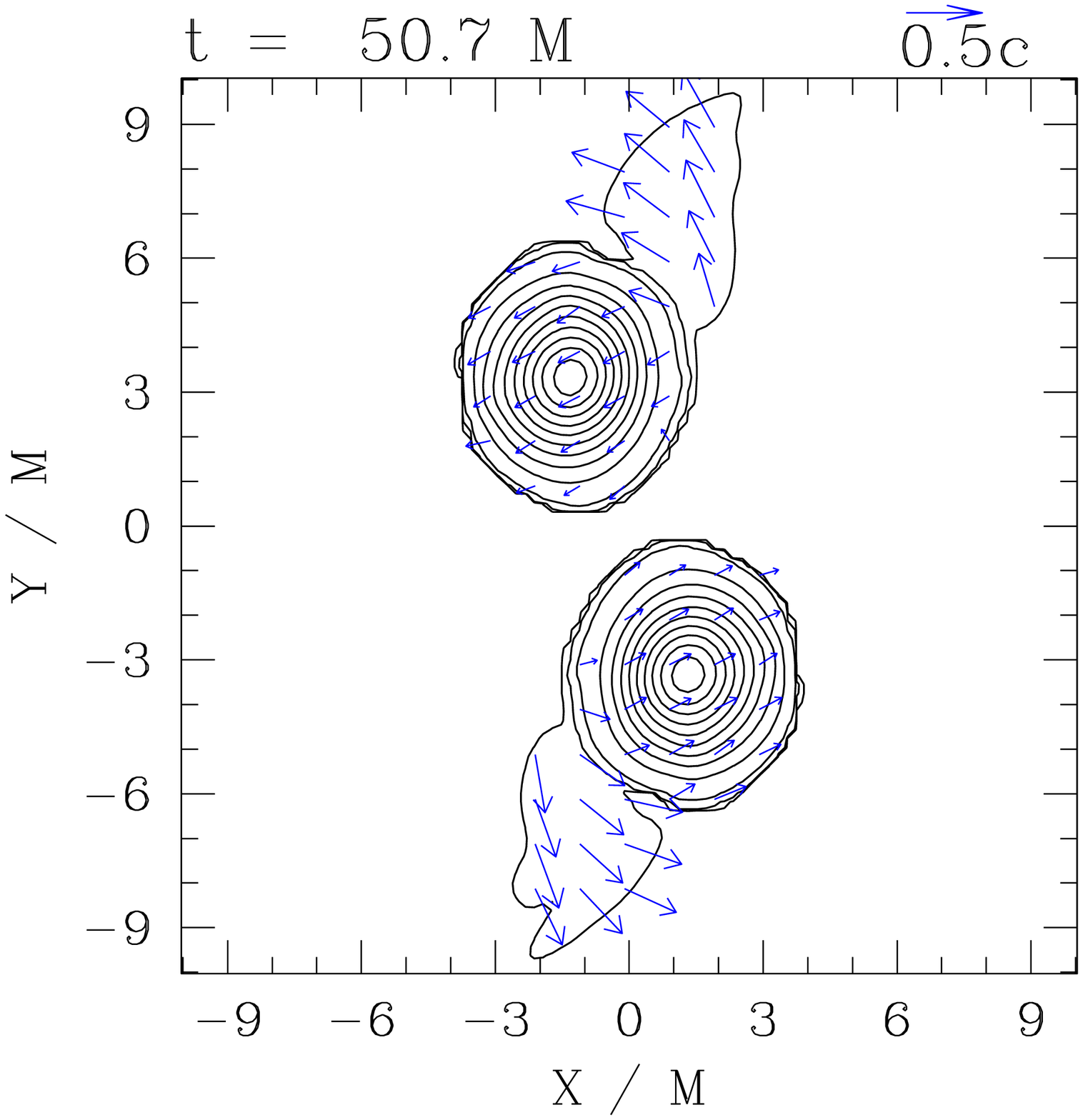}
\epsfxsize=2.15in
\leavevmode
\epsffile{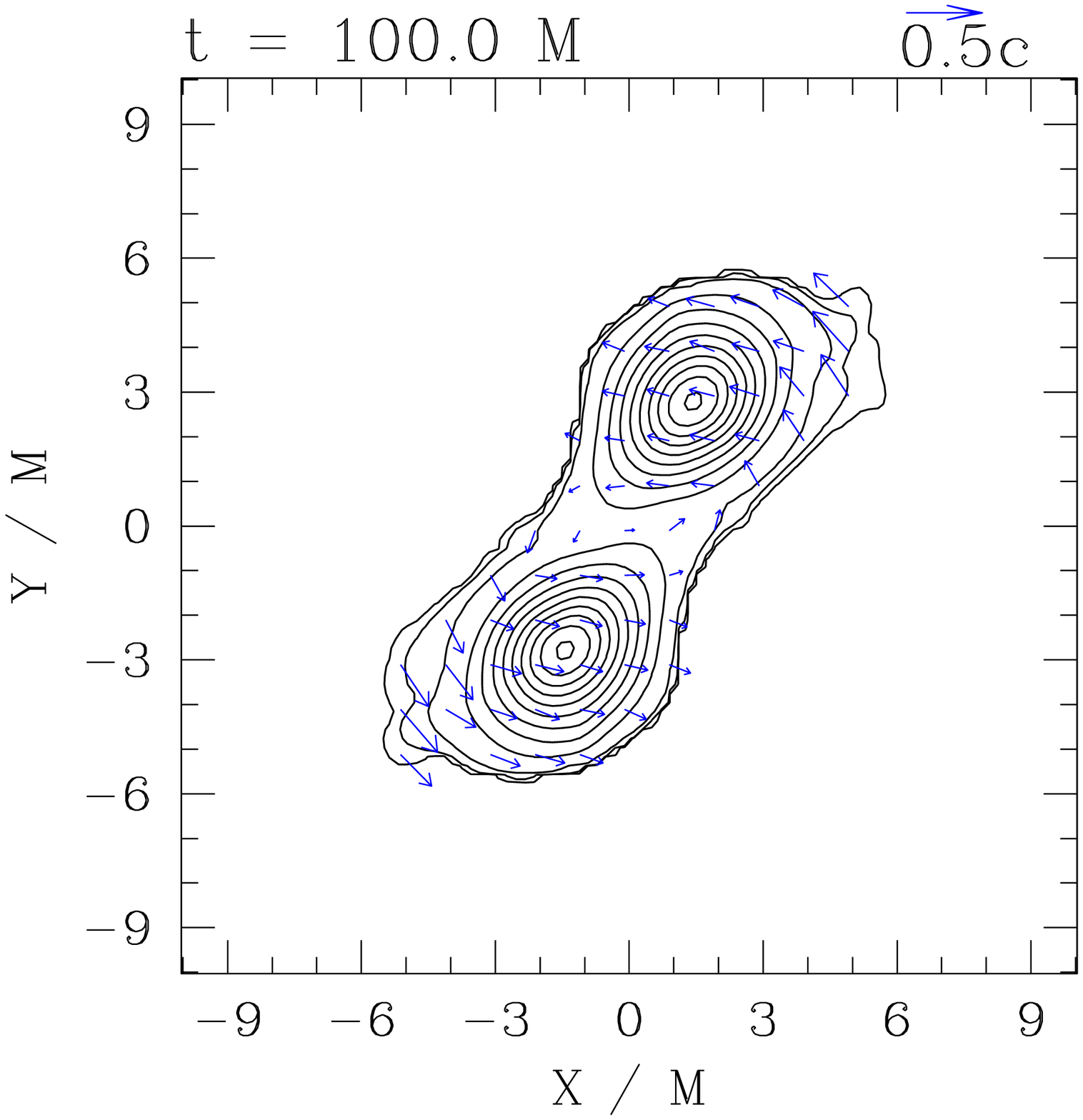}\\
\epsfxsize=2.15in
\leavevmode
\epsffile{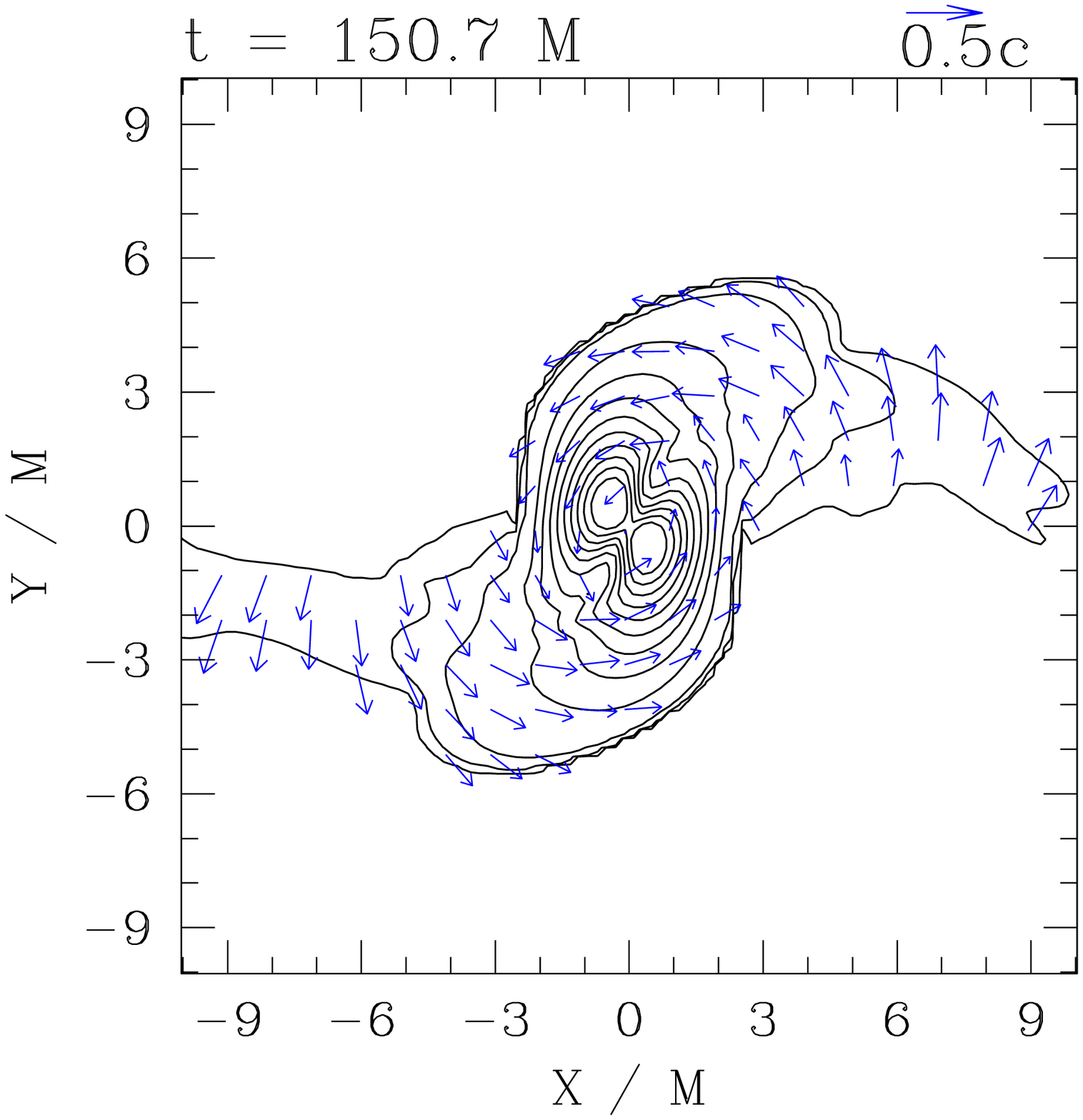}
\epsfxsize=2.15in
\leavevmode
\epsffile{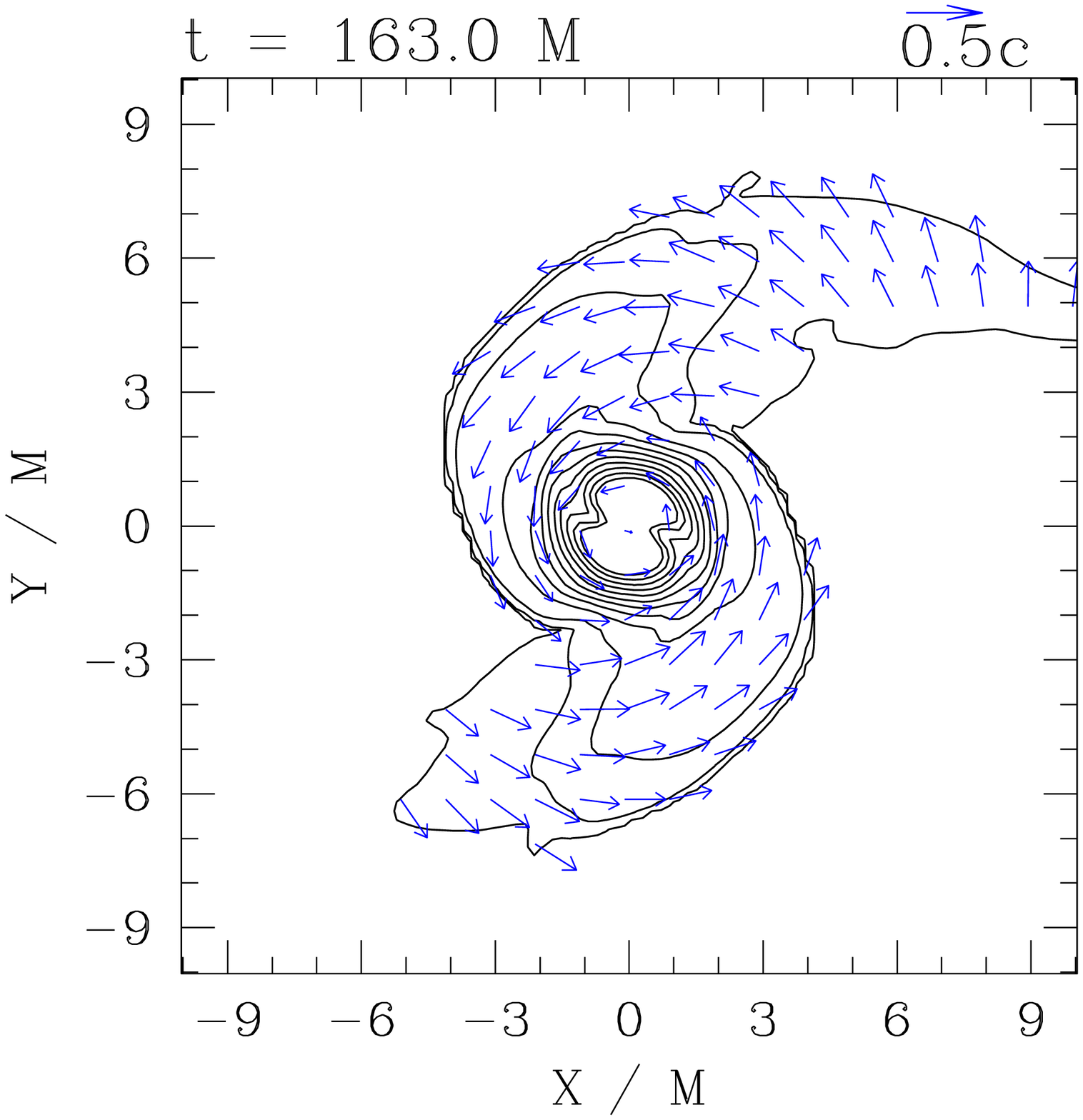}
\epsfxsize=2.15in
\leavevmode
\epsffile{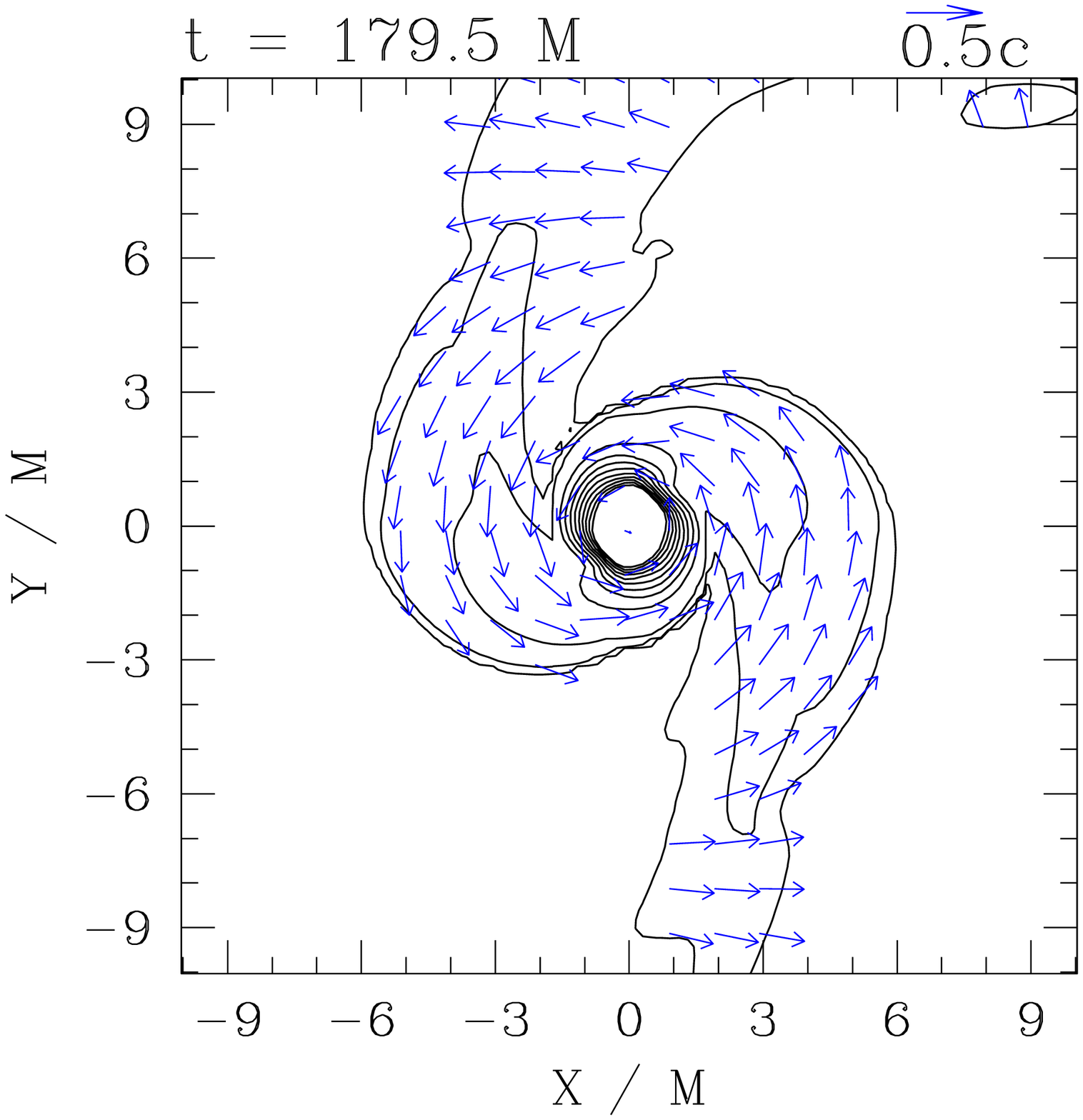}\\
\epsfxsize=2.15in
\leavevmode
\epsffile{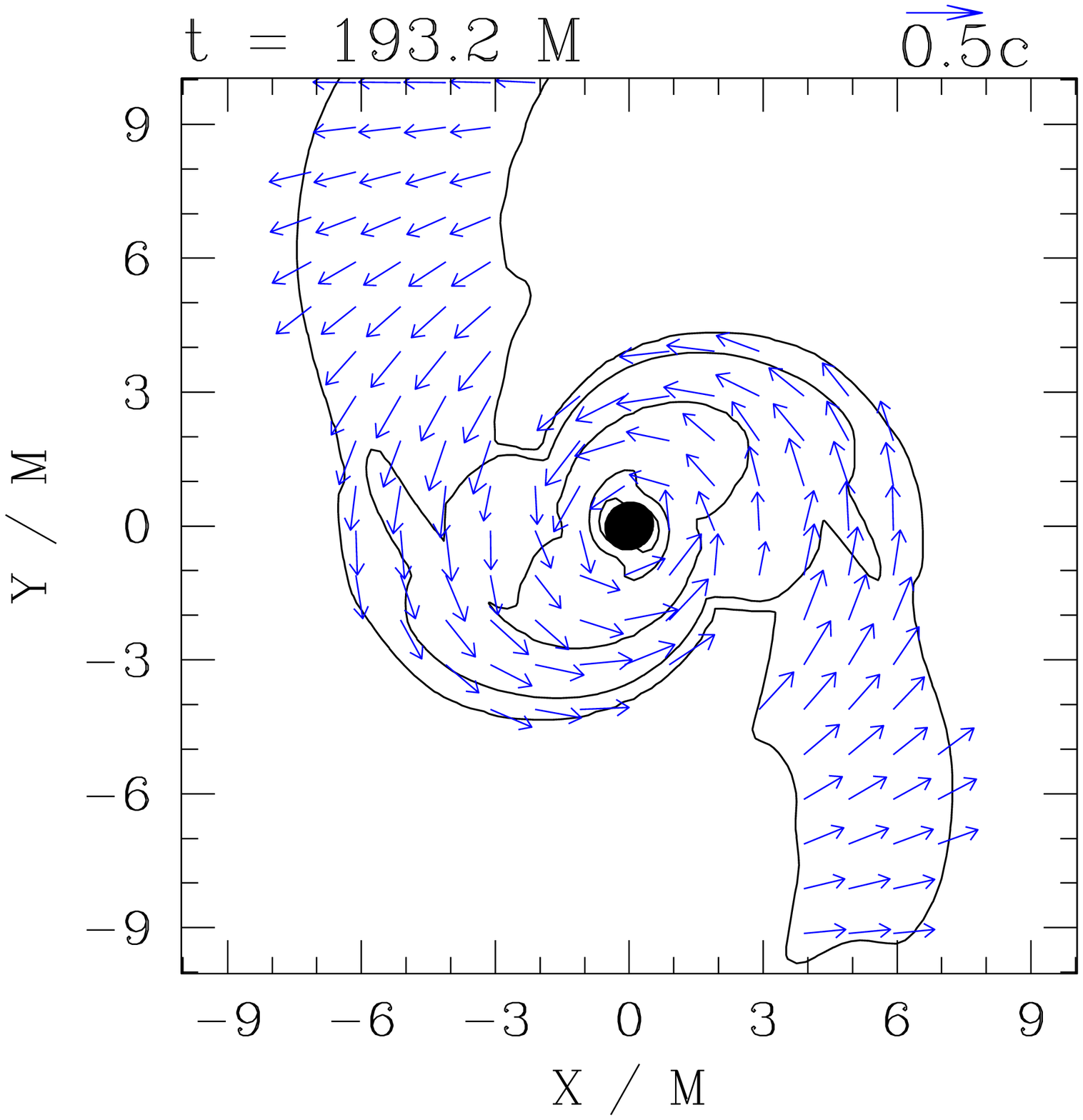}
\epsfxsize=2.15in
\leavevmode
\epsffile{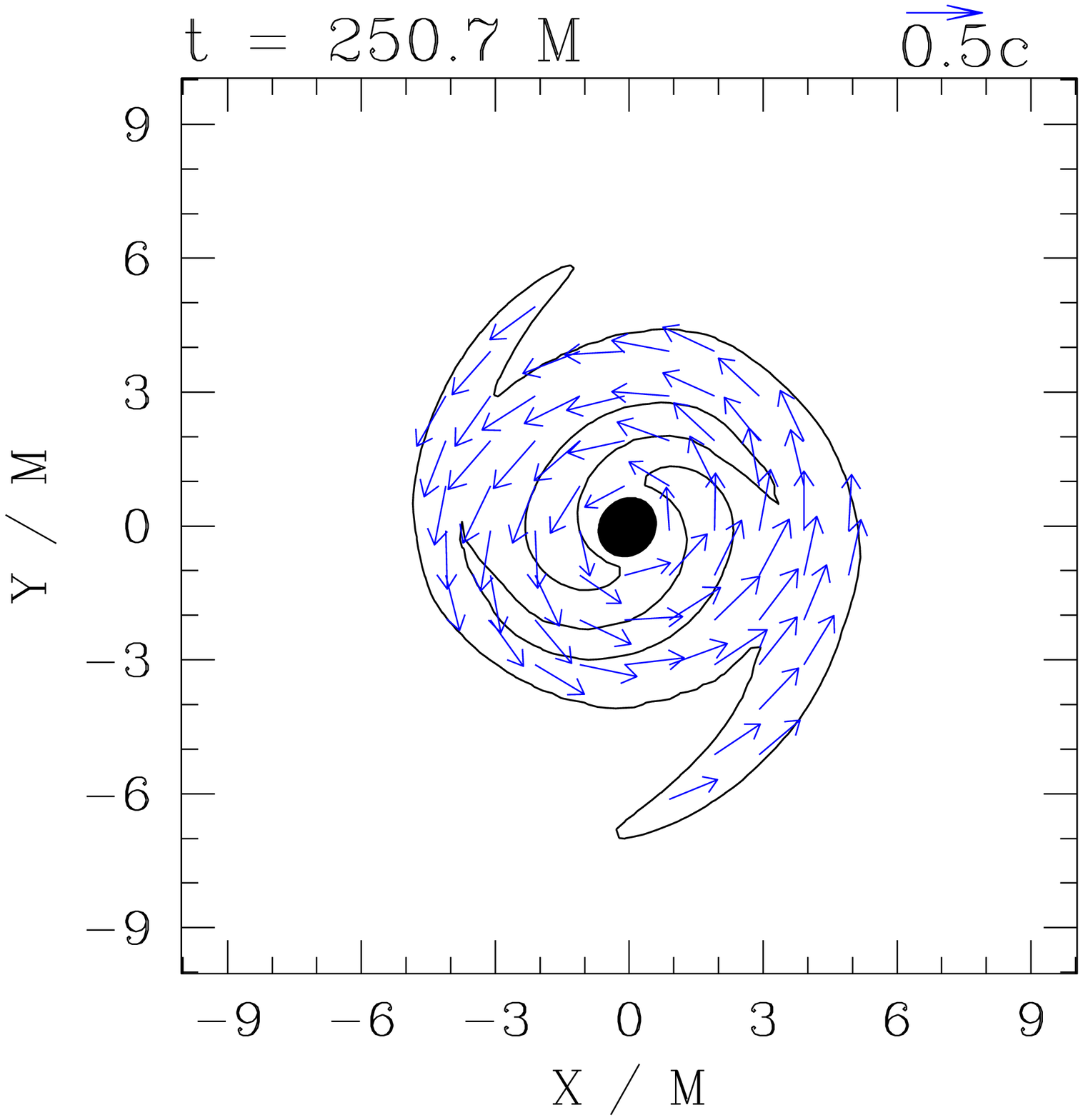}
\epsfxsize=2.15in
\leavevmode
\epsffile{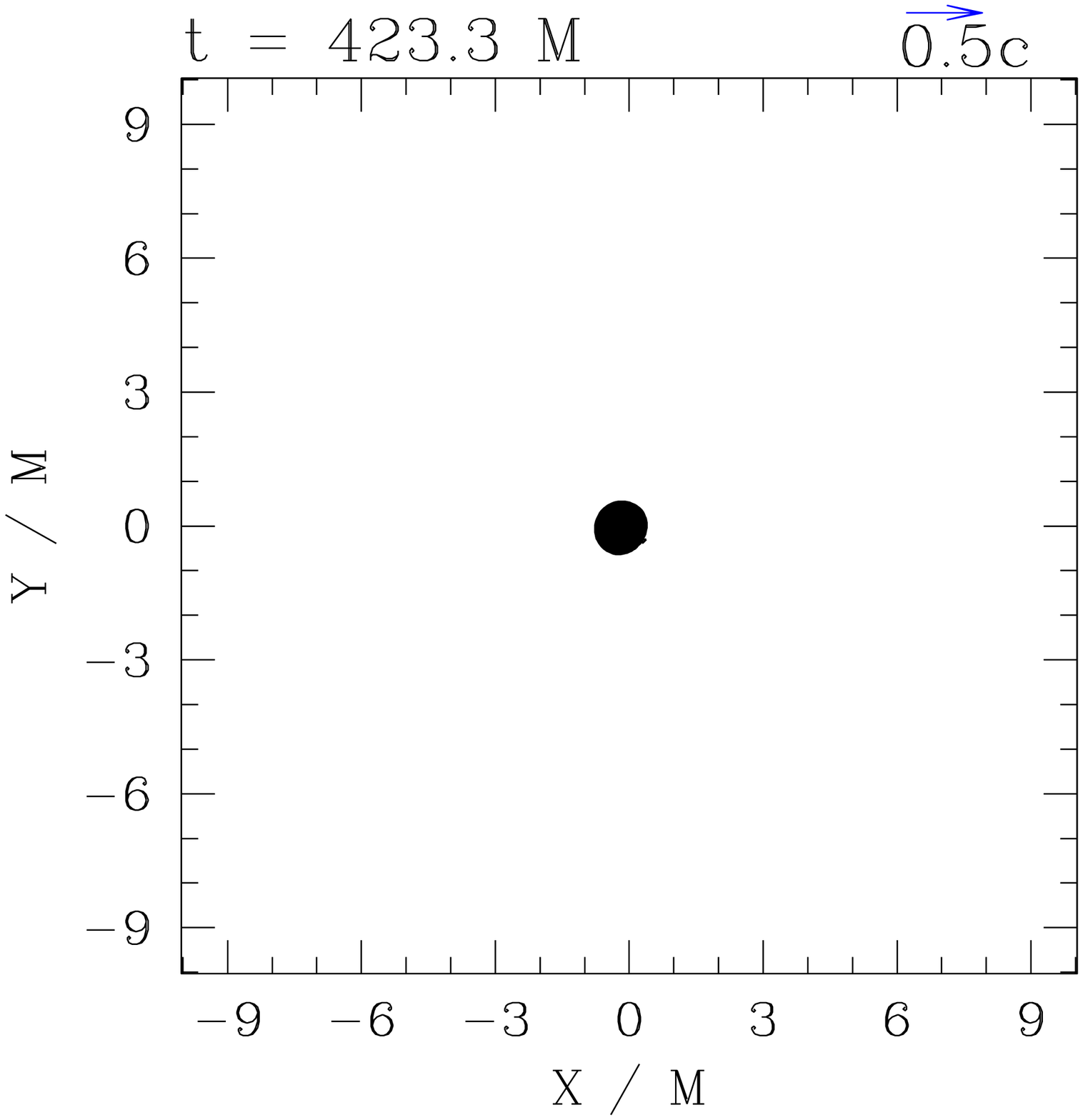}
\caption{Same as Fig.~\ref{fig:M1616_m0_snapshots} but for run
M1616B1 (magnetized run).}
\label{fig:M1616_m1_snapshots}
\end{center}
\end{figure*}

\begin{figure*}
\vspace{-4mm}
\begin{center}
\epsfxsize=2.15in
\leavevmode
\epsffile{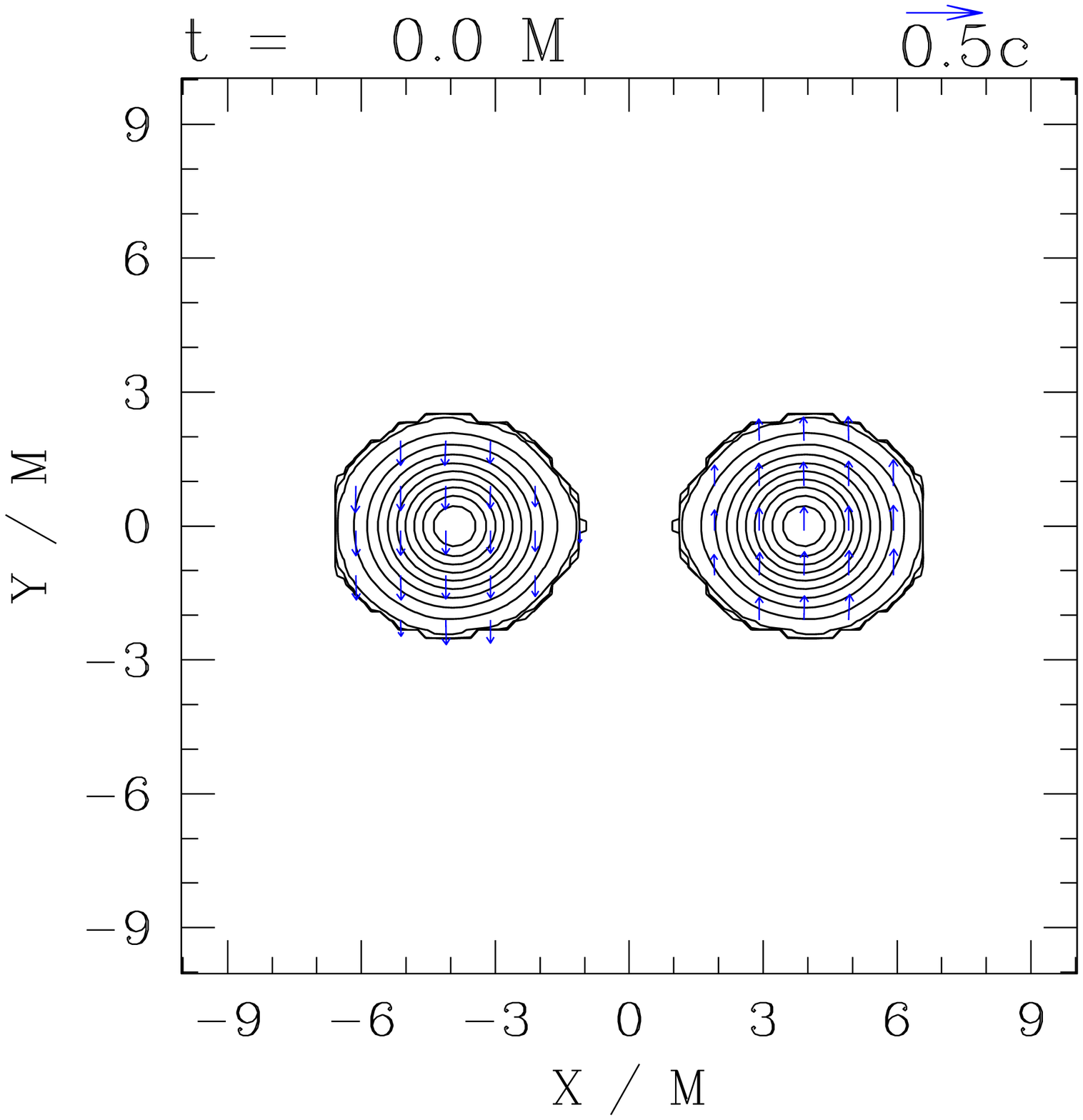}
\epsfxsize=2.15in
\leavevmode
\epsffile{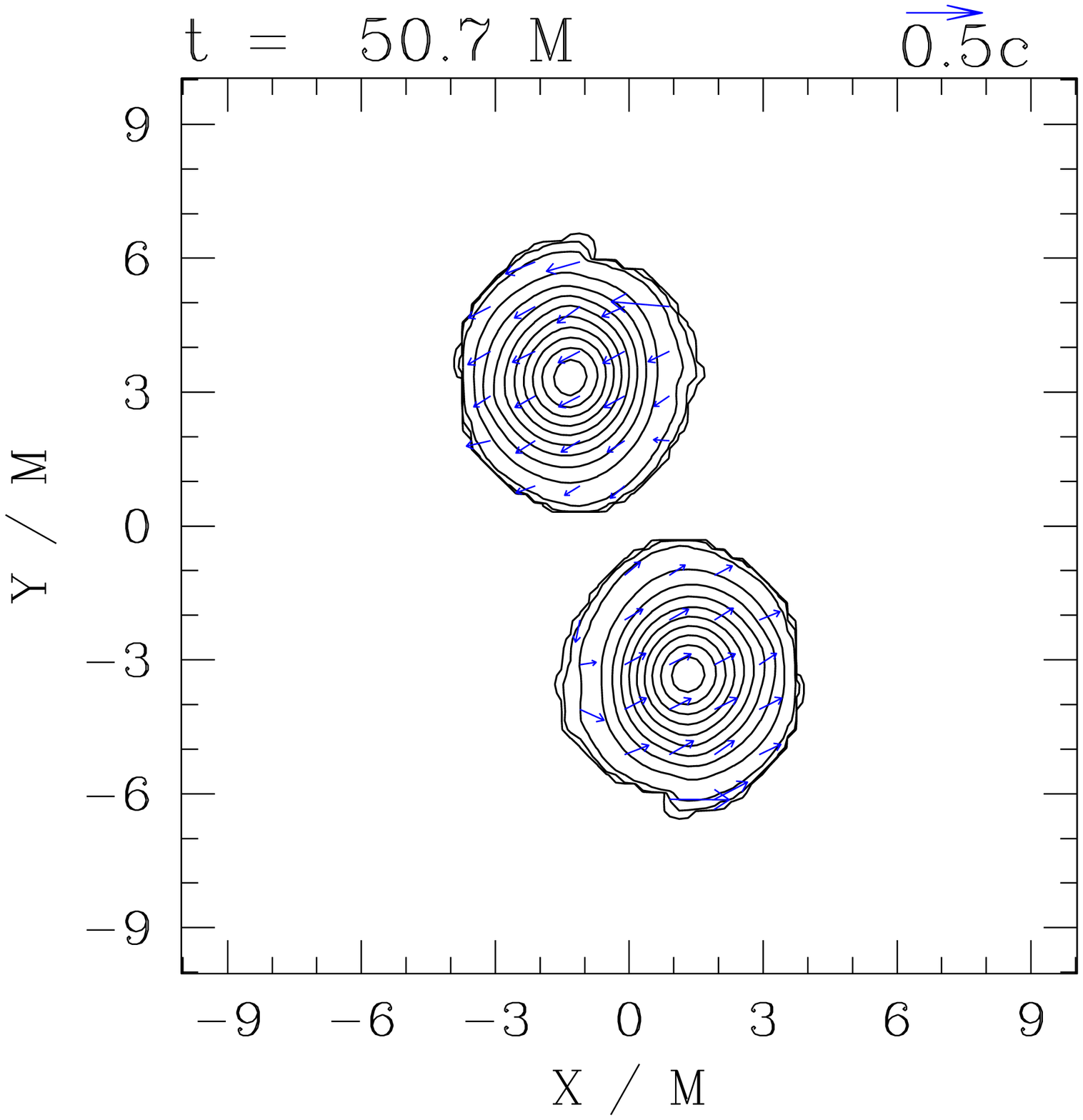}
\epsfxsize=2.15in
\leavevmode
\epsffile{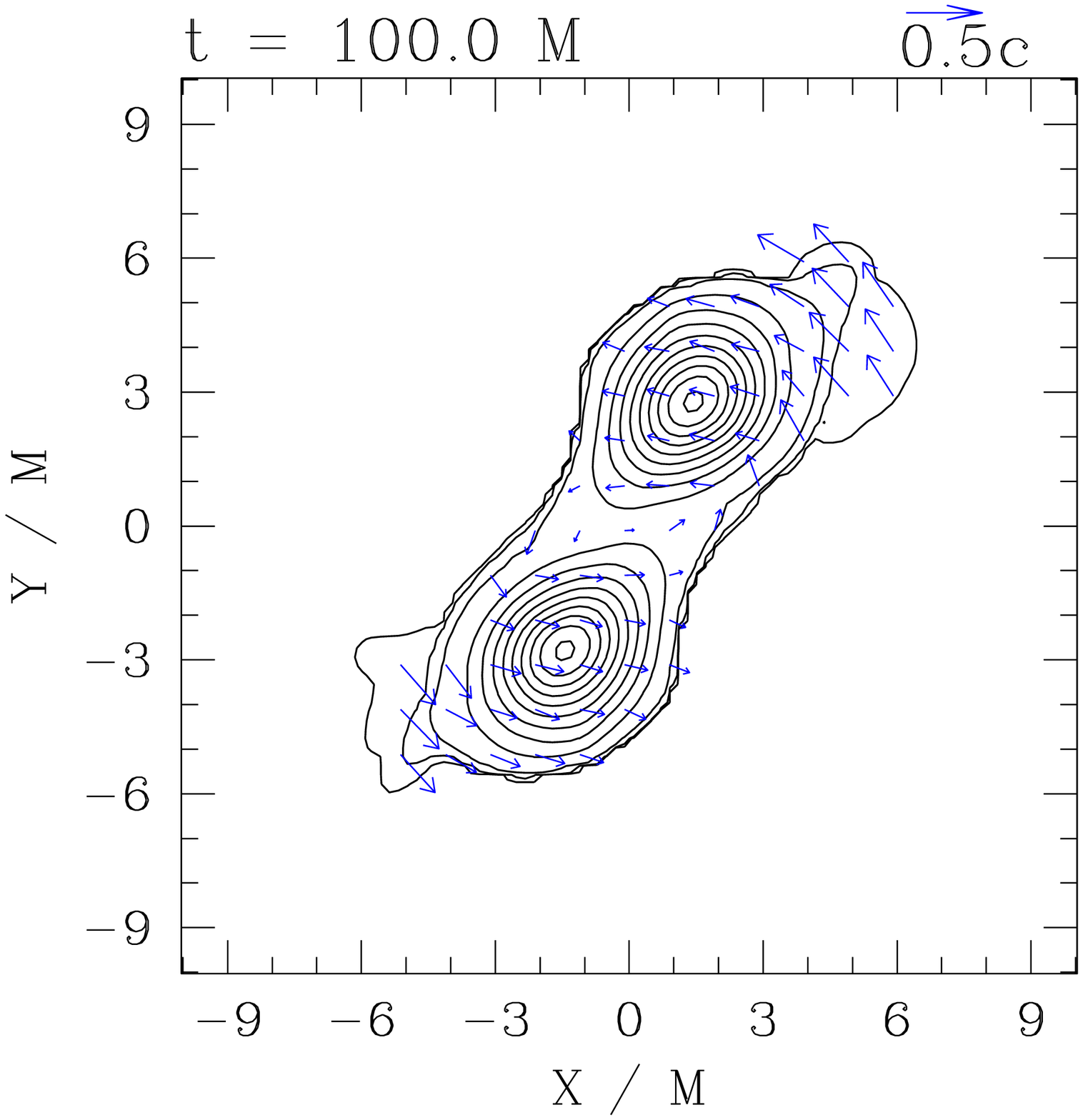}\\
\epsfxsize=2.15in
\leavevmode
\epsffile{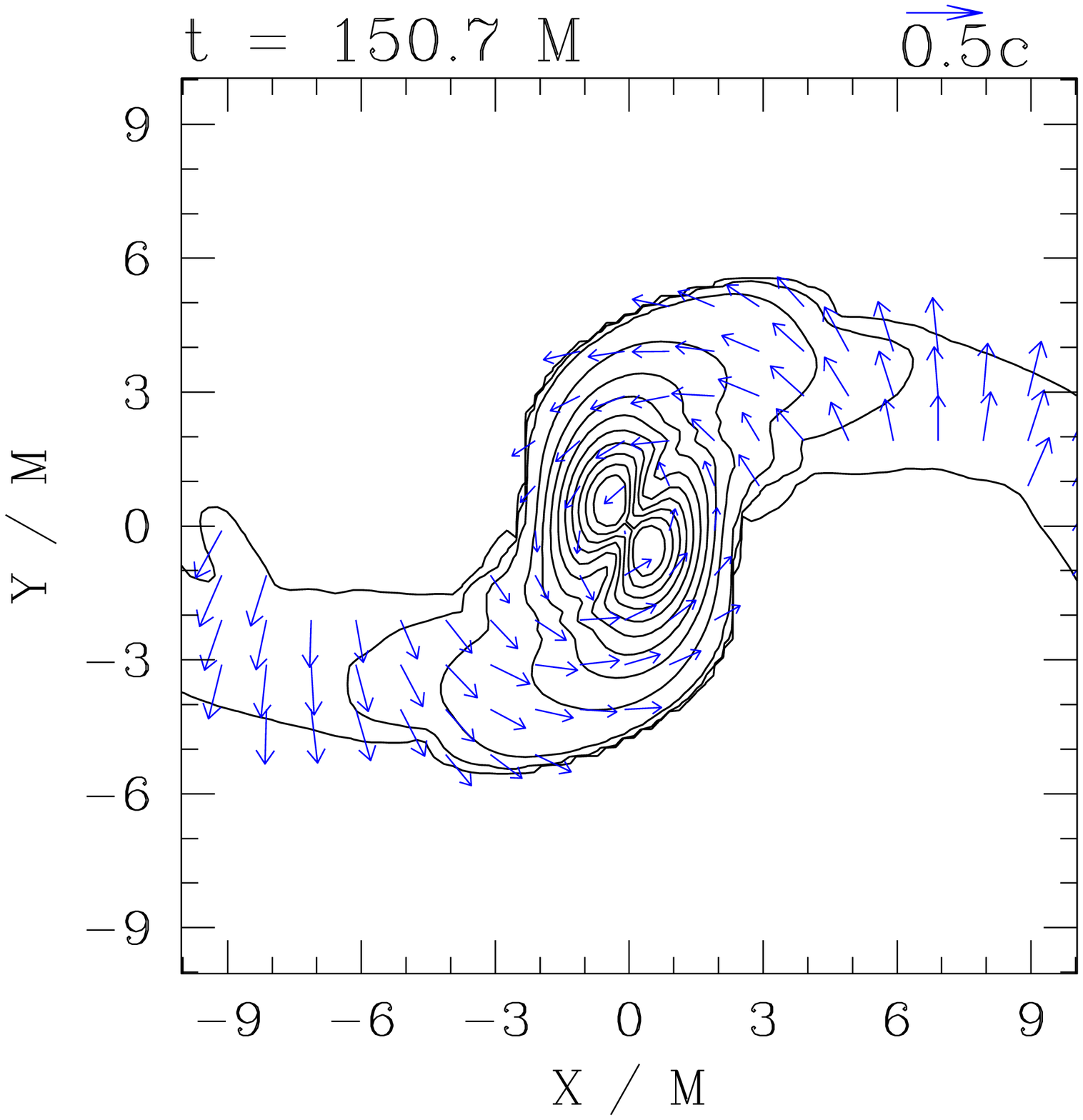}
\epsfxsize=2.15in
\leavevmode
\epsffile{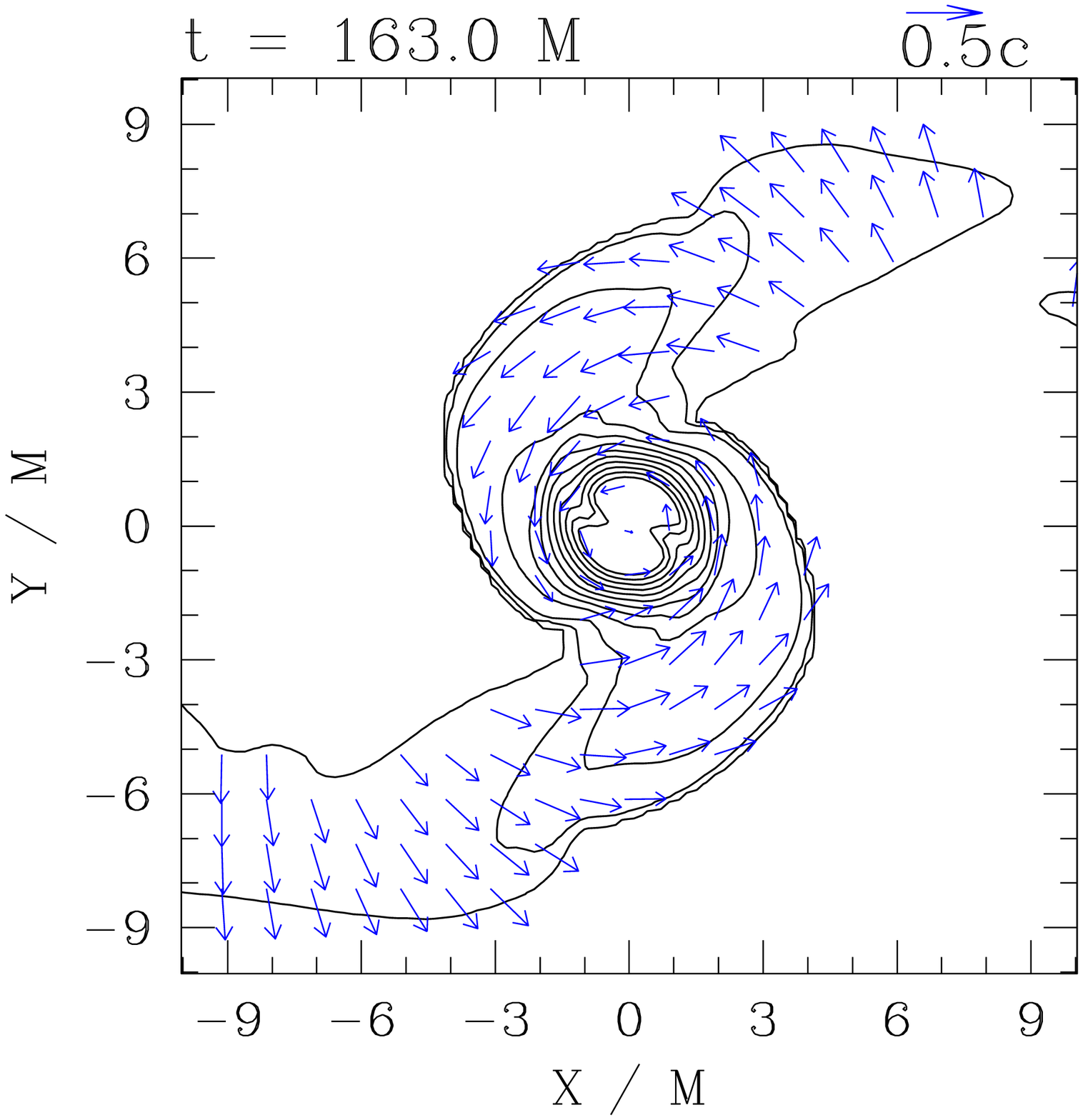}
\epsfxsize=2.15in
\leavevmode
\epsffile{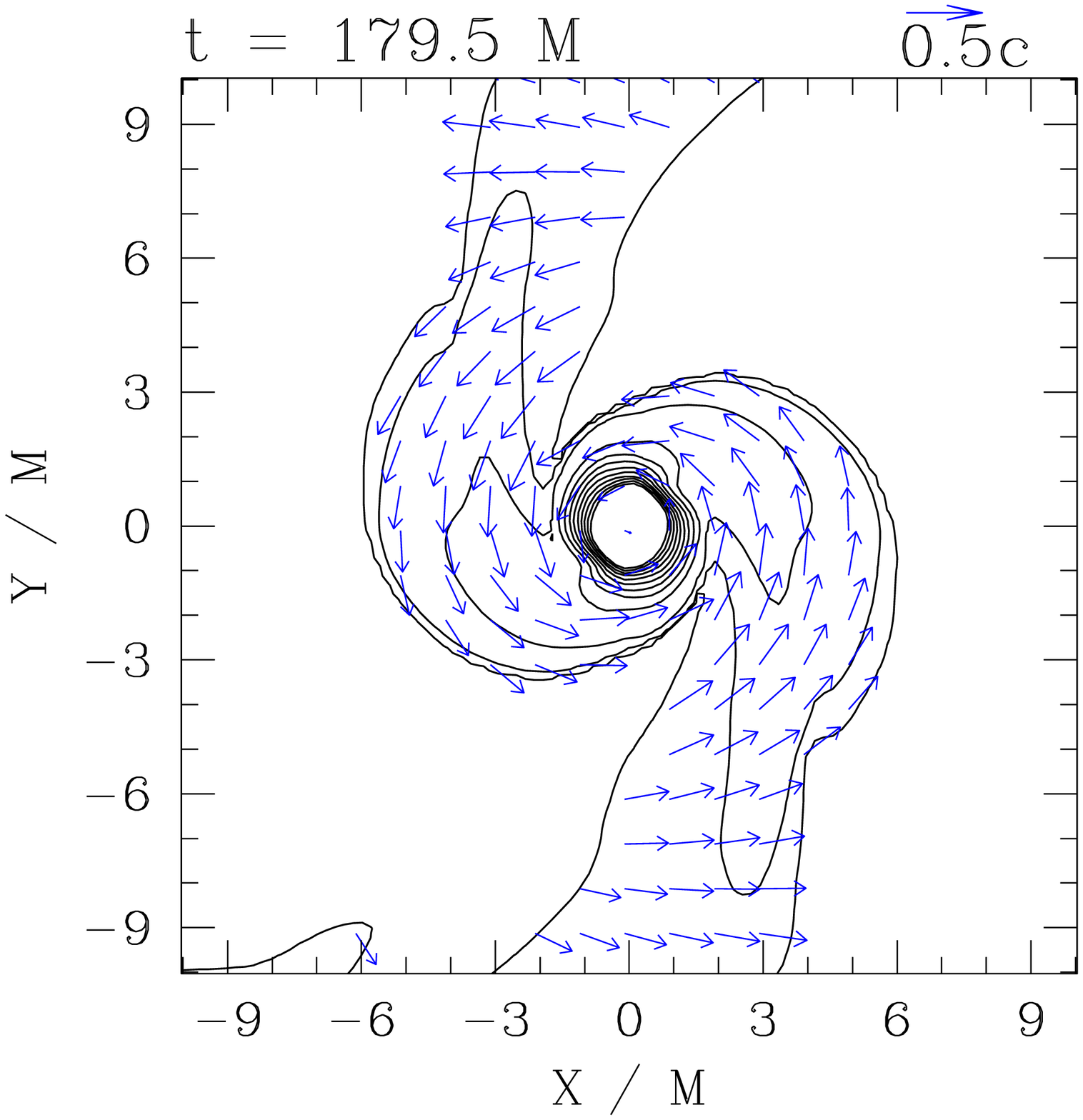}\\
\epsfxsize=2.15in
\leavevmode
\epsffile{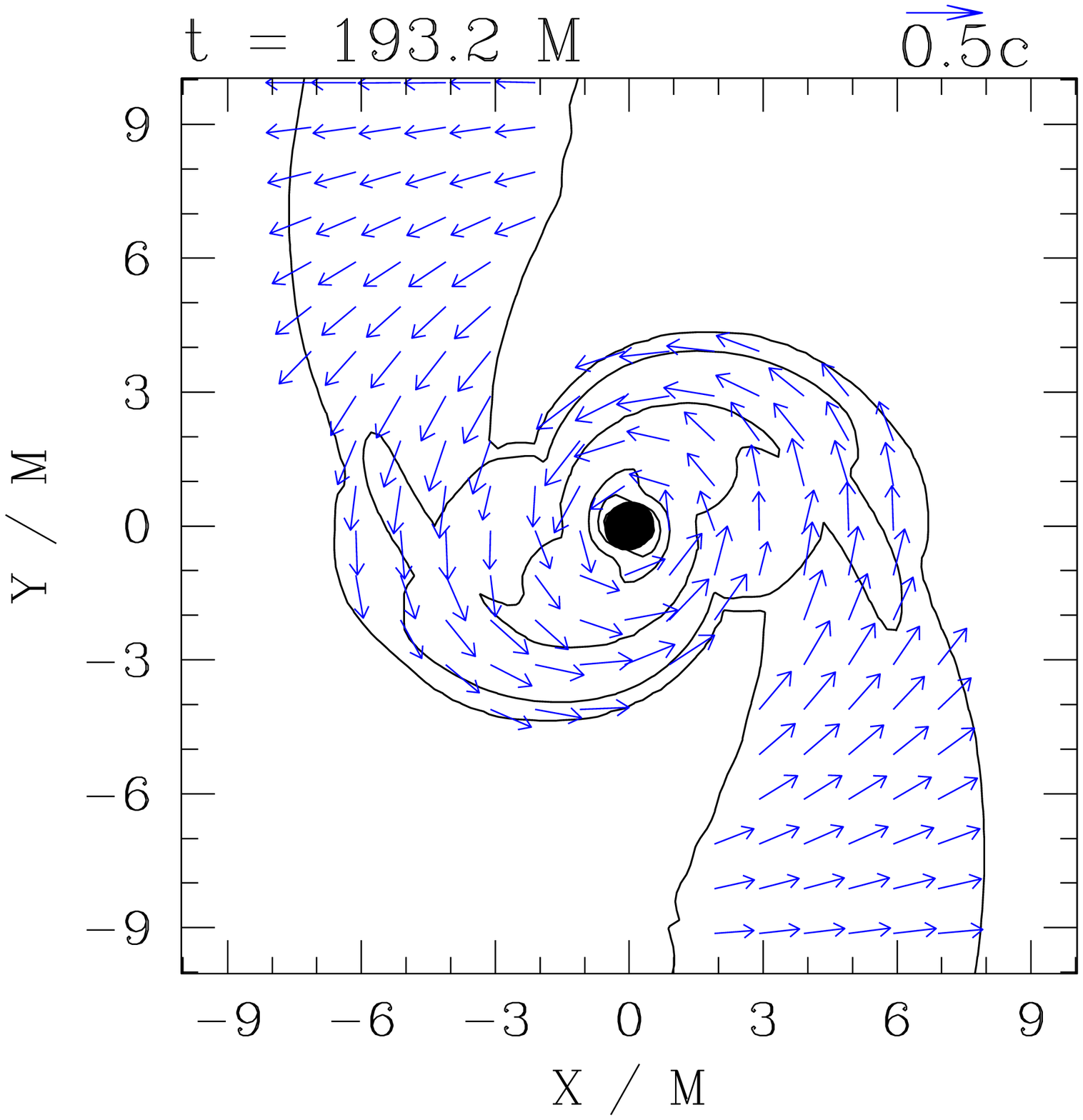}
\epsfxsize=2.15in
\leavevmode
\epsffile{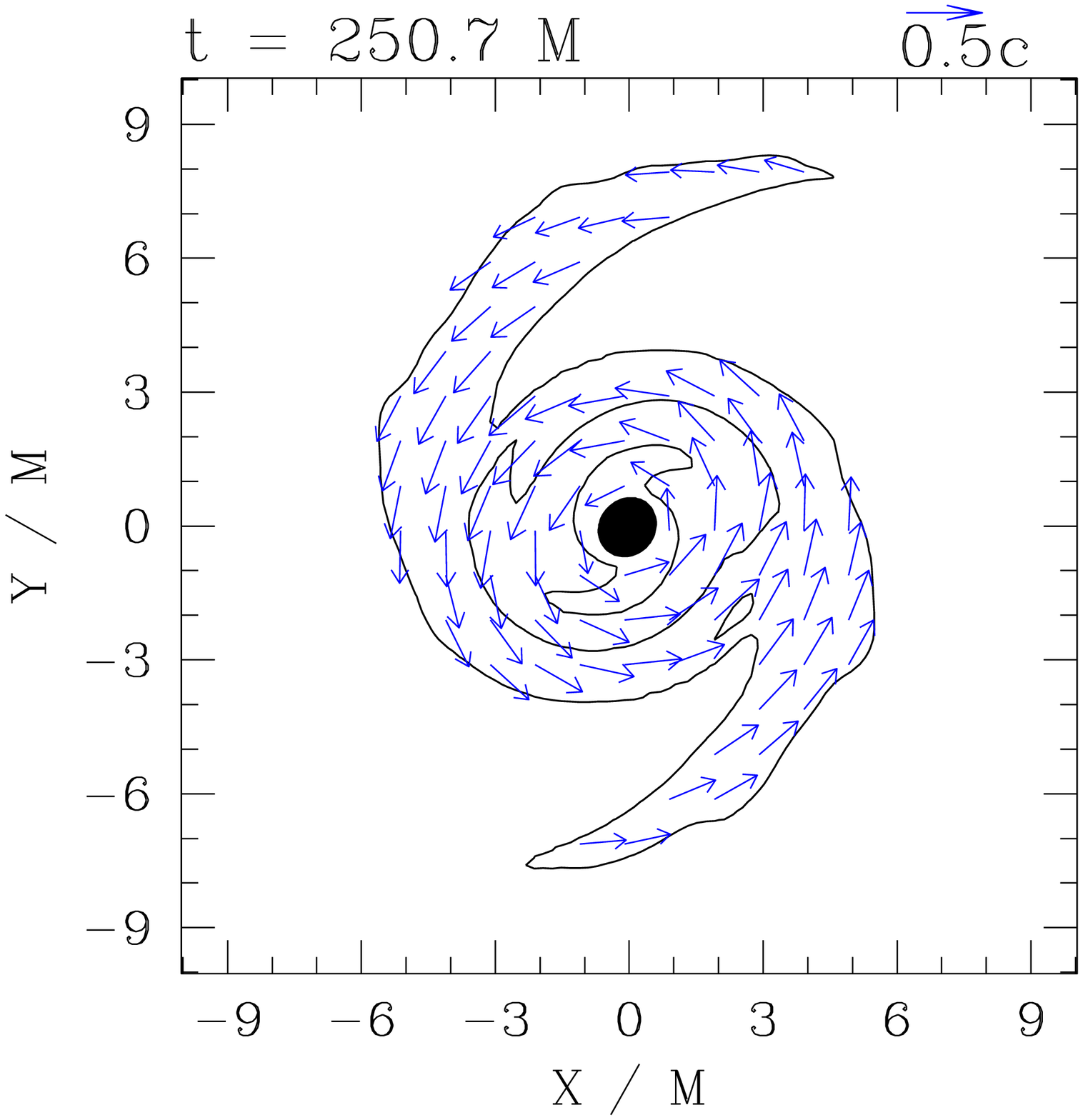}
\epsfxsize=2.15in
\leavevmode
\epsffile{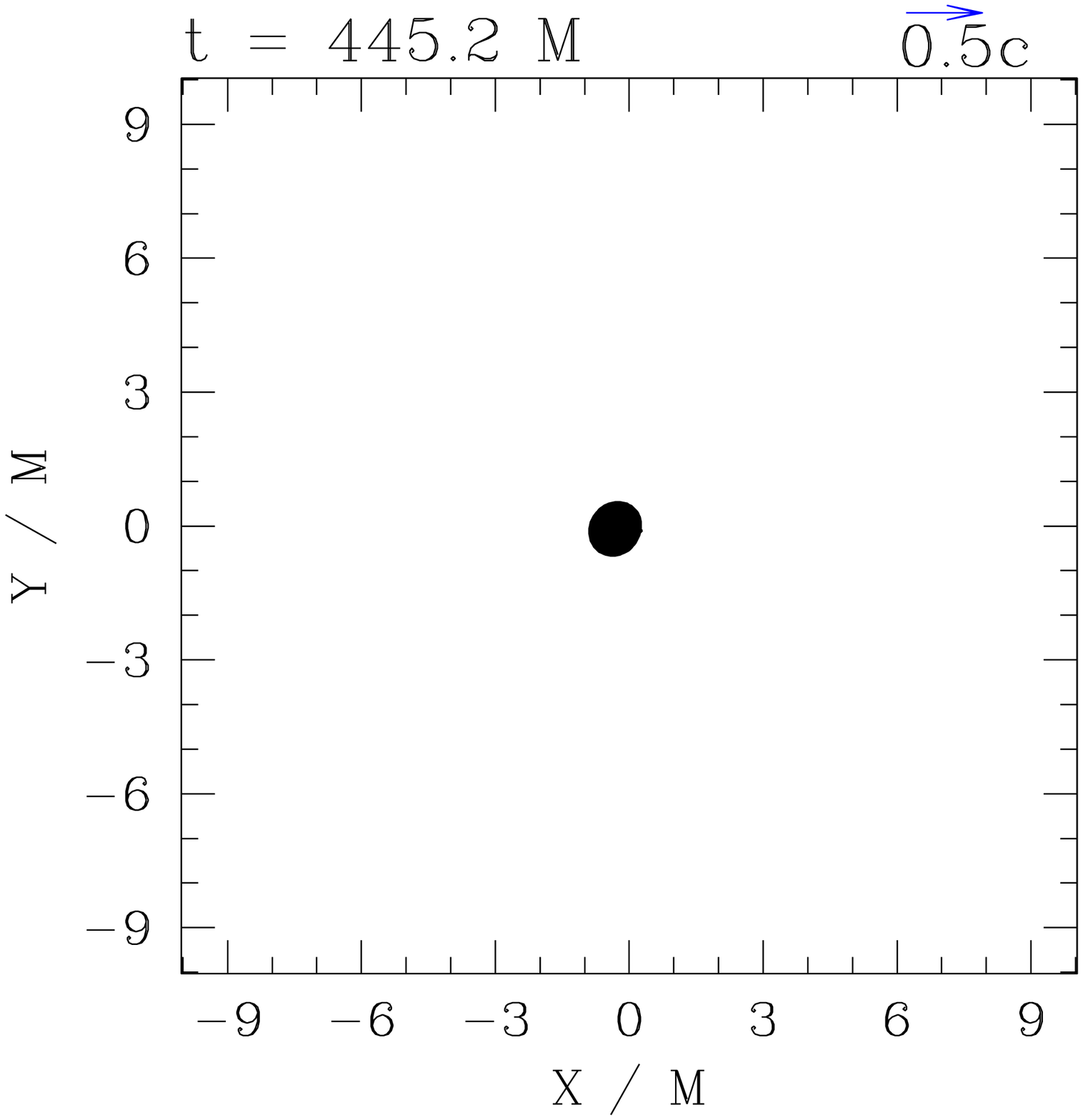}
\caption{Same as Fig.~\ref{fig:M1616_m0_snapshots} but for run
M1616B2 (magnetized run).}
\label{fig:M1616_m2_snapshots}
\end{center}
\end{figure*}

\begin{figure}
\includegraphics[width=8cm]{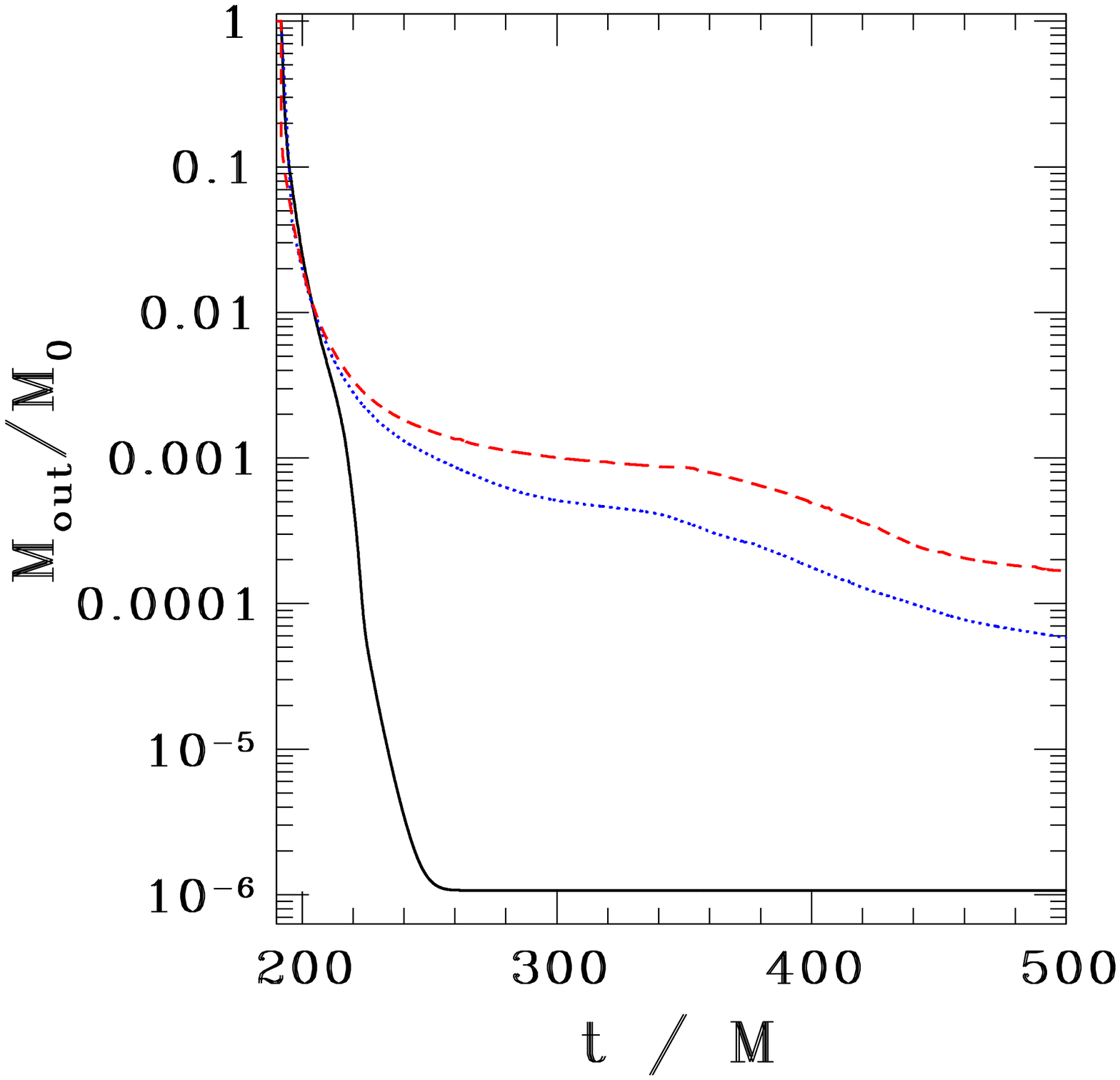}
\caption{Rest mass of the material outside the apparent horizon $M_{\rm out}$
for runs M1616B0 (black solid line), M1616B1 (blue dotted line) and 
M1616B2 (red dash line). }
\label{fig:M1616_mass_out}
\end{figure}

Model M1616 is also an equal-mass NSNS binary. The total rest mass
of the system is $M_0=1.78M_0^{\rm (TOV)}$. In the absence of magnetic
field, the system merges at $t \approx 150M$ ($\approx 0.9P_0$) and 
the star promptly collapses to a black hole. An apparent horizon forms 
at $t=192M$. Figure~\ref{fig:M1616_rho_alp} shows
the evolution of the maximum density $\rho_0^{\max}$ and minimum
lapse $\alpha_{\rm min}$. Figure~\ref{fig:M1616_m0_snapshots} shows 
snapshots of the 
equatorial density contours and velocity field. Our result again 
agrees with~\cite{stu03}. The simulation in~\cite{stu03} is 
terminated soon after the formation of an apparent horizon 
because of the grid stretching. Hence \cite{stu03} can only give an estimate 
of the upper bound of 0.002$M_0$ for the amount of material that 
can form a disk. 
We are able to use the puncture technique to continue the evolution 
until the system settles down to a stationary state. 
Figure~\ref{fig:M1616_mass_out} shows the
rest mass of the material outside the apparent horizon $M_{\rm out}$.
We see that all the material falls into the black hole. The small
residual value of $M_{\rm out} \approx 10^{-6}M_0$ at late times 
is due to the presence of our (artificial) atmosphere. 
After $t > 250M$, the system settles down to a vacuum rotating 
Kerr black-hole spacetime.

We perform two simulations for the magnetized cases with different
initial magnetic field profiles (see Table~\ref{tab:runs}). Run
M1616B1 has the same profile as M1414B1. In run M1616B2, more magnetic
field is placed in the outer layers of the neutron stars, and hence it could
counteract the gravitational pull of the black hole more effectively. 
We see from Fig.~\ref{fig:M1616_rho_alp} that runs M1414B1 and M1616B2 
are qualitatively the same as run M1616B0 (unmagnetized run). They both 
collapse promptly to a black hole. The apparent horizon appears at about the 
same time ($t=192M$) in all three cases. 
Figures~\ref{fig:M1616_m1_snapshots} and \ref{fig:M1616_m2_snapshots} show 
snapshots of equatorial density contours and velocity field. 
In Fig.~\ref{fig:M1616_mass_out}, we see that 
$M_{0\rm disk}/M_0\lesssim 10^{-4}$ for both M1616B1 and M1616B2. 
We see that magnetic fields cause a substantial delay in the time at 
which material in the low-density region falls into the black hole. 
The effect is 
more pronounced for the case M1616B2 when more magnetic field is in 
the low-density region. The simulations of 
M1616B1 and M1616B2 are terminated at $t\approx 500M$ 
when constraint violations start to become large 
(see Fig.~\ref{fig:M1616_cons_m1}). 
However, our current results already indicate
that even in the presence of magnetic fields the amount of material 
outside the black hole is very small. 

\begin{figure}
\includegraphics[width=8cm]{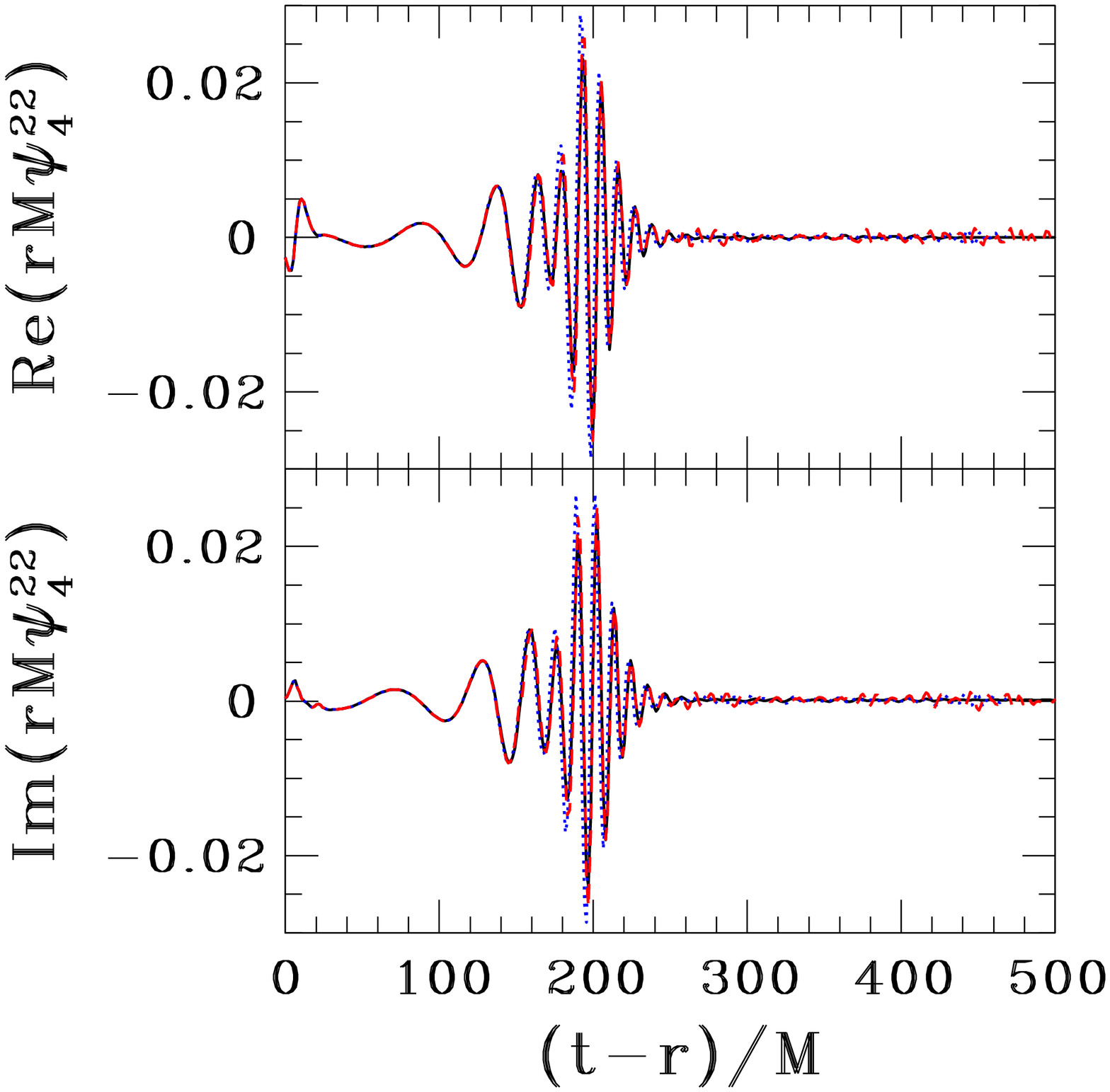}
\caption{Gravitation waveform $\psi_4^{22}(t-r)$ for runs 
M1616B0 (black solid line), M1616B1 (blue dotted line) and 
M1616B2 (red dash line). Gravitation
waves are extracted at radius $r=43M$.}
\label{fig:M1616_psi4}
\end{figure}

\begin{figure}
\includegraphics[width=8cm]{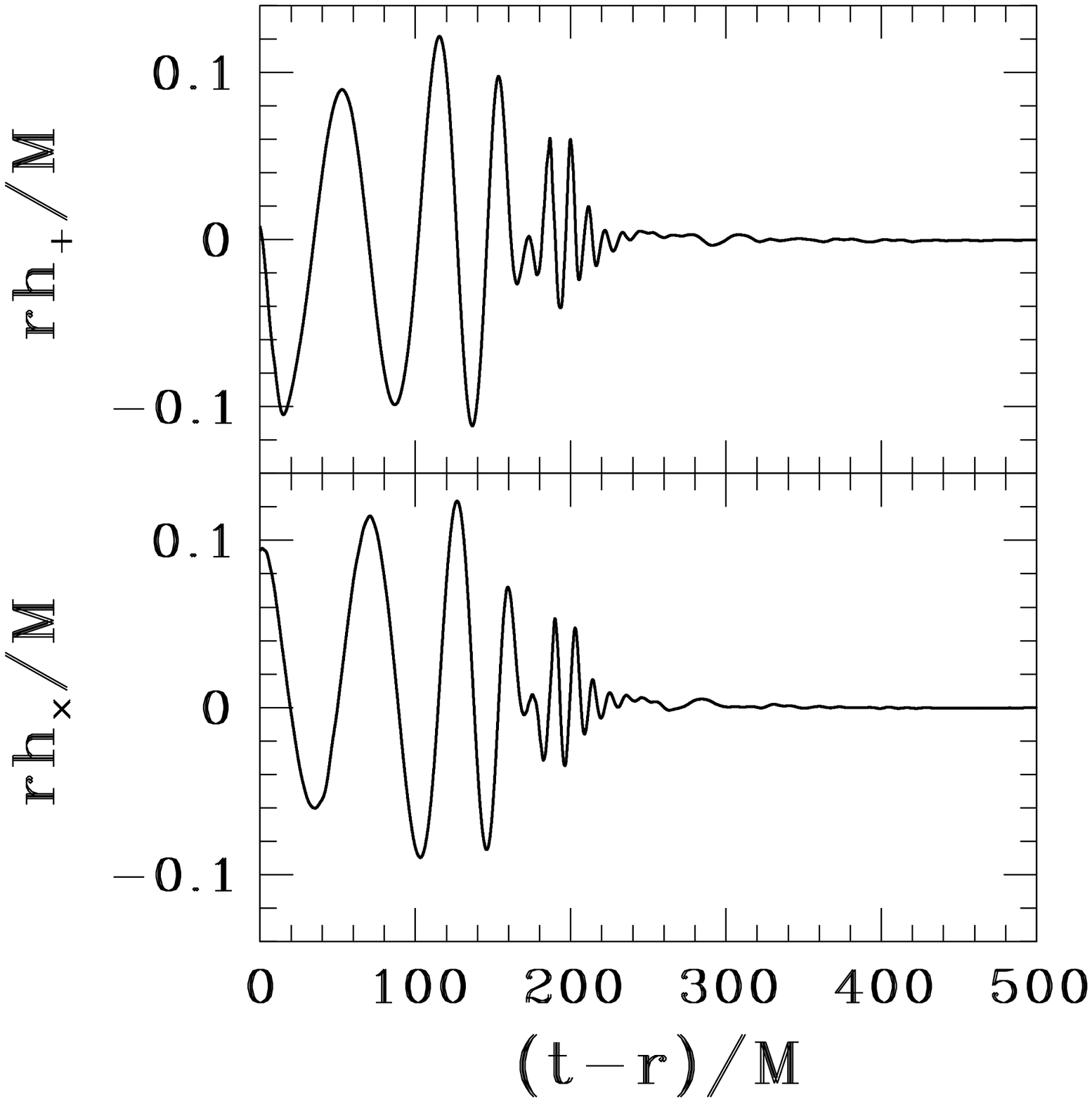}
\caption{Gravitation waveform $h_+$ and $h_{\times}$ for run
M1616B0 observed in the direction $45^\circ$ to the $z$-axis.}
\label{fig:M1616_hphc}
\end{figure}

Figure~\ref{fig:M1616_psi4} shows the gravitational waveforms for 
all M1616 models. 
We see that the waveforms for the three runs are very close. 
This is expected because magnetic fields can affect the dynamics 
substantially only well after the merger. However, the merged remnants
quickly collapse 
to black holes before the magnetic fields have enough time to change 
the fluid's motion significantly. 
We do see, however, that the amplitude of the waves during the collapse 
are slightly larger for the magnetized cases. 
Figure~\ref{fig:M1616_hphc} shows the two polarizations 
$h_+$ and $h_{\times}$ of run 
M1616B0 as observed in the direction $45^\circ$ to the $z$-axis axis. 

\begin{figure}
\includegraphics[width=8cm]{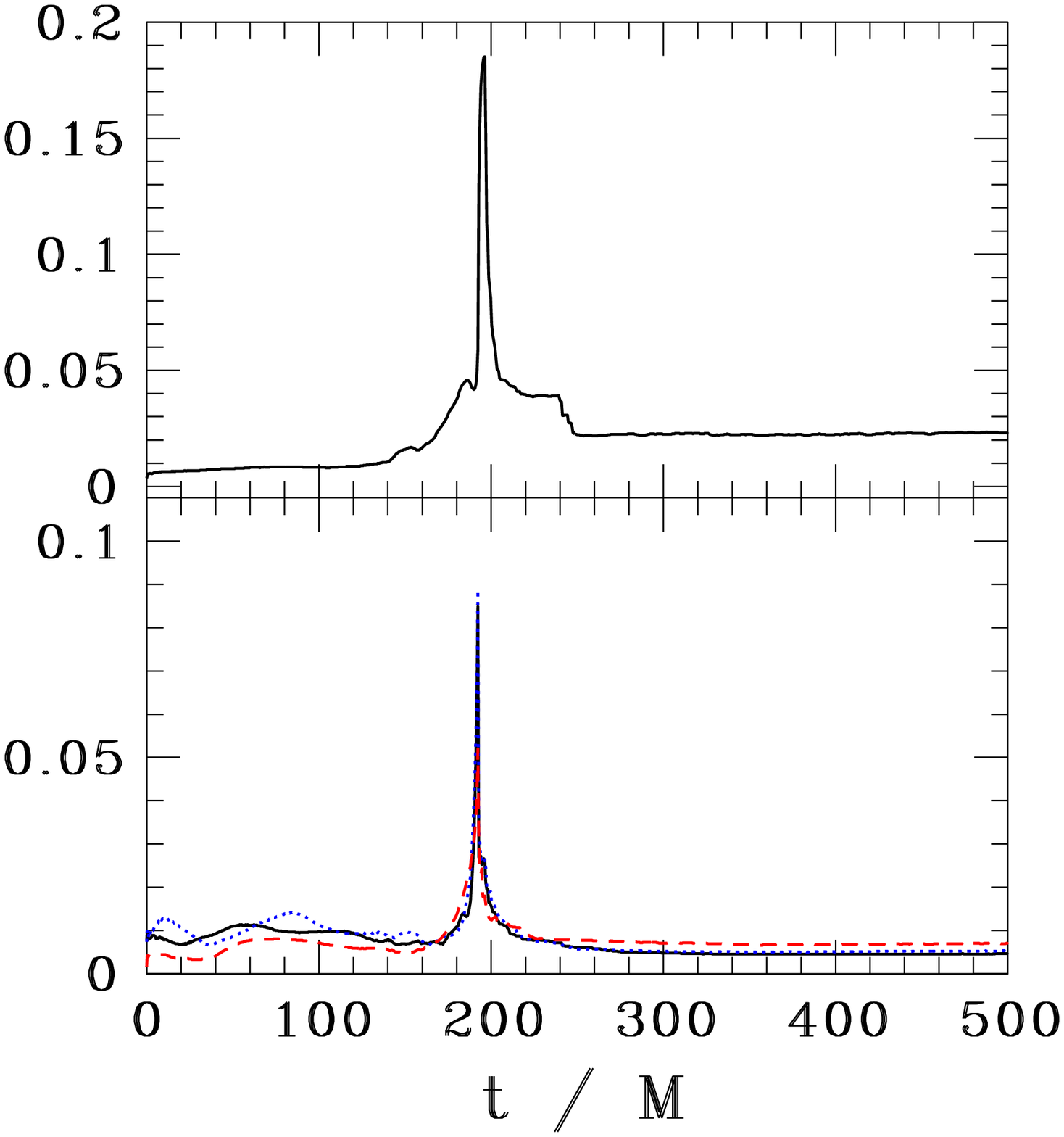}
\caption{Constraint violations for the unmagnetized run M1414B0.
Upper panel: Normalized L2 norm of the Hamiltonian constraint.
Lower panel: Normalized L2
norm of the $x$ (black solid line), $y$ (blue dotted line) and $z$ (red
dash line) components of the momentum constraint. Note that the large 
violations corresponding to the peaks at $t=192M$ are trapped inside 
the event horizon. When the apparent
horizon is detected ($t>192M$), the constraints are computed only in
the region outside the horizon.}
\label{fig:M1616_cons_m0}
\end{figure}

\begin{figure}
\includegraphics[width=8cm]{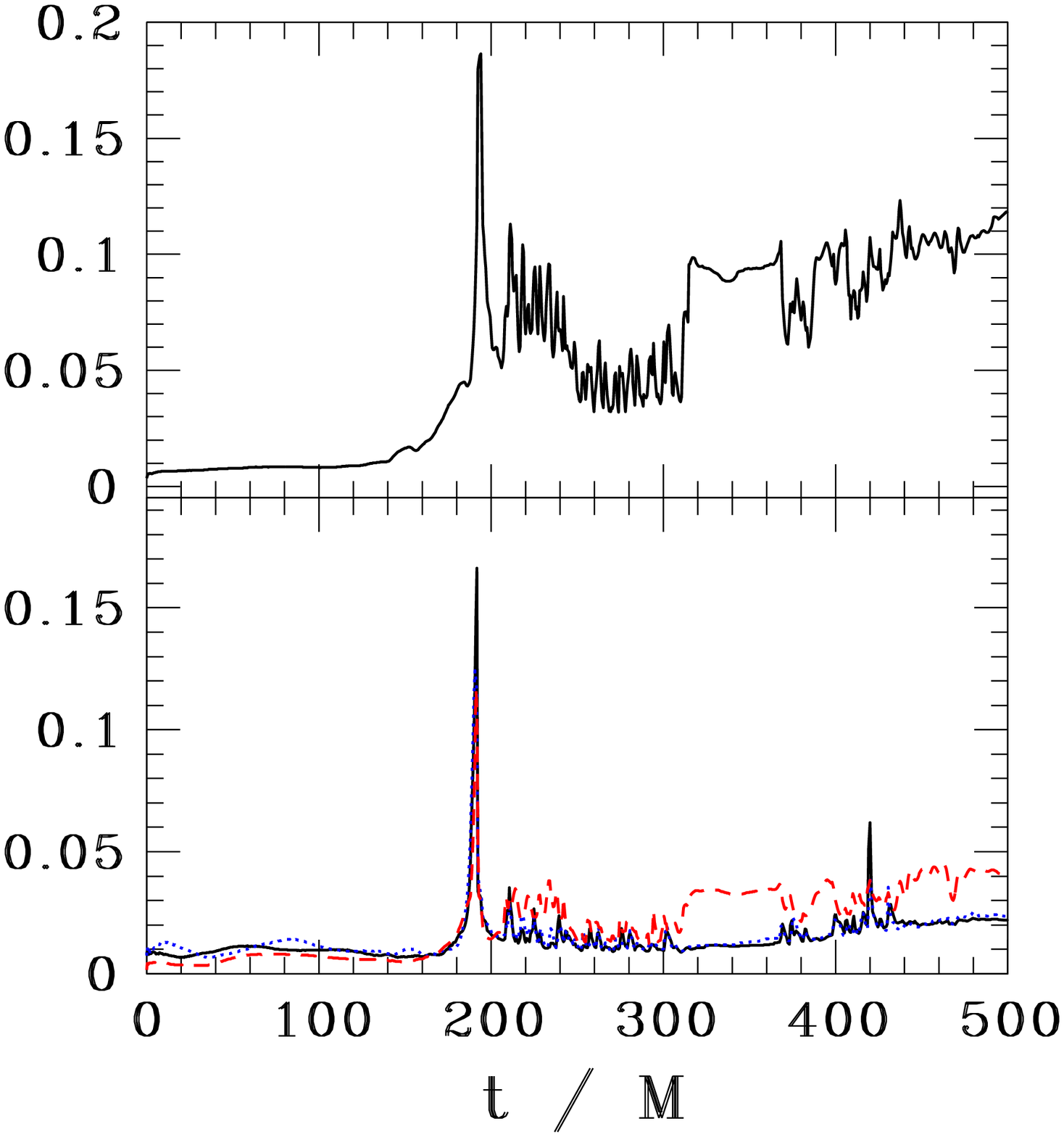}
\caption{Same as Fig.~\ref{fig:M1616_cons_m0} but for the 
magnetized run M1414B1.}
\label{fig:M1616_cons_m1}
\end{figure}

Figures~\ref{fig:M1616_cons_m0} and \ref{fig:M1616_cons_m1} show the 
L2 norms of the constraint violations for runs M1616B0 and M1616B1. 
The plots for M1616B2 are similar to those of M1616B1 and so are not 
shown here. The peaks at $t=192M$ are due to the formation 
of a central singularity and are contained inside the event horizon. 
After the apparent horizon appears at $t=192M$, 
the constraints are computed only in the region outside the apparent horizon, 
and we see the constraint violations drop to much lower values. This result 
confirms that the large constraint violations in the strong-field region 
are trapped inside the event horizon. For run M1616B0, the 
constraint violations are less than 3\% most of the time. 
For run M1616B1, the momentum constraint violations are less than 5\% 
most of the time. The Hamiltonian constraint violation is less than 
5\% before the collapse, around 5\%--10\% after the apparent horizon 
forms, and gradually increases to 12\% at $t=500M$, after which the 
simulation is terminated.

\subsection{Model M1418}

\begin{figure}
\includegraphics[width=8cm]{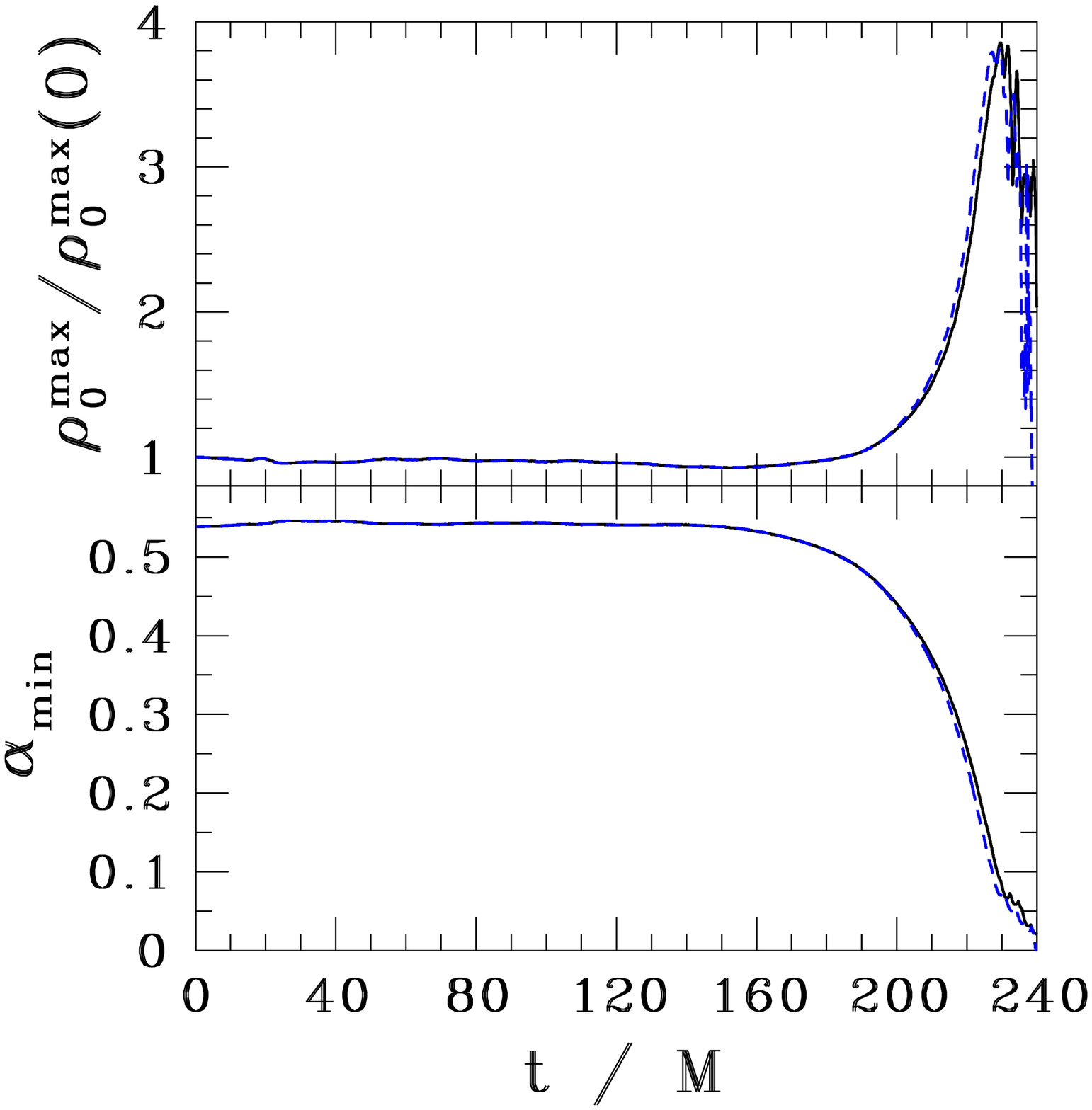}
\caption{Evolution of maximum density $\rho_0^{\max}$ and minimum
lapse $\alpha_{\rm min}$ for unmagnetized (black solid line) and magnetized
(blue dash line) runs of model M1418. The merger occurs at $t \approx 180M$, 
and an apparent horizon appears at $t=232M$ for both magnetized 
and unmagnetized cases.}
\label{fig:M1418_rho_alp}
\end{figure}

\begin{figure*}
\vspace{-4mm}
\begin{center}
\epsfxsize=2.15in
\leavevmode
\epsffile{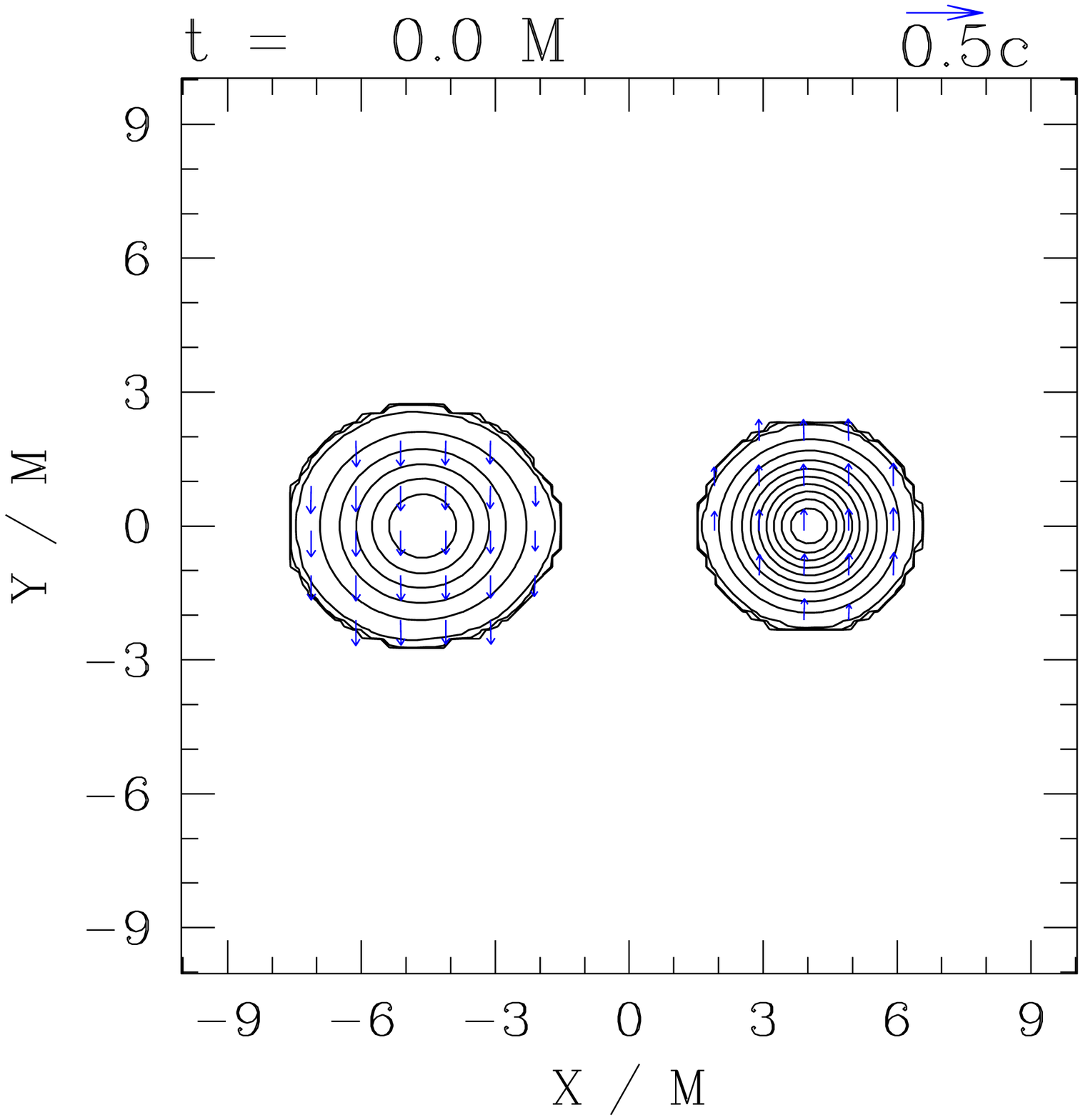}
\epsfxsize=2.15in
\leavevmode
\epsffile{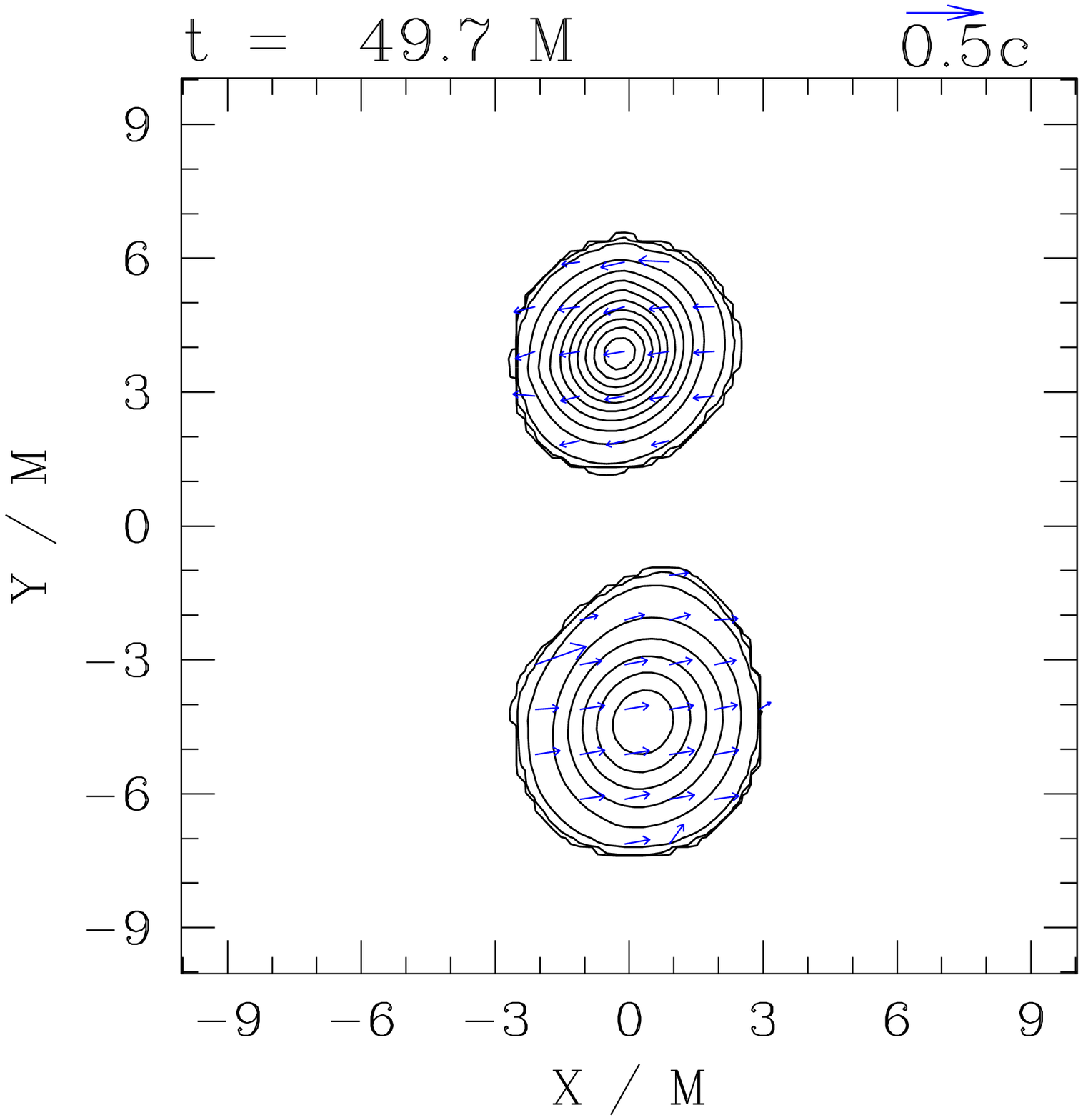}
\epsfxsize=2.15in
\leavevmode
\epsffile{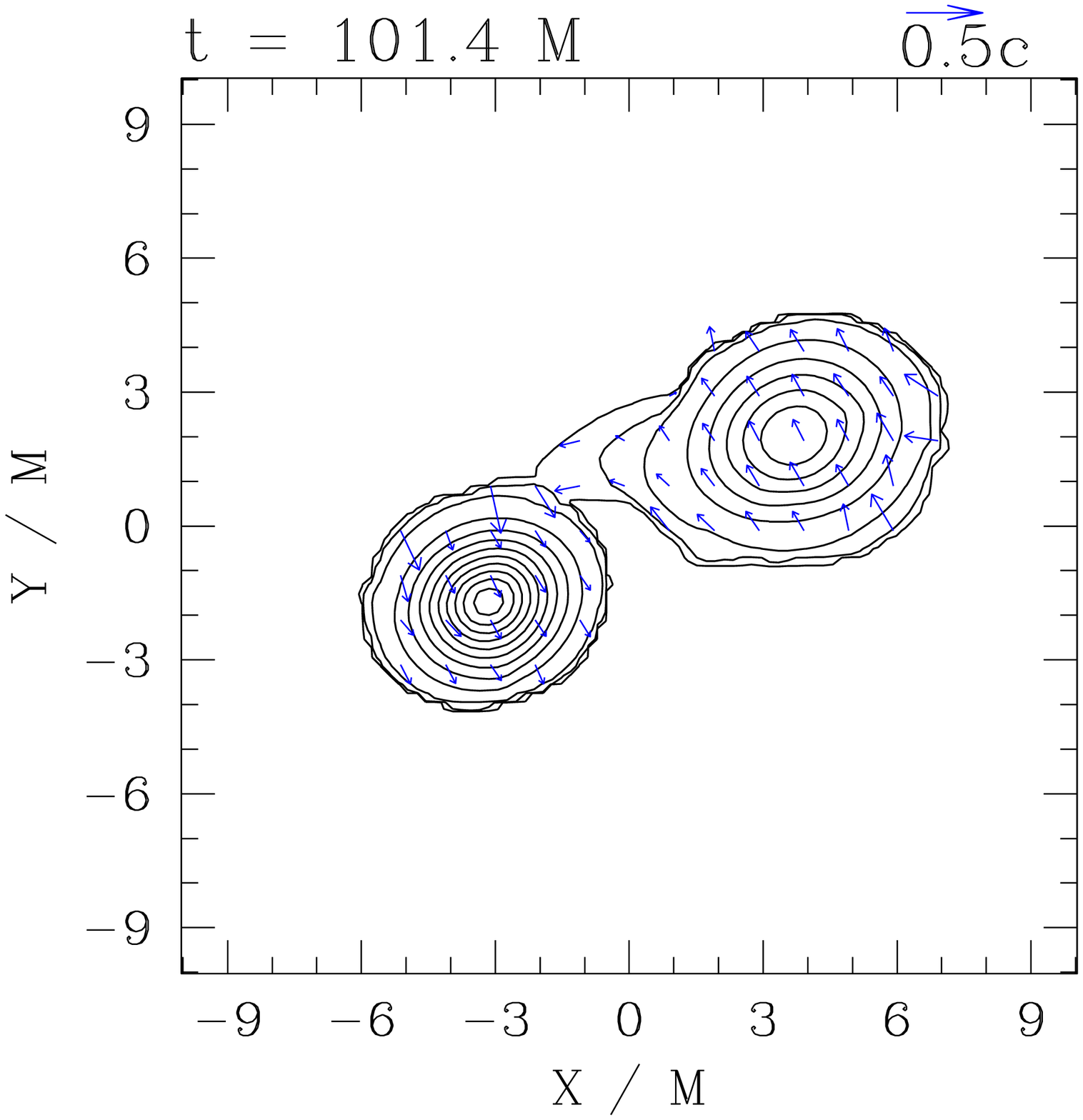}\\
\epsfxsize=2.15in
\leavevmode
\epsffile{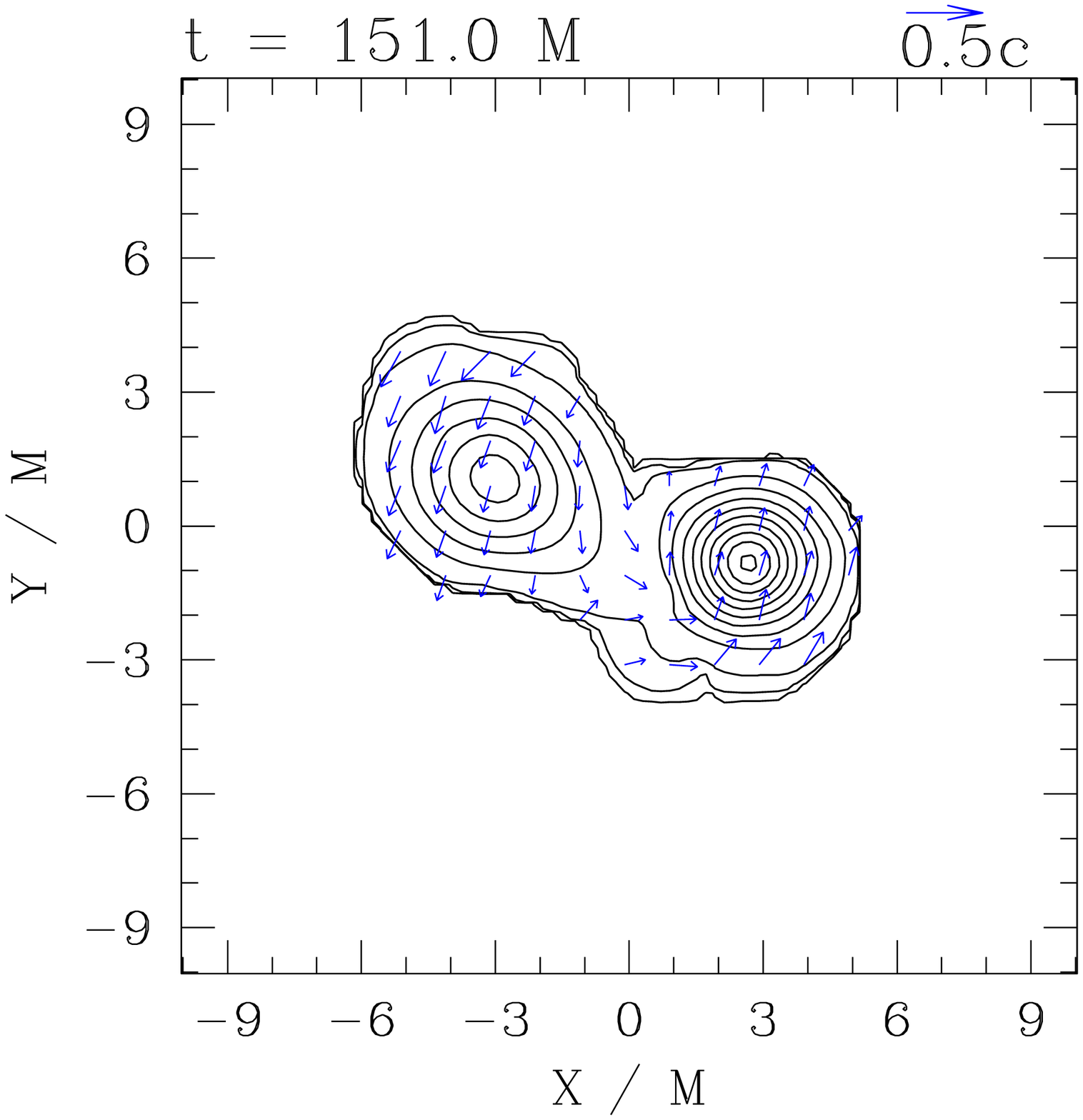}
\epsfxsize=2.15in
\leavevmode
\epsffile{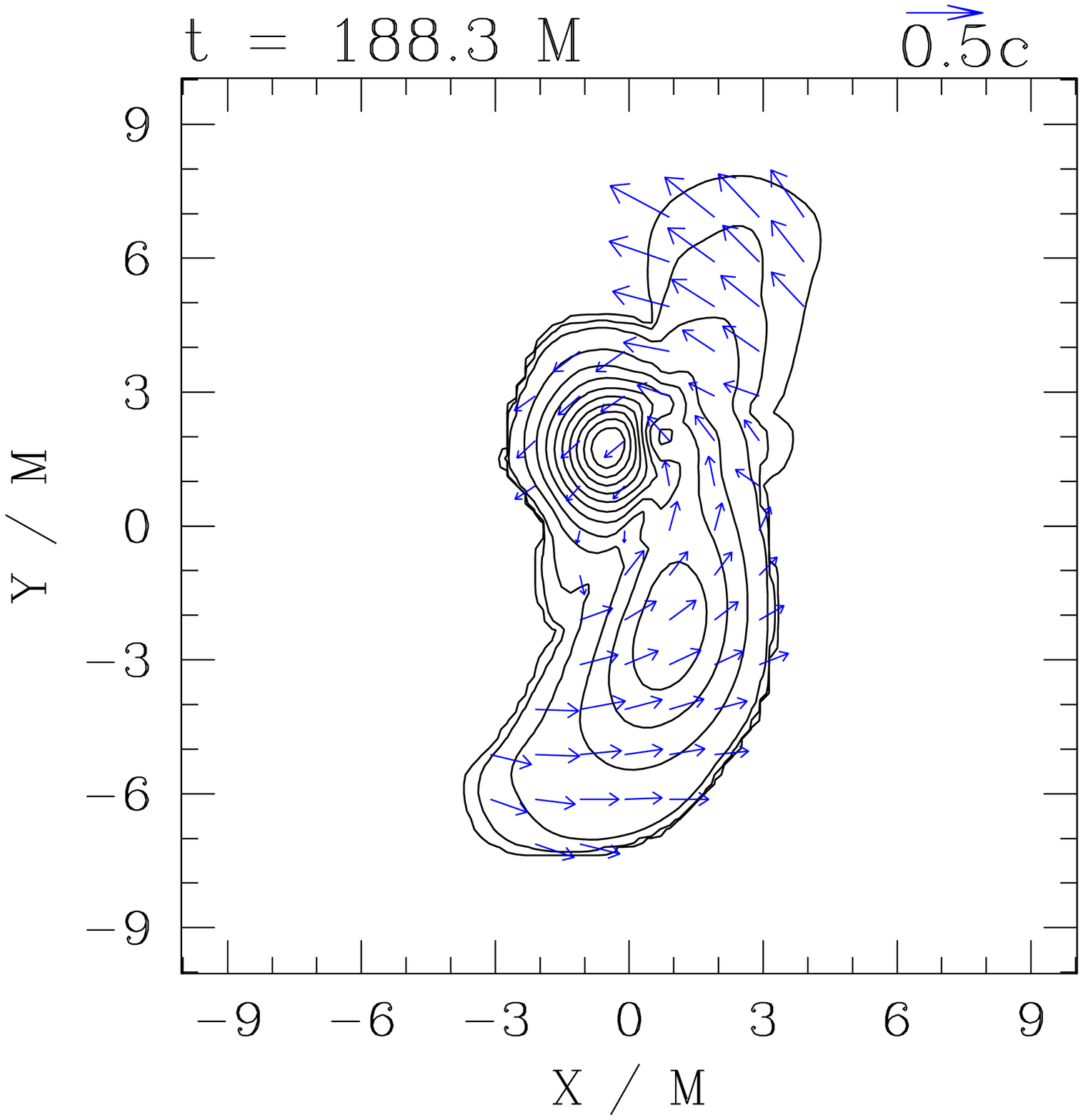}
\epsfxsize=2.15in
\leavevmode
\epsffile{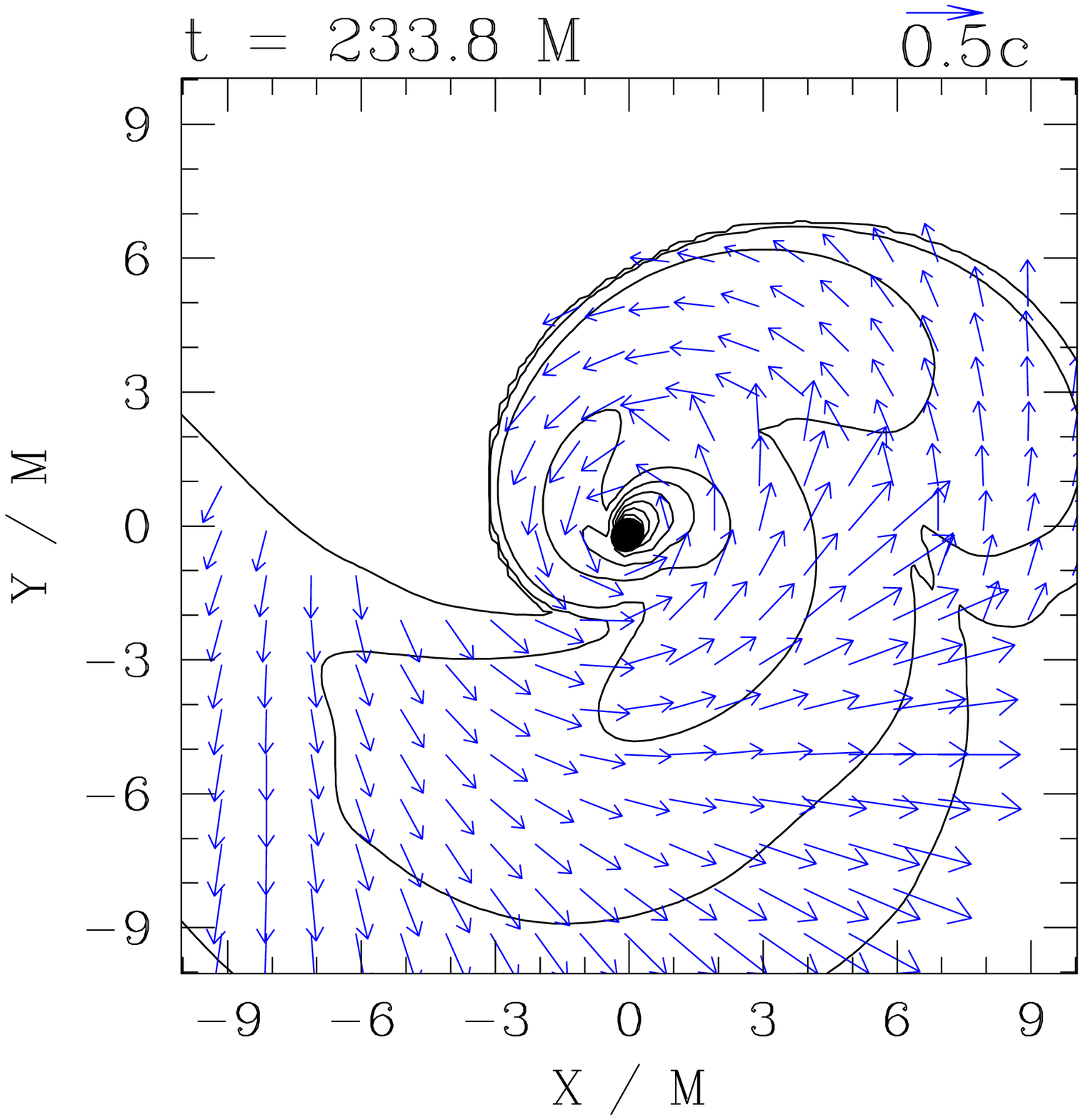}\\
\epsfxsize=2.15in
\leavevmode
\epsffile{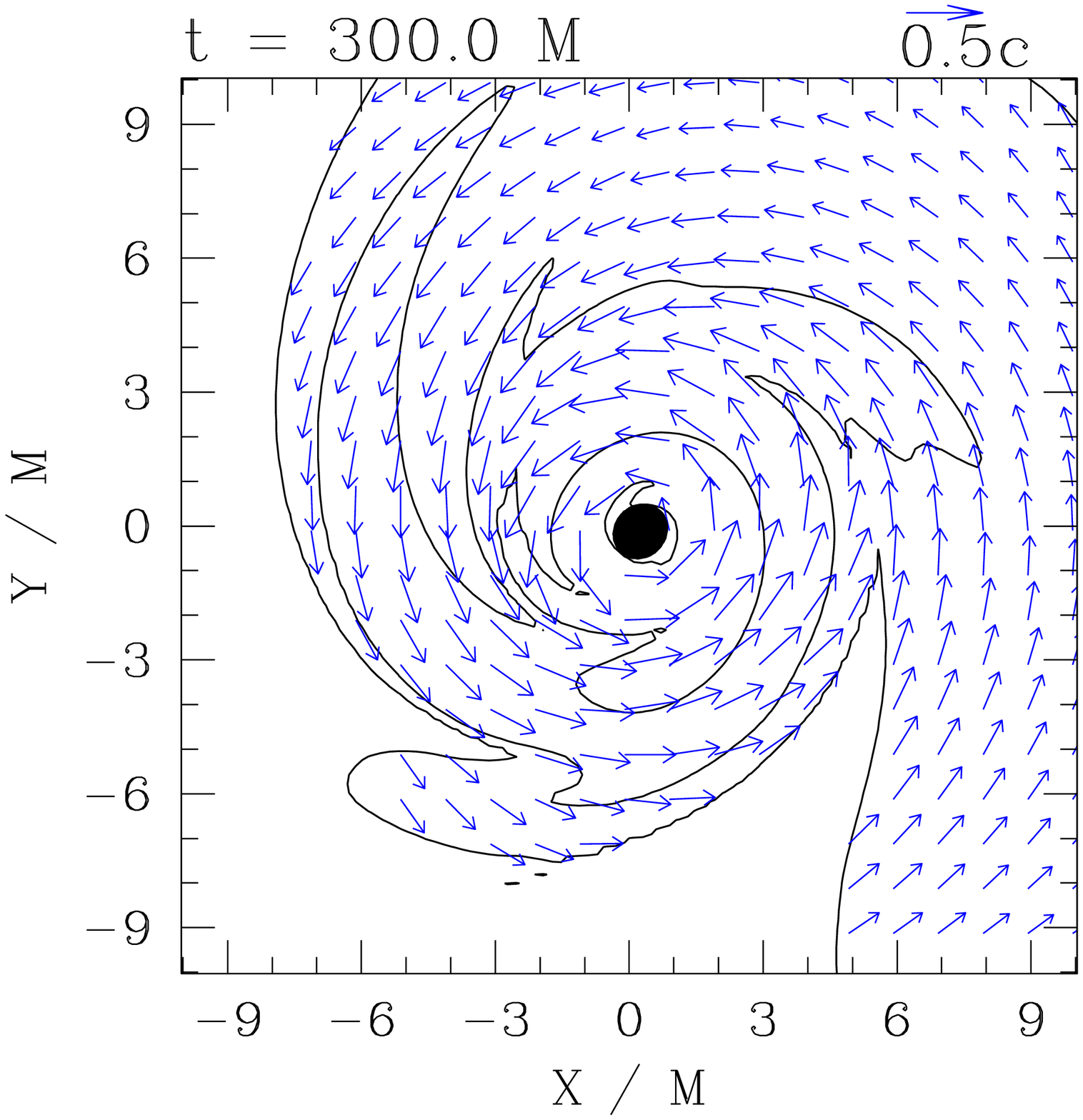}
\epsfxsize=2.15in
\leavevmode
\epsffile{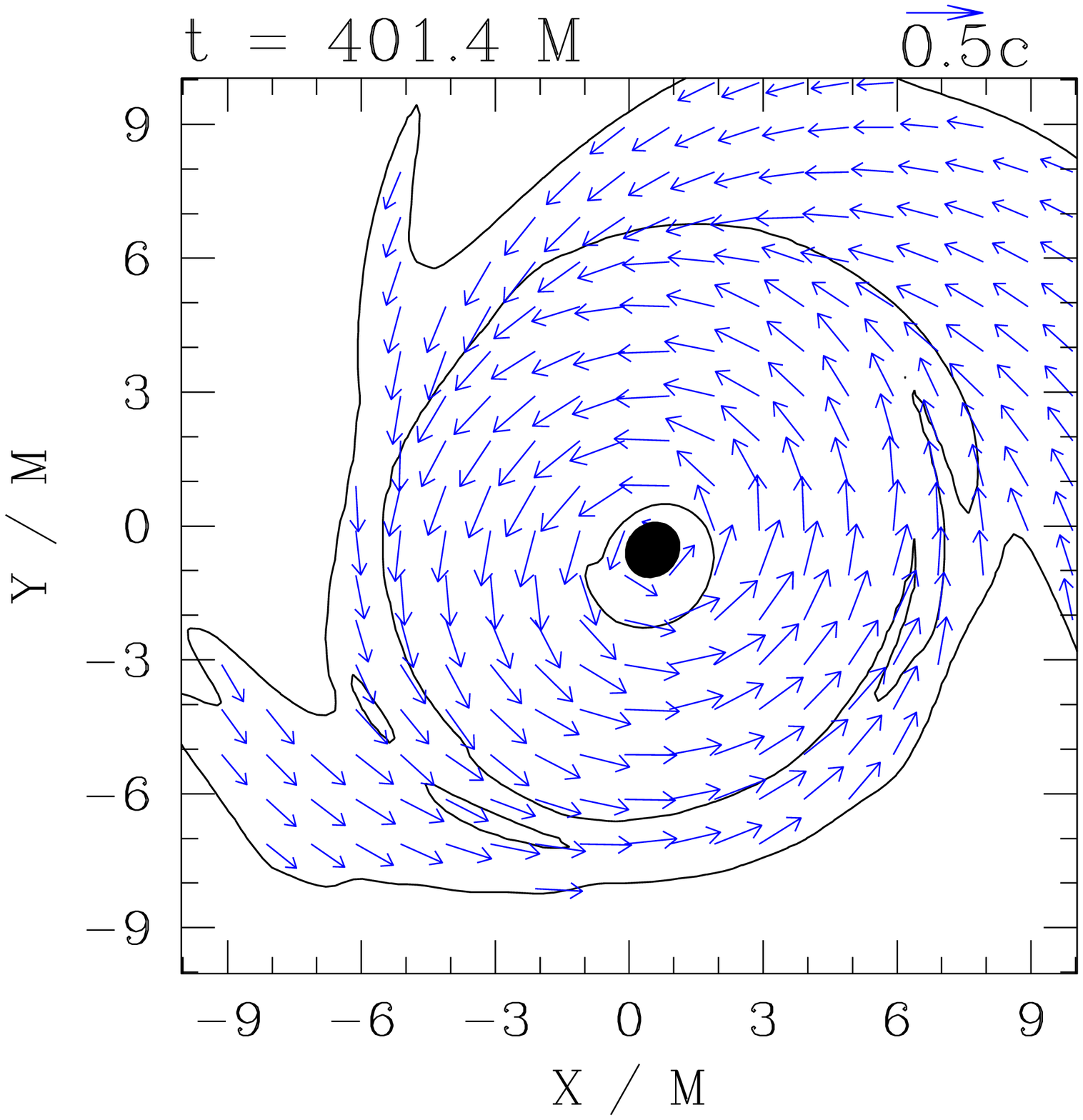}
\epsfxsize=2.15in
\leavevmode
\epsffile{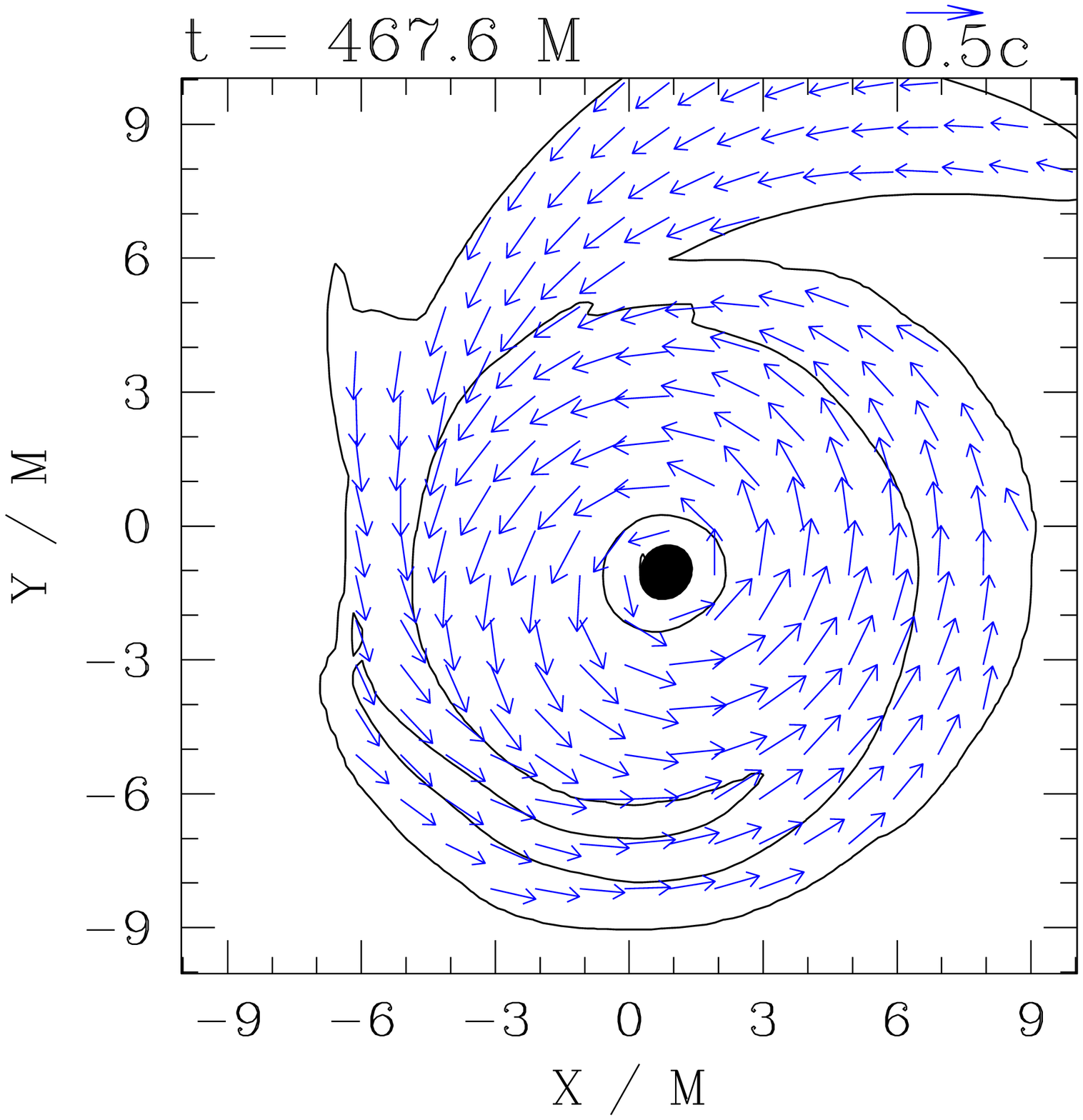}
\caption{Same as Fig.~\ref{fig:M1616_m0_snapshots} but for run
M1418B0 (unmagnetized run).}
\label{fig:M1418_m0_snapshots}
\end{center}
\end{figure*}

\begin{figure*}
\vspace{-4mm}
\begin{center}
\epsfxsize=2.15in
\leavevmode
\epsffile{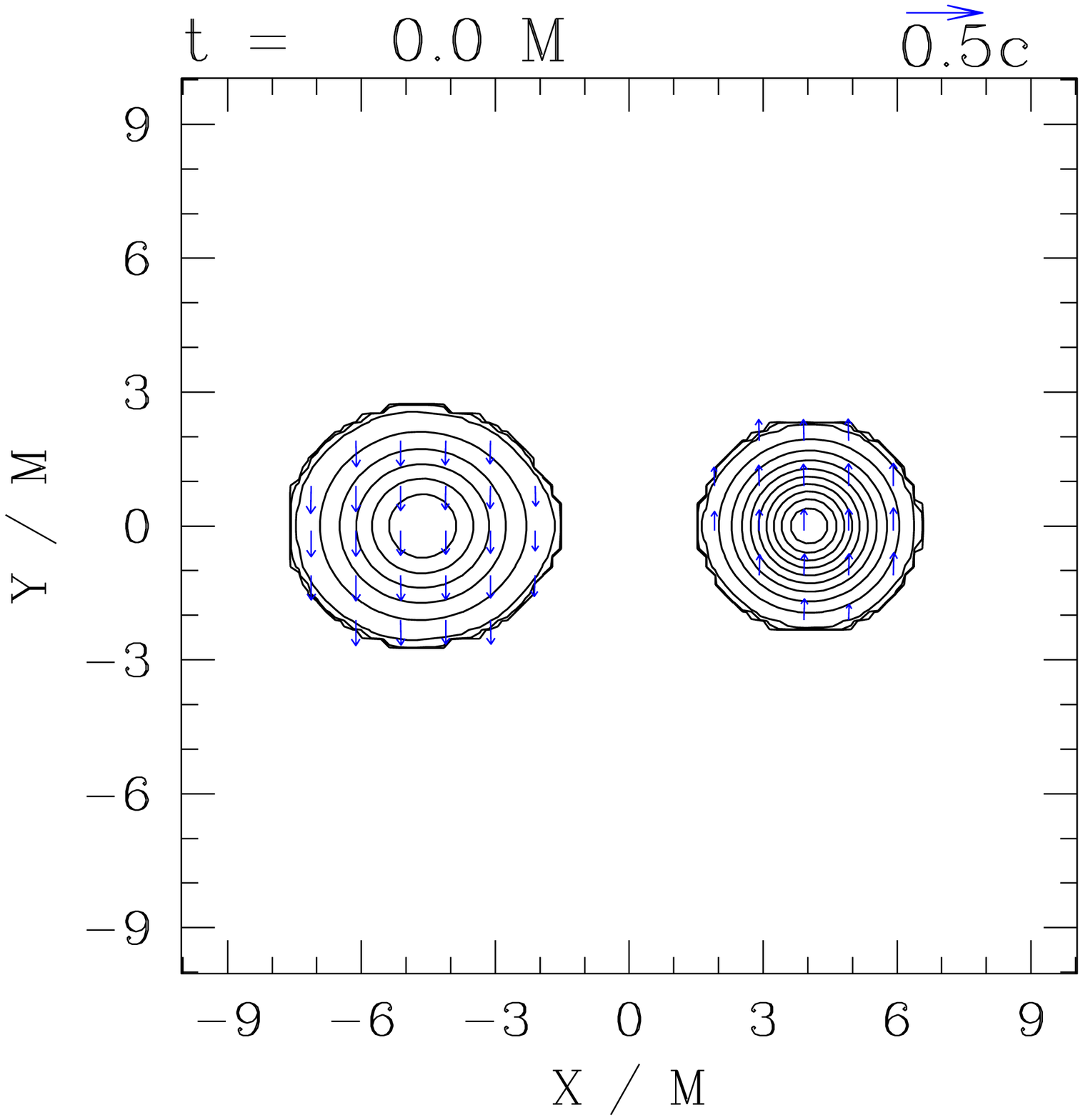}
\epsfxsize=2.15in
\leavevmode
\epsffile{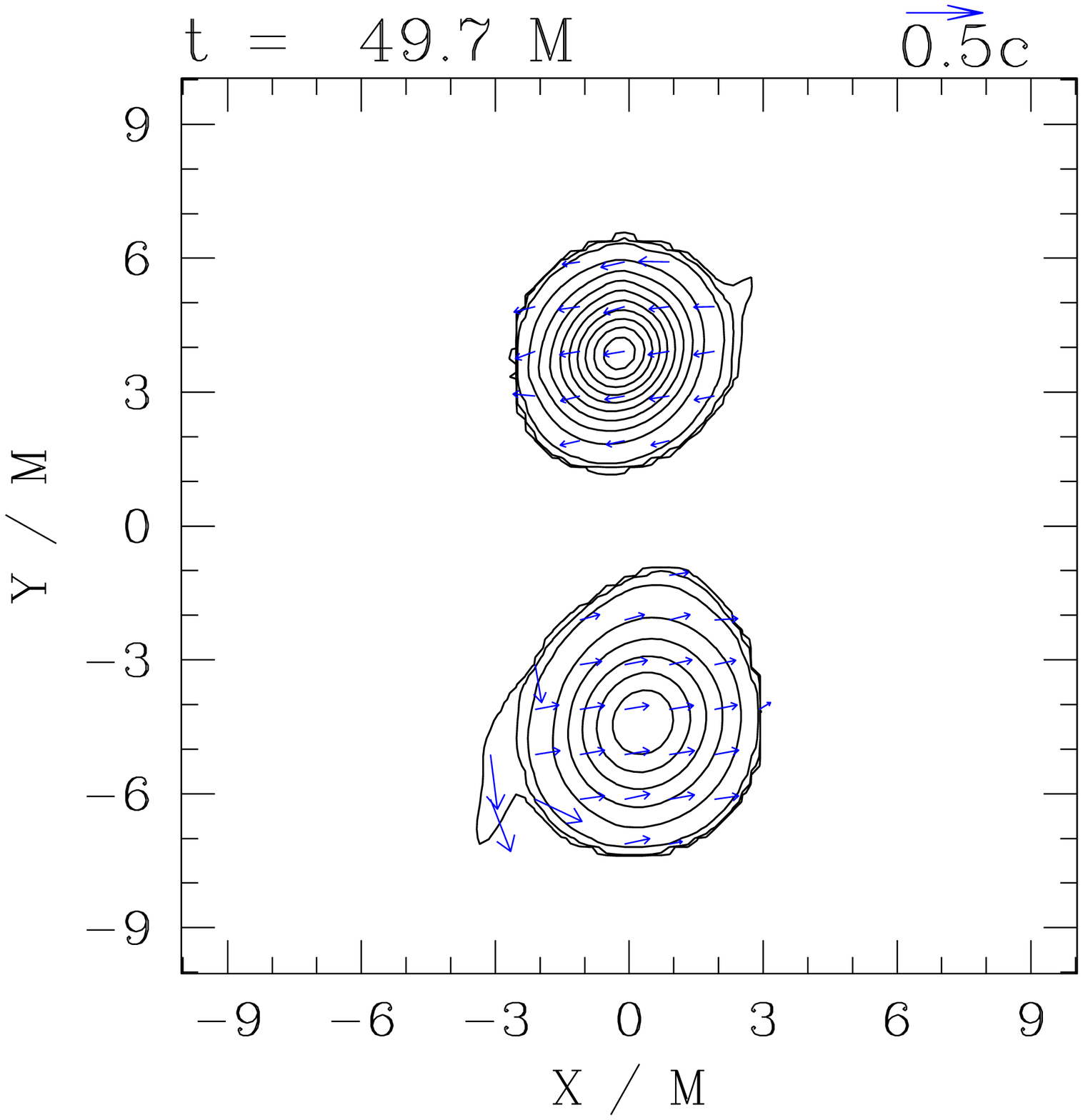}
\epsfxsize=2.15in
\leavevmode
\epsffile{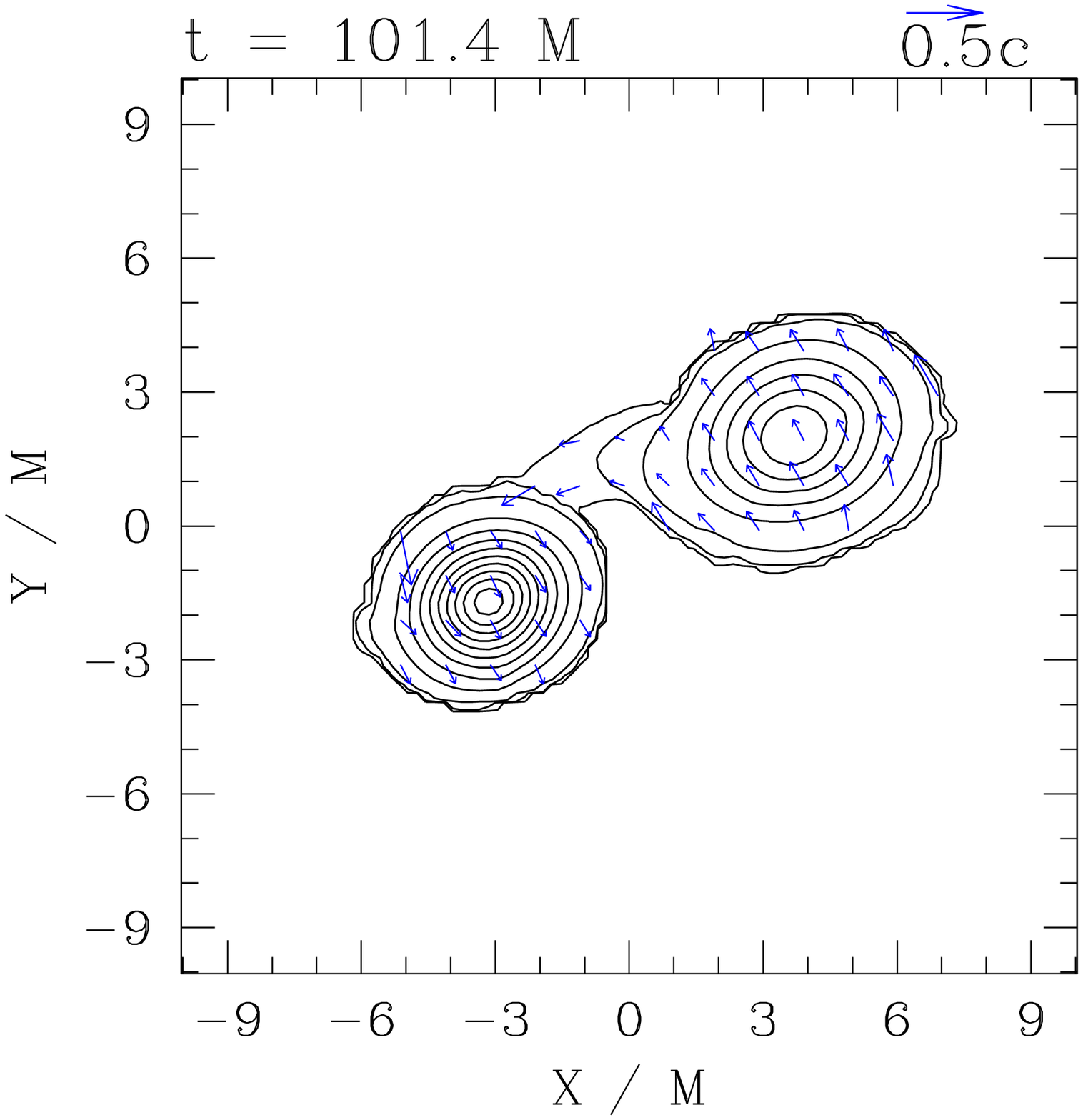}\\
\epsfxsize=2.15in
\leavevmode
\epsffile{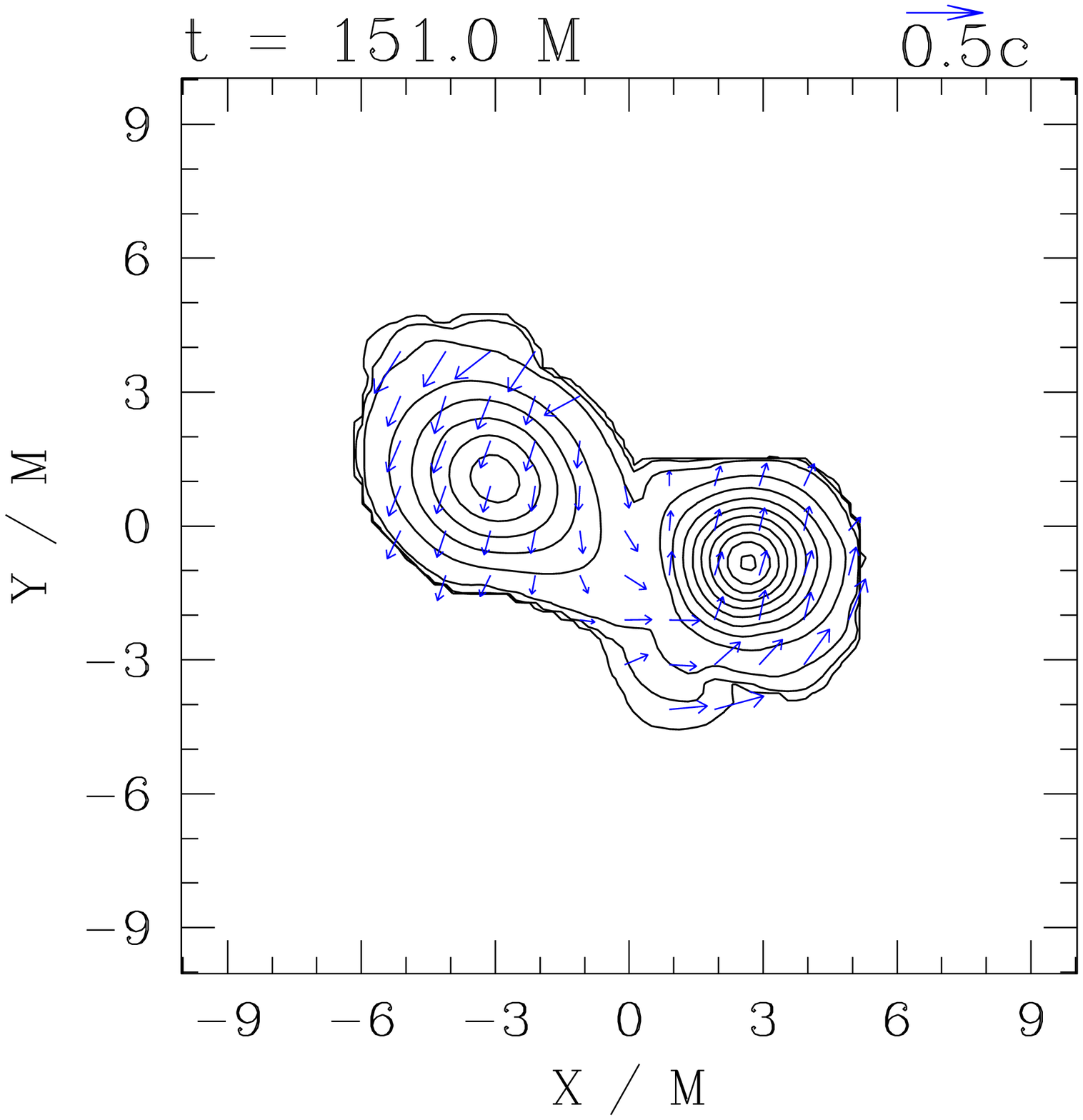}
\epsfxsize=2.15in
\leavevmode
\epsffile{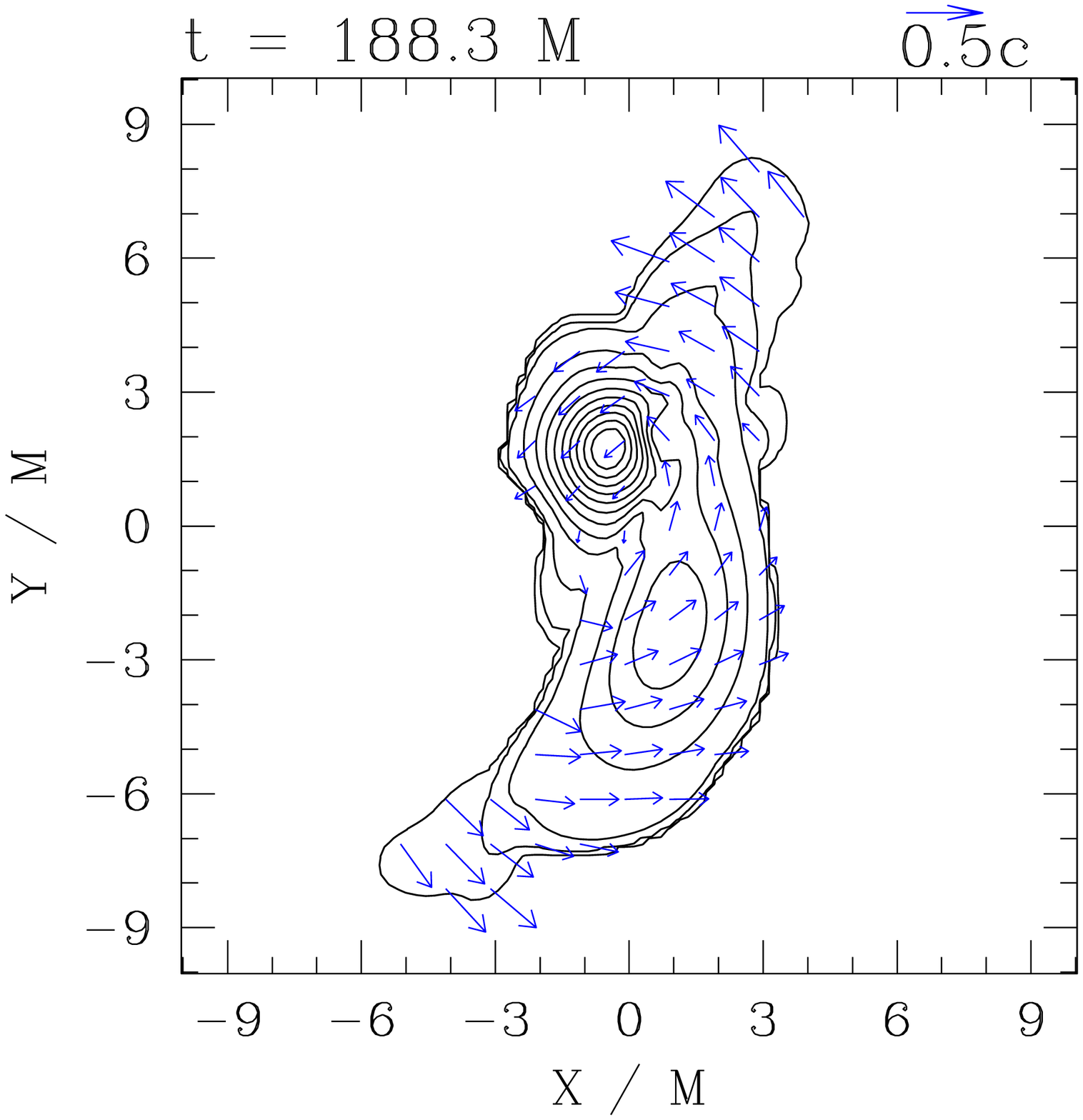}
\epsfxsize=2.15in
\leavevmode
\epsffile{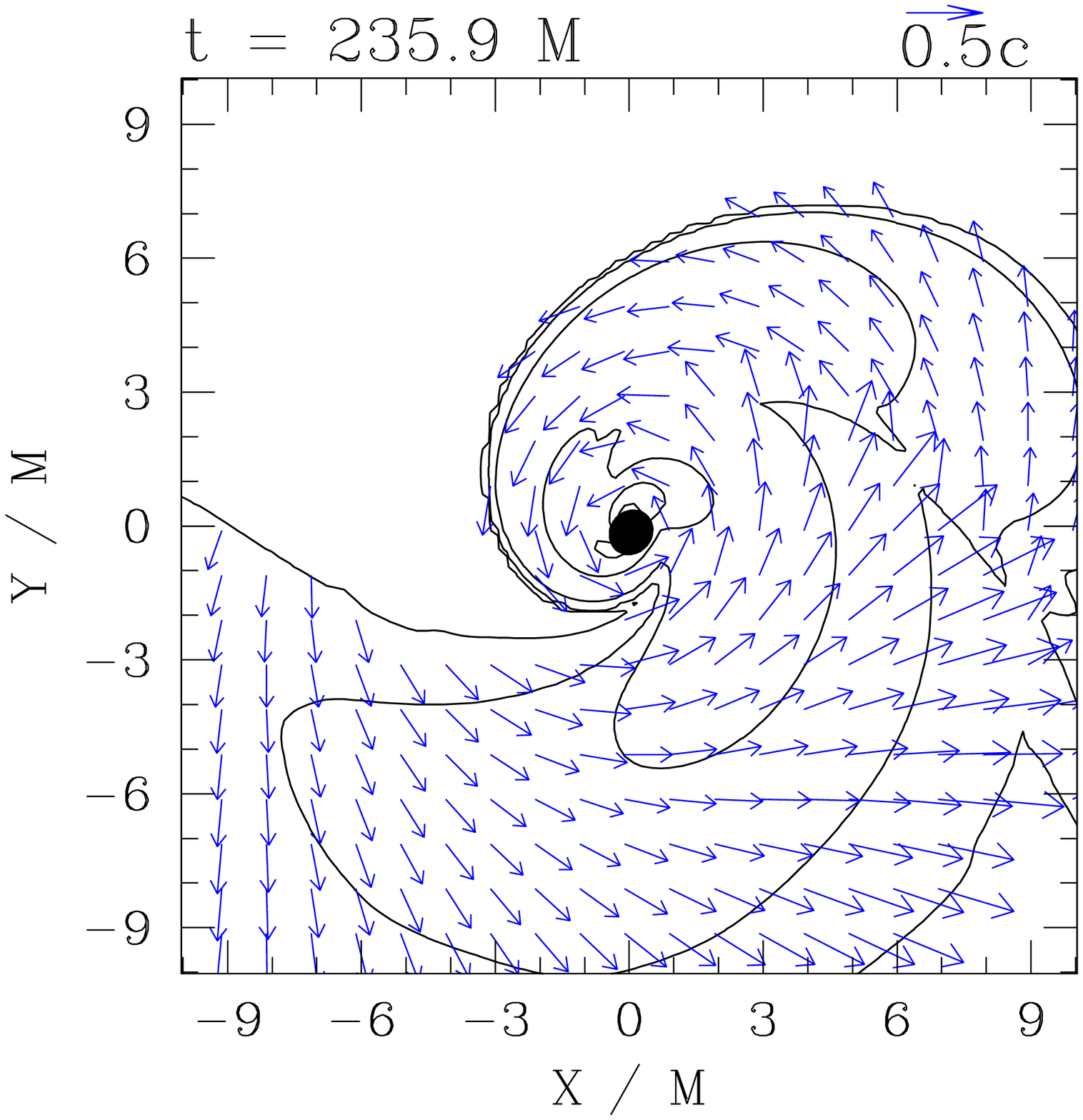}\\
\epsfxsize=2.15in
\leavevmode
\epsffile{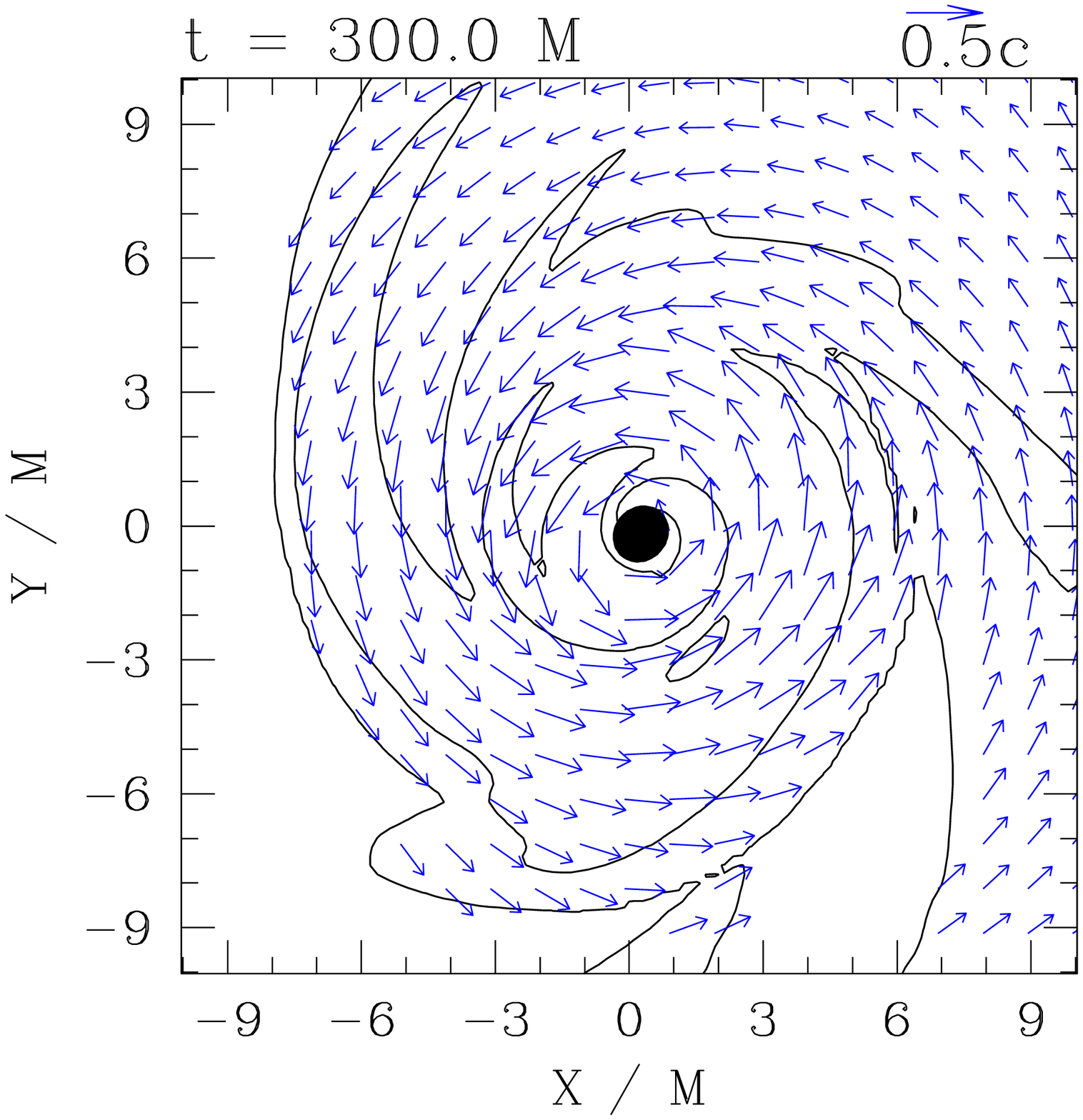}
\epsfxsize=2.15in
\leavevmode
\epsffile{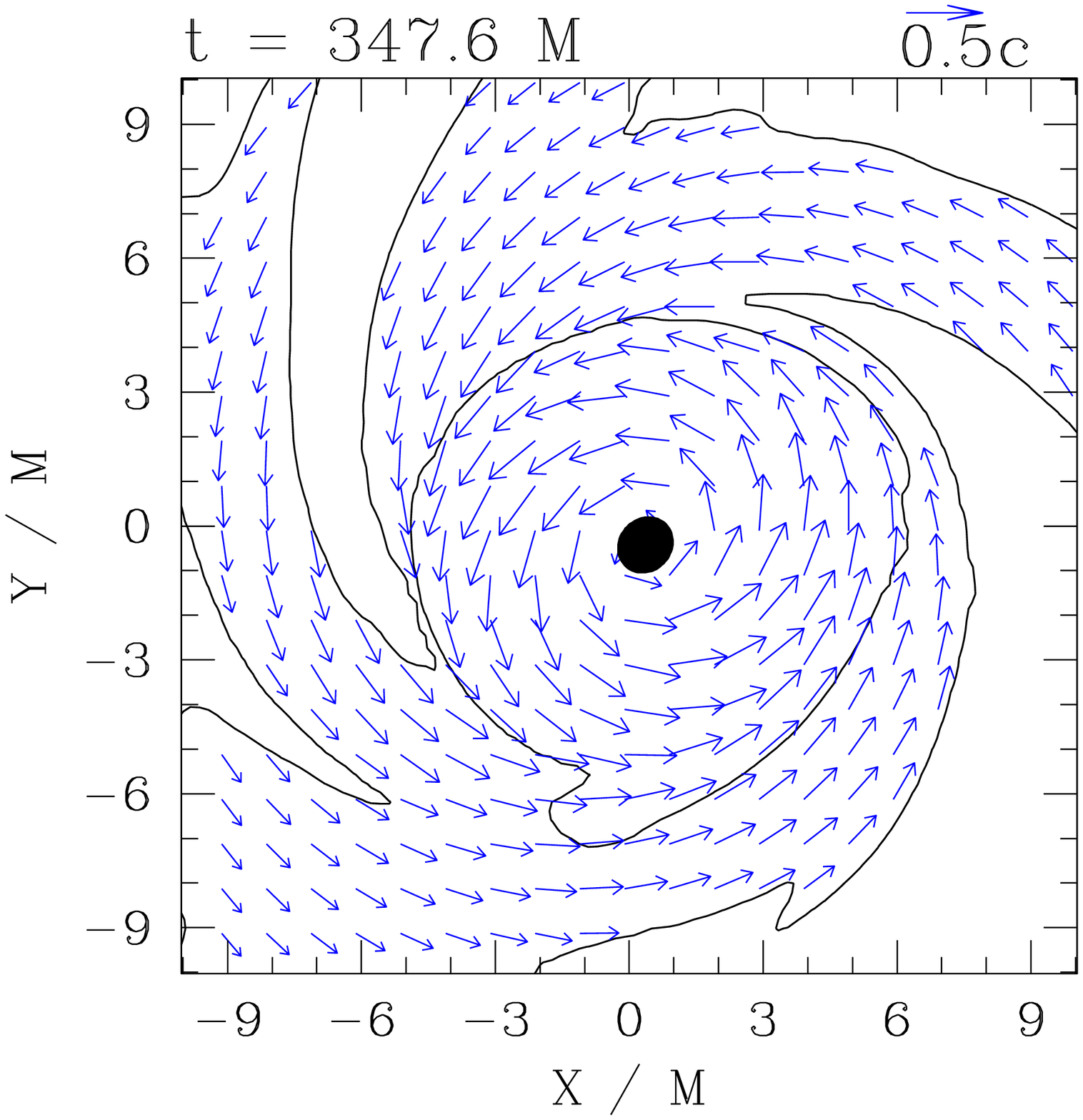}
\epsfxsize=2.15in
\leavevmode
\epsffile{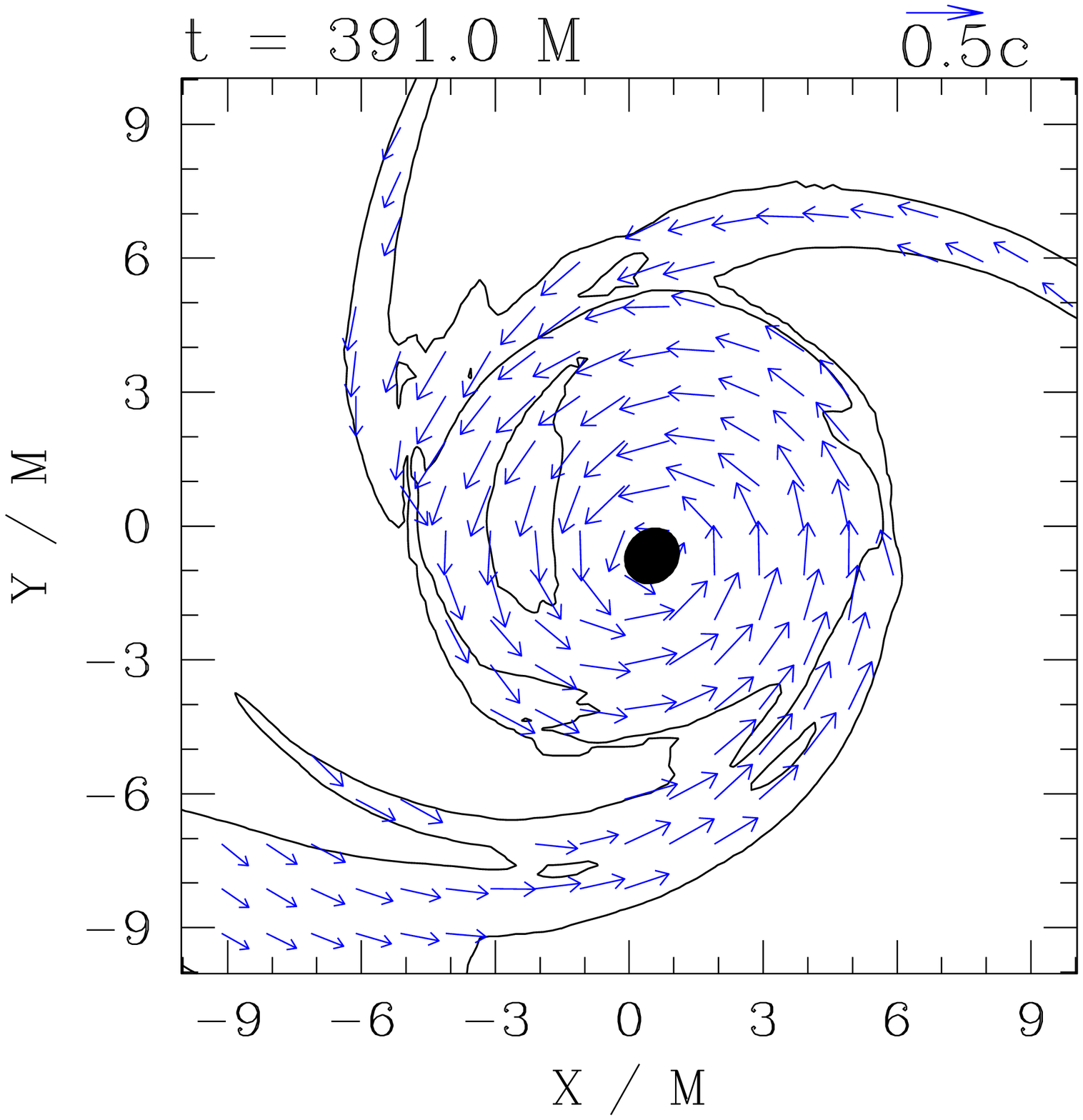}
\caption{Same as Fig.~\ref{fig:M1616_m0_snapshots} but for run
M1418B1 (magnetized run).}
\label{fig:M1418_m1_snapshots}
\end{center}
\end{figure*}

\begin{figure}
\includegraphics[width=8cm]{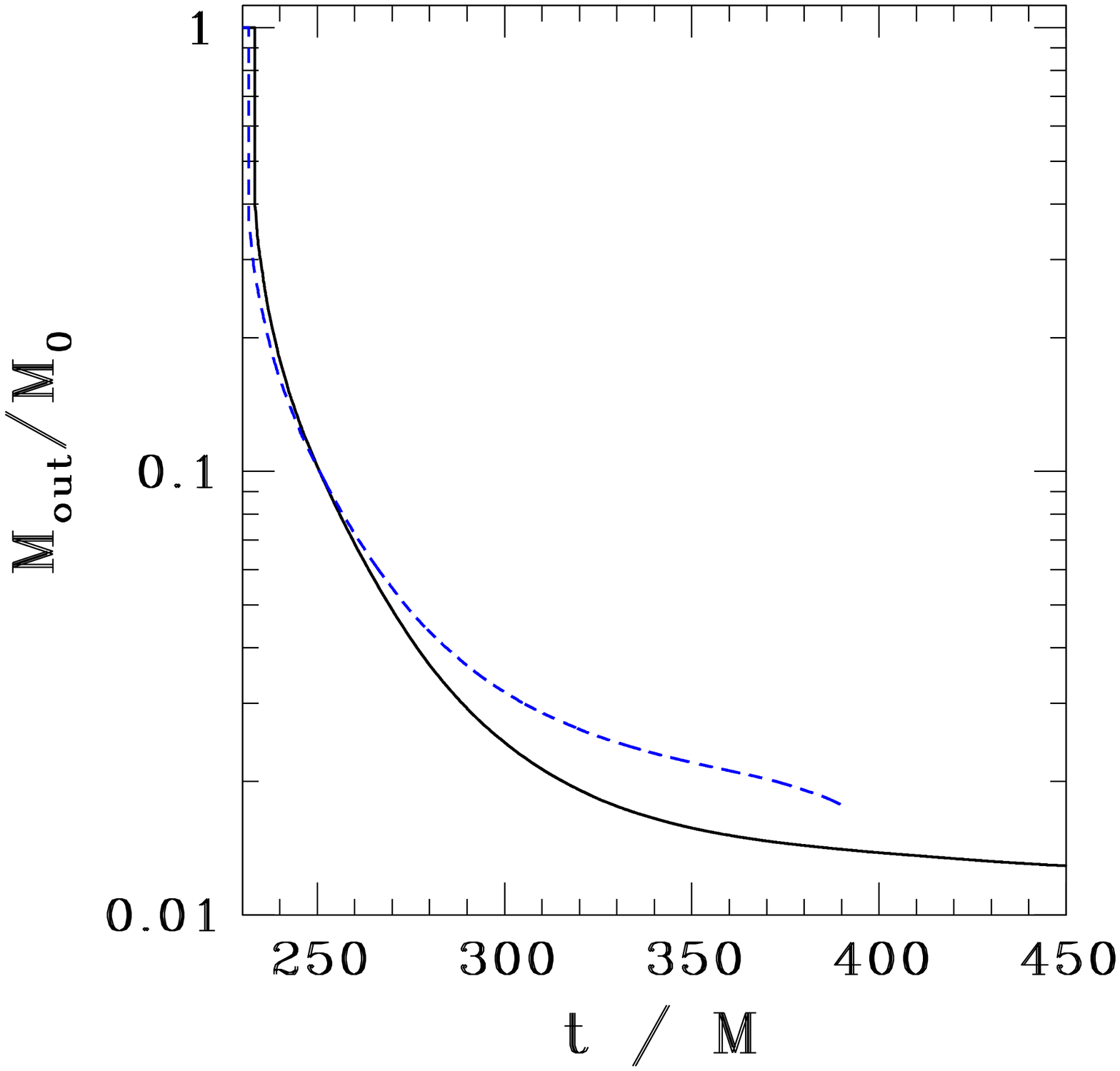}
\caption{Rest mass of the material outside the apparent horizon $M_{\rm out}$ 
for the unmagnetized (black solid line) and magnetized (blue dash line) 
runs of model M1418.}
\label{fig:M1418_mass_out}
\end{figure}

For model M1418, the ratio of the rest masses of the neutron stars 
are $q=M_0^{(1)}/M_0^{(2)}=0.855$. The total rest mass of the system is 
$M_0 = 1.76M_0^{\rm (TOV)}$, 
which is about the same as model M1616. As in model M1414, we perform 
an unmagnetized (run M1418B0) and a magnetized (run M1418B1) simulation.  
The merger occurs at 
$t\approx 180M \approx 1P_0$. The merged remnant collapses to a black hole. An 
apparent horizon forms at $t=232M$ for both cases. 
Figure~\ref{fig:M1418_rho_alp} shows the evolution of the maximum 
density $\rho_0^{\max}$ and minimum lapse $\alpha_{\rm min}$. 
Figures~\ref{fig:M1418_m0_snapshots} and \ref{fig:M1418_m1_snapshots} 
show snapshots of equatorial density contours and the velocity 
vector field.

Figure~\ref{fig:M1418_mass_out} shows the rest 
mass of the material outside the apparent horizon $M_{\rm out}$. 
We see that for the unmagnetized case, $M_{\rm out}/M_0$ is settling down 
to an equilibrium value $\approx 0.013$, which is consistent with the upper 
bound 0.04 reported in~\cite{stu03}. The simulation is terminated 
at $t \approx 450M$ since the constraint violations outside the 
apparent horizon increase to 
more than 15\% and so the evolution becomes inaccurate. We suspect that 
the problem can be solved by increasing the grid resolution near the 
central singularity. We plan to investigate this issue in the near future. 
For the magnetized case, $M_{\rm out}/M_0$ drops to 0.018 at the end of our 
simulation. The simulation is terminated at $t\approx 380M$ when 
the constraint violations exceed 15\%. The magnetized run is qualitatively 
very similar to unmagnetized run at this stage. Unlike model M1616, however, 
there is a substantial amount of material left to form 
a disk in this case. Magnetic fields are expected to play an important 
role in the subsequent secular evolution of the disk. The field could drive MHD 
turbulence and generate ultra-relativistic jets. However, the disk 
mass is probably not large enough to produce a short-hard GRB 
in this case~\cite{piran05b}. We have already studied axisymmetric, 
magnetized disk evolution around black holes in~\cite{ssl08}.

\begin{figure}
\includegraphics[width=8cm]{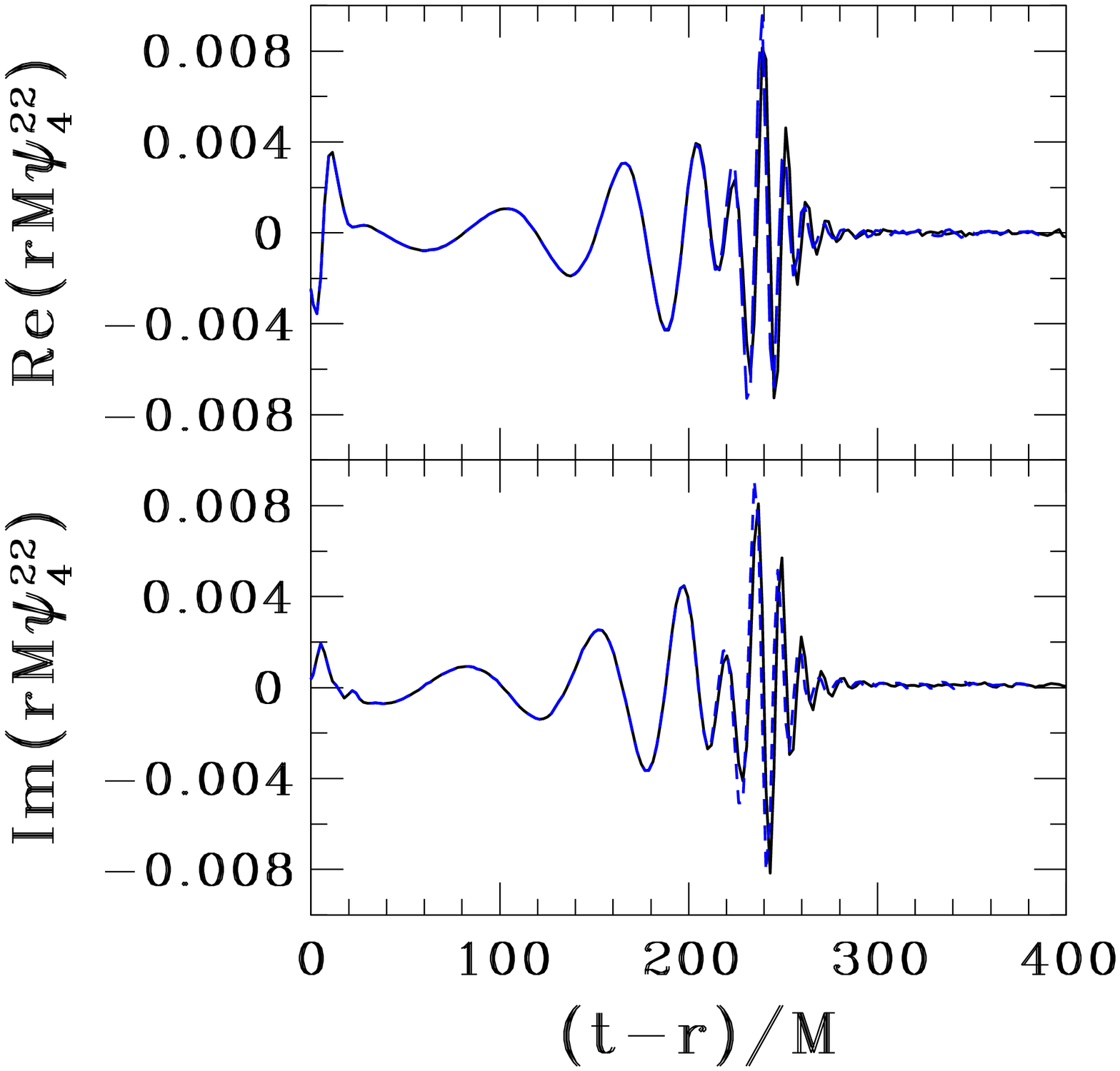}
\caption{Gravitation waveform $\psi_4^{22}(t-r)$ for unmagnetized (black solid 
line) and magnetized (blue solid line) runs of model M1418. Gravitation 
waves are extracted at radius $r=43M$.}
\label{fig:M1418_psi4}
\end{figure}

Figure~\ref{fig:M1418_psi4} shows the gravitational waveforms. We see 
that the waveforms of the unmagnetized and magnetized simulations 
are very close, as we have found for model M1616.

\section{Summary}
\label{sec:discussion}

We have performed a series of new simulations involving the coalescence of 
NSNS binaries in full general relativity, using the CTS quasiequilibrium 
NSNS initial data. We considered three models M1414, M1616 and M1418 
previously studied by Shibata, Taniguchi \& Ury\=u~\cite{stu03}. 
We performed unmagnetized and magnetized simulations for each model. 

We find that our results for the unmagnetized runs agree with those 
in~\cite{stu03}. In particular, the remnant of the merger is 
a hypermassive neutron star for model M1414, a rotating black hole 
with negligible disk for model M1616, 
and a black hole surrounded by a disk with a rest mass less than 
2\% of the total rest mass of the system for model M1418. 
Given our good agreement with the results of Shibata, Taniguchi \& 
Ury\=u~\cite{stu03} for unmagnetized binaries, the results of Anderson et 
al.~\cite{ahllmnpt08a,ahllmnpt08b} remain somewhat puzzling for such 
configurations. 
Our simulations on magnetized NSNSs indicate that the magnetic fields 
can cause observable differences in the dynamics and gravitational 
waveforms after the merger (especially for model M1414). 
However, we find that the magnetic effects are less significant
than those reported in~\cite{ahllmnpt08b}.

For model M1414, the merged remnant consists of a double core rotating 
around the center, and the star pulsates. The motion in this remnant emits 
gravitational radiation. We see observable difference between the magnetized 
and unmagnetized cases in the amplitude of the pulsations.
Gravitational waveforms also show differences in 
amplitude and phase after the merger. We expect that the magnetic 
field will be amplified by differential rotation on a (secular) timescale 
of $\sim 10$ rotation periods. 

For the more massive model M1616, prompt collapse to a black hole 
occurs following 
the merger. For the unmagnetized case, all the material falls into 
the black hole. For the magnetized cases, we try two different initial 
magnetic-field profiles and find that the final result is about the same as 
the unmagnetized case. We find that $\lesssim 10^{-4}$ 
of the total rest mass resides outside the black hole at the end of our 
simulations ($500M$). We see only a slight difference in the amplitude 
of the gravitational radiation between the magnetized and unmagnetized 
cases during the collapse.

For model M1418 consisting of unequal masses, the merged remnant 
also collapses promptly to a black hole 
after the merger, but there is a substantial amount of material left 
to form a disk. We find that the disk mass is $\lesssim 0.013M_0$ for the 
unmagnetized case and $\lesssim 0.018M_0$ for the magnetized case, where 
$M_0$ is the total rest mass of the system. Magnetic fields are crucial 
for the subsequent, secular evolution of the disk. We have previously 
performed long-term, axisymmetric simulations of magnetized disks around 
black holes resulting from the collapse of hypermassive neutron 
stars~\cite{ssl08}. We find that the magnetic fields can cause 
outflows, depending on the EOS and the magnetic field configuration.

In summary, we find that the effects of magnetic fields during and 
shortly after the merger phase are significant but not dramatic. 
We believe that the most important role of magnetic fields are 
on the long-term, secular evolution of the merged remnants 
consisting of a hypermassive neutron star or a black hole surrounded
by a disk, as we have demonstrated in our axisymmetric simulations 
in previous publications~\cite{dlsss06a,dlsss06b,ssl08}. 
This is not to say that a long-term evolution
in 3+1 dimensions is not necessary. For one thing, a 3+1 simulation 
of the remnants will
evolve self-consistently from the NSNS initial data. For the other,
MHD turbulence is expected to be more prominent in three dimensions
than in axisymmetry~\cite{hawley00}, and hence may affect the
dynamics in the long-term evolution. We therefore plan to follow
the long-term evolution of the remnants in 3+1 dimensions in the future.
We thus need to overcome the difficulty 
of growing constraint violations we observe in our black hole evolutions. 
We suspect the 
constraint violations can be controlled by increasing the spatial 
resolution near the singularity. We currently use a resolution of about $M/7$ 
near the puncture in our simulations, which is a much lower resolution than the 
resolutions used in most recent binary black hole simulations. Higher 
resolution is also required in the case of model M1414 to resolve the MHD 
instabilities. We also plan to perform simulations of NSNS binaries with 
larger initial separation. This will allow us to compute the gravitational 
waveforms with more cycles so that they can be matched with the 
post-Newtonian waveforms.

\acknowledgments

Numerical computations
were performed on the {\tt abe} cluster at the National Center 
for Supercomputing Applications at
the University of Illinois at Urbana-Champaign (UIUC).
This work was supported in part by NSF
Grants PHY02-05155, PHY03-45151, and PHY06-50377 as well as NASA
Grants NNG04GK54G and NNX07AG96G at UIUC.

\bibliography{paper}

\begin{thebibliography}{66}
\bibitem[{fn()}]{fn}
\bibinfo{note}{Note that it is the total rest mass that is conserved during the
  evolution. The rest mass of each NS in the binary is obtained by solving the
  Tolman-Oppenheimer-Volkov (TOV) equations with the same EOS and central
  density as those in~\cite{ahllmnpt08a}.}
\end{thebibliography}

\end{document}